\documentclass[traditabstract]{aa}  

\usepackage{graphicx}
\usepackage{txfonts}
\usepackage{natbib}
\usepackage{url}
\usepackage{multirow}
\usepackage{rotating}

\begin{document}

\newcommand{\zabs}{\ensuremath{z_{\rm abs}}}
\newcommand{\zem}{\ensuremath{z_{\rm em}}}
\newcommand{\HH}{\mbox{H$_2$}}
\newcommand{\HD}{\mbox{HD}}
\newcommand{\DD}{\mbox{D$_2$}}
\newcommand{\CO}{\mbox{CO}}
\newcommand{\dla}{damped Lyman-$\alpha$}
\newcommand{\Dla}{damped Lyman-$\alpha$}
\newcommand{\lya}{Ly\,$\alpha$}
\newcommand{\lyb}{Ly\,$\beta$}
\newcommand{\Ha}{H\,$\alpha$}
\newcommand{\Hb}{H\,$\beta$}
\newcommand{\lyg}{Ly\,$\gamma$}
\newcommand{\ArI}{\ion{Ar}{i}}
\newcommand{\CaII}{\ion{Ca}{ii}}
\newcommand{\CI}{\ion{C}{i}}
\newcommand{\CII}{\ion{C}{ii}}
\newcommand{\CIV}{\ion{C}{iv}}
\newcommand{\ClI}{\ion{Cl}{i}}
\newcommand{\ClII}{\ion{Cl}{ii}}
\newcommand{\CrII}{\ion{Cr}{ii}}
\newcommand{\CuII}{\ion{Cu}{ii}}
\newcommand{\DI}{\ion{D}{i}}
\newcommand{\FeI}{\ion{Fe}{i}}
\newcommand{\FeII}{\ion{Fe}{ii}}
\newcommand{\HI}{\ion{H}{i}}
\newcommand{\MgI}{\ion{Mg}{i}}
\newcommand{\MgII}{\ion{Mg}{ii}}
\newcommand{\MnII}{\ion{Mn}{ii}}
\newcommand{\NI}{\ion{N}{i}}
\newcommand{\NII}{\ion{N}{ii}}
\newcommand{\NV}{\ion{N}{v}}
\newcommand{\NiII}{\ion{Ni}{ii}}
\newcommand{\OI}{\ion{O}{i}}
\newcommand{\OII}{\ion{O}{ii}}
\newcommand{\OIII}{\ion{O}{iii}}
\newcommand{\OVI}{\ion{O}{vi}}
\newcommand{\PII}{\ion{P}{ii}}
\newcommand{\PbII}{\ion{Pb}{ii}}
\newcommand{\SI}{\ion{S}{i}}
\newcommand{\SII}{\ion{S}{ii}}
\newcommand{\SiII}{\ion{Si}{ii}}
\newcommand{\SiIV}{\ion{Si}{iv}}
\newcommand{\TiII}{\ion{Ti}{ii}}
\newcommand{\ZnII}{\ion{Zn}{ii}}
\newcommand{\AlII}{\ion{Al}{ii}}
\newcommand{\AlIII}{\ion{Al}{iii}}
\newcommand{\Ho}{\mbox{H$_0$}}
\newcommand{\angstrom}{\mbox{{\rm \AA}}}
\newcommand{\abs}[1]{\left| #1 \right|} 
\newcommand{\avg}[1]{\left< #1 \right>} 
\newcommand{\kms}{\ensuremath{{\rm km\,s^{-1}}}}
\newcommand{\cmsq}{\ensuremath{{\rm cm}^{-2}}}
\newcommand{\qso}{J1135$-$0010}
\newcommand{\qsolong}{SDSS\,J113520.39$-$001053.56}
\newcommand{\nhi}{n_{\rm HI}}

\newcommand{\iap}{UPMC-CNRS, UMR7095, Institut d'Astrophysique de Paris, F-75014 Paris, France --  
\email{noterdaeme@iap.fr}}
\newcommand{\iucaa}{Inter-University Centre for Astronomy and Astrophysics, 
Post Bag 4, Ganeshkhind, 411\,007 Pune, India} 
\newcommand{\uchile}{Departamento de Astronom\'ia, Universidad de Chile, 
Casilla 36-D, Santiago, Chile}
\newcommand{\dark}{Dark Cosmology Centre, Niels Bohr Institute, Copenhagen University, 
Juliane Maries Vej 30, 2100 Copenhagen O, Denmark}
\newcommand{\eso}{European Southern Observatory, Alonso de C\'ordova 3107, 
Vitacura, Casilla 19001, Santiago 19, Chile}
\newcommand{\inaf}{Instituto Nazionale di Astrofisica, Osservatorio Astronomico 
di Brera, Via Bianchi 46 I-23807 Merate, Italy}
\newcommand{\yale}{Department of Astronomy, Yale University, P.O. Box 208101, New Haven, CT 06520-8101, USA}
\newcommand{\okc}{Oskar Klein Centre, Dept. of Astronomy, Stockholm University, SE-10691 AlbaNova, Stockholm, Sweden}

\title{Discovery of a compact gas-rich DLA galaxy at $z=2.2$: \\
evidences for a starburst-driven outflow
\thanks{Based on data obtained with MagE at the Clay telescope of the Las Campanas Observatory (CNTAC Prgm.~ID~2011B-90) 
and X-shooter/UVES at the Very Large Telescope of the European Southern Observatory 
(Prgm.~ID~286.A-5044 and 385.A-0778).}}

\author{P. Noterdaeme\inst{1,2}, P. Laursen\inst{3,4}, P. Petitjean\inst{1}, S.~D. Vergani\inst{5}, 
M.-J. Maureira\inst{2,6}, C. Ledoux\inst{7},\\ J.~P.~U. Fynbo\inst{4}, S. L\'opez\inst{2} and R. Srianand\inst{8}}
\authorrunning{PN et al.}

\institute{\iap \and \uchile \and \okc \and \dark \and \inaf \and \yale \and \eso \and \iucaa}

\date{Received /Accepted}

\abstract{
We present the detection of \lya, [\OIII] and \Ha\ emission associated with an extremely 
strong damped Lyman-$\alpha$ (DLA) system ($N(\HI)=10^{22.10}$~\cmsq) at $z=2.207$ towards the 
quasar \qsolong. This is the largest \HI\ column density ever measured along a QSO line of sight, 
though typical of what is seen in GRB-DLAs. This absorption system also classifies as ultra-strong 
\MgII\ system with $W_{\rm r}^{\lambda2796}\simeq3.6$~{\AA}. 
The mean metallicity of the gas ($[$Zn/H$]$~=~$-$1.1) and 
dust depletion factors ($[$Zn/Fe$]$~=~0.72, $[$Zn/Cr$]$~=~0.49) are 
consistent with (and only marginally larger than) the mean values found in the general QSO-DLA population. 

The [\OIII]-H$\alpha$ emitting region has a very small impact parameter with respect to the QSO line 
of sight, $b\approx 0.1$\arcsec\ (0.9~kpc proper distance),
and is unresolved.
From the \Ha\ line, we measure a significant star formation rate SFR~$\approx$~25~M$_{\odot}$\,yr$^{-1}$ (uncorrected for dust). 
The shape of the \lya\ line is double-peaked, which is the signature of resonant scattering of 
\lya\ photons, and the \lya\ emission is spatially extended.
More strikingly, the blue and red \lya\ peaks arise from distinct regions extended over 
a few kpc on either side of the star-forming region. We propose that this is the consequence of \lya\ transfer 
in outflowing gas. The presence of starburst-driven outflows is also in agreement with the large SFR 
together with a small size and low mass of the galaxy ($M_{\rm vir}\sim10^{10}$~M$_{\odot}$).
From putting constraints on the stellar UV continuum luminosity of the galaxy, we estimate 
an age of at most a few $10^7$ yr, again consistent with a recent starburst scenario. 

We interpret the data as the observation of a young, gas rich, compact starburst galaxy, 
from which material is expelled through collimated winds powered by the vigorous star formation activity. 
We substantiate this picture by modelling the radiative transfer of \lya\ photons in the galactic 
counterpart. Though 
our model (a spherical galaxy with bipolar outflowing jets) is a simplistic representation of 
the true gas distribution and velocity field, the agreement between the observed and simulated properties is 
particularly good (spectral shape and width of the Lyman-$\alpha$ emission, spatial configuration, escape fraction 
as well as absorption kinematics, \HI\ column density and dust reddening).

Finally, we propose that selecting DLAs with very high \HI\ column density may be an efficient way to 
detect star-forming galaxies at small impact parameters from the background QSO lines of sight.}

\keywords{quasars: absorption lines -- galaxies: ISM -- galaxies: high-redshift -- 
          galaxies: star formation -- quasars: individual: \qsolong}
\maketitle

\section{Introduction}

Detecting and performing detailed studies of high-$z$ (i.e. $z>2$) galaxies is observationally a 
challenging task. Several 
observational strategies have emerged in the past 15 years, each of them targeting a subset of the 
population of high-$z$ galaxies --which is named after the selection technique-- 
with some overlap in their properties. 
For example, because of their selection in flux-limited surveys, 
Lyman-break galaxies \citep[LBGs,][]{Steidel03} tend to probe the bright end of the 
high-redshift galaxy luminosity function. A fraction of LBGs are also Lyman-$\alpha$ emitters 
\citep[LAEs, see e.g.][]{Cowie98}. However, as 
most LAEs are selected from the \lya\ emission line, 
they tend to sample lower luminosities better than colour-selected LBGs \citep[][]{Fynbo03,Kornei10}. 
Indeed, recent deep searches of LAEs seem to have reached the faint end of the high-$z$ luminosity function 
\citep[e.g.][]{Rauch08,Grove09}. 

In turn, the detection of damped Lyman-$\alpha$ systems (DLAs) in absorption against 
bright background sources (such as quasars or gamma ray burst (GRB) afterglows) has the 
major advantage that it does not depend on the luminosity 
of the associated galaxies but on the cross-section of the neutral gas. These systems are
selected from their large neutral hydrogen column densities, $N(\HI)\ge 2\times 10^{20}$~\cmsq\ 
\citep[][]{Wolfe86} and represent the main reservoir of neutral hydrogen at high redshift 
\citep[e.g.][]{Noterdaeme09dla}. 
The presence of associated heavy elements \citep[e.g.][]{Prochaska02} and molecules 
\citep[e.g.][]{Petitjean00,Ledoux03,Noterdaeme08}, the evolution with 
redshift of the \HI\ mass density in DLAs \citep{Peroux03, Prochaska05, Noterdaeme09dla} and 
the detectability of DLAs over a wide range of redshift make them the appropriate laboratories for 
studying the cosmological evolution of star formation activity in a luminosity-unbiased way. 

Although observational studies of DLAs have been pursued over 25 
years, an important question that remains unanswered yet is the connection 
between DLAs and star-forming galaxies. Because of their selection, most DLA galaxies probably 
lay on the faint end of the luminosity function \citep[e.g.][]{Fynbo08}. Nevertheless, DLA
galaxies can provide substantial contributions to the global SFR density at high redshifts 
\citep{Wolfe03,Srianand05,Rauch08,Rahmani10}. 
In addition, it has long been debated whether the kinematics of the absorbing gas, studied through the velocity 
profiles of metal absorption lines, is more representative of that of large rotating galactic discs 
\citep{Prochaska97} or small low-mass galactic clumps \citep{Ledoux98} that build up hierarchically 
galaxies known today \citep[e.g.][]{Haehnelt98}. The detection of DLA galactic 
counterparts is therefore of first importance to shed light on the nature of DLAs.

Until recently, searches for direct emission from high redshift intervening DLA galaxies 
have resulted mostly in non-detections \citep[e.g.][as well as several unpublished works]
{Bunker99,Kulkarni00, Kulkarni06, Lowenthal95,Christensen09} with few cases spectroscopically 
confirmed through the detection of Ly-$\alpha$ emission \citep{Moller02,Moller04}. 
Thankfully, improved selection strategies have emerged from the understanding that 
correlations between metallicity, mass and luminosity observed for field galaxies
also apply to DLA galaxies \citep{Moller04,Ledoux06a,Fynbo08}.
This led to several new detections (\citealt{Fynbo10,Fynbo11}, Krogager et al., in prep). 

In this paper, we present an interesting case with the detection of \lya, \Ha\ and [\OIII] emission associated 
with an extremely strong DLA system at $z=2.207$ towards the quasar \qsolong\ (hereafter \qso). In Sect.~\ref{obs}, 
we detail our observations performed with MagE and X-shooter and describe UVES data retrieved from 
the archive. 
The properties of the absorbing gas along the QSO line of sight are studied in Sect.~\ref{abs}. We analyse 
and discuss the emission properties of the galaxy in Sect.\ref{em}. 
We use radiative transfer (RT) modelling of the galactic \lya\ emission to test our results in Sect.~\ref{model}.
Finally, we present our conclusions in Sect.~\ref{concl}. In this work, we use standard $\Lambda$CDM cosmology with 
$\Omega_{\Lambda}=0.73$, $\Omega_m=0.27$ and \Ho~=~70~\kms\,Mpc$^{-1}$ \citep{Komatsu11}.

\section{Observations and data reduction \label{obs}}

\subsection{Clay/MagE}
\begin{figure}
\centering
\includegraphics[bb=133 234 479 334,clip=,width=0.95\hsize]{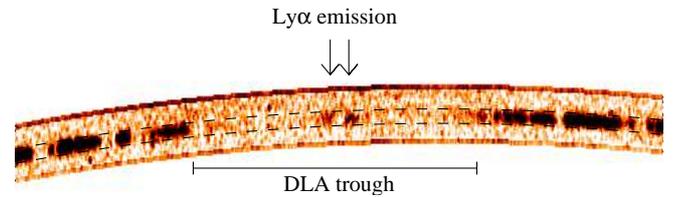}
\caption{Portion of the 2D MagE spectrum. The double peaked Lyman-$\alpha$ emission is clearly seen in the DLA 
trough, where the flux of the quasar is completely absorbed. The dashed lines show the interpolated QSO trace 
in this region. \label{lya_mage}}
\end{figure}

We observed \qso\ ($\zem=2.89$) on February 12, 2011 with the Magellan Echellette 
spectrograph \citep[MagE;][]{Marshall08} 
mounted on the 6.5\,m Clay telescope at Las Campanas Observatory. MagE is a moderate-resolution long-slit 
spectrograph covering at once the range 300\,nm to 1\,$\mu$m. The spectrograph has been designed to have 
excellent throughput in the blue with an overall efficiency (telescope plus instrument) higher than 15\% at 3900~{\AA}. 
In spite of very bad weather conditions, we could obtain a 1\,h spectrum of the quasar with a slit width of 2 arcsec 
(the length is fixed to 10 arcsec). 
This allowed us to cover a 17$\times$85~kpc region around the QSO (proper distance at $z=2.207$) in a single exposure. 
We detected a double-peaked Lyman-$\alpha$ emission at the systemic velocity of metal absorption lines 
(see Fig.~\ref{lya_mage}). 
The significance of the detection (using a 30$\times$15 pixels aperture) is above 7$\sigma$ for each peak independently. 
From Fig.~\ref{lya_mage} it is also clear that the emission falls well within the extent of the QSO trace, 
indicating a small impact parameter ($<1$ arcsec).
Unfortunately, the bad observing conditions prevented us from obtaining additional spectra of this 
target at other slit angles and/or 
narrower slit width.
Nevertheless, the detection in a single exposure demonstrates that MagE is particularly well suited to search for faint \lya\ 
emission at high redshift. The length of the slit and the good spatial sampling (0.3~arcsec/pixel) 
allows one to efficiently cover the expected position of DLA galaxies as well as to measure their impact parameter using 
spectra obtained at 2-3 slit position angles \citep[see e.g. Fig.~1 of][]{Fynbo10}. Similar searches with MagE are probably 
also possible even further in the blue, close to the atmospheric cutoff.

\subsection{VLT/X-shooter}
Follow-up observations were performed in service mode with X-shooter \citep{Vernet11} at the 
European Southern Observatory (ESO) Very Large Telescope (VLT). 
The observations were performed under good seeing ($0.6-0.8$~arcsec), at low airmass, and in dark 
time (new moon) on May 1$^{\rm st}$ and 2$^{\rm nd}$, 2011 using Director Discretionary Time. 
X-shooter covers the full wavelength range from 300~nm to 2.5\,$\mu$m at intermediate spectral resolution 
thanks to the simultaneous use of three spectroscopic arms (UVB, VIS and NIR). 
Therefore, X-shooter is a suitable instrument not only for detecting \lya\ down 
to the atmospheric cutoff, but also for detecting rest-frame optical lines 
(this work and \citealt{Fynbo10,Fynbo11}).
We used the long-slit mode to observe the quasar at three position 
angles (PAs) for slightly more than 1\,h each. To optimise the sky subtraction in the NIR, the nodding mode was used 
following a ABBA scheme with a nod throw of 5~arcsec and an additional jitter box of 1~arcsec. The log of the 
X-shooter observations is given in Table~\ref{log}.

The 2D and 1D spectra were extracted using the X-shooter pipeline v1.3.7 \citep{Goldoni06}. 
We remarked a constant 0.5 pixel shift for exposures from the VIS arm (resp. UVB) towards redder 
(resp. bluer wavelengths)\footnote{This could originate from a thermal drift 
between calibration and science exposures, combined with possible flexure residuals or from a problem 
with the current X-shooter pipeline.}.
The shift was clearly apparent in the overlapping UVB-VIS region and 
by comparing both UVB and VIS spectra with a high-resolution UVES spectrum (see below) of the same object.
This was further confirmed by comparing the observed (VLT rest-frame) wavelengths of telluric absorption lines 
(at $\sim$~7000, 7600 and 9400~{\AA}) with a synthetic spectrum from HITRAN \citep{Rothman05}. Although 
this has little consequence on the science results, we corrected the wavelength scales accordingly. 

The 1D spectra were then combined altogether using a sliding window and weighting each 
pixel by the inverse of its variance. 
The spectrum has been corrected for relative spectral response of the instrument using standard 
stars as reference. Subsequent constant rescaling of the UVB and VIS spectra was necessary to 
match the SDSS spectrum. The NIR spectrum was scaled to match the VIS spectrum in the overlapping region around $1 \mu$m. 
We then corrected the overall spectrum for Galactic extinction using dust map from \citet{Schlegel98}.
The observed flux at $\lambda_{\rm eff}=1.235$\,$\mu$m corresponds very well (within 10\%, see also Fig.~\ref{sed}) to the 
J-band magnitude extracted from 2MASS images \citep[J~=~17.19,][]{Schneider10}. 

The spectral resolution of the combined spectrum in the visible, $R\approx9700$, was measured directly from the 
width of telluric absorption lines. This 
is higher than the nominal resolution expected for the setting used ($R_{\rm nom}=6700$) and is due to the seeing 
being smaller than the slit width. We estimated the UVB and NIR resolution following \citet{Fynbo11} and 
obtained $R\approx5200$ and $R\approx7100$, respectively. We checked that the resolution in the UVB is consistent 
with the width of the narrowest lines in the spectrum.

\begin{table*}[!ht]
\centering
\caption{Log of X-shooter observations \label{log}}
\begin{tabular}{r c c c c c c}
\hline
\hline
{\large \strut} PA       & Observing date & DIMM seeing    & Air Mass & \multicolumn{3}{c}{Exposure time (s)} \\
         &                &   (arcsec)     &          & UV           & Vis          & NIR                   \\ 
\hline
0\degr   & 01/05/2011     &  0.76$\pm$0.06 & 1.1      & 4$\times$1000 & 4$\times$965 & 4$\times$2$\times$480 \\
60\degr  & 01/05/2011     &  0.59$\pm$0.07 & 1.2      & 4$\times$1000 & 4$\times$965 & 4$\times$2$\times$480 \\
$-$60\degr & 02/05/2011   &  0.80$\pm$0.09 & 1.1      & 4$\times$1000 & 4$\times$965 & 4$\times$2$\times$480 \\
\hline
\end{tabular}
\end{table*}

\subsection{VLT/UVES}
Finally, we retrieved a single 4760\,s exposure high-resolution spectrum from the ESO 
archive (Prgm. ID.: 385.A-0778(A), PI: C. P\'eroux), 
which was obtained using the VLT Ultraviolet and Visual Echelle Spectrograph \citep[UVES;][]{Dekker00} with 
both blue and red arms with Dichroic 1 and a binning $1\times1$. We reduced the spectrum using the UVES 
pipeline v4.9.5 based on the ESO Common Pipeline Library system. The seeing varied between 1.01 and 1.37\arcsec\ 
during the exposure, yielding a resolving power of $R\approx 47\,250$ in the blue and $R\approx 44\,300$ in the red with 
the slit width of 1\arcsec. The signal-to-noise ratio of the UVES spectrum is S/N$\approx$6 per pixel at 5000\,{\AA}, i.e. 
more than a factor of 10 lower than the SNR in the X-shooter spectrum at the same wavelength. However, its spectral resolution 
is about 5 times higher. While most of the paper is based on X-shooter data, the availability 
of a UVES spectrum will help us to check the velocity spread and structure of the absorption lines as 
well as detect saturation of the latter.
 


\section{Properties of the absorbing gas \label{abs}}

In this section, we study the properties of the absorbing gas along the QSO line of sight. 
We analysed the DLA system at $z=2.207$ towards \qso\ using standard Voigt-profile fitting 
techniques to derive the column densities of different species. Atomic data were taken from \citet{Morton03} 
and abundances are given with respect to solar values \citep{Asplund09} with 
[X/H]~$\equiv \log ($X/H$)-\log ($X/H$)_{\odot}$.  

\subsection{Neutral gas content}

\begin{figure}
\centering
\includegraphics[angle=90,bb=70 70 350 760,clip=,width=\hsize]{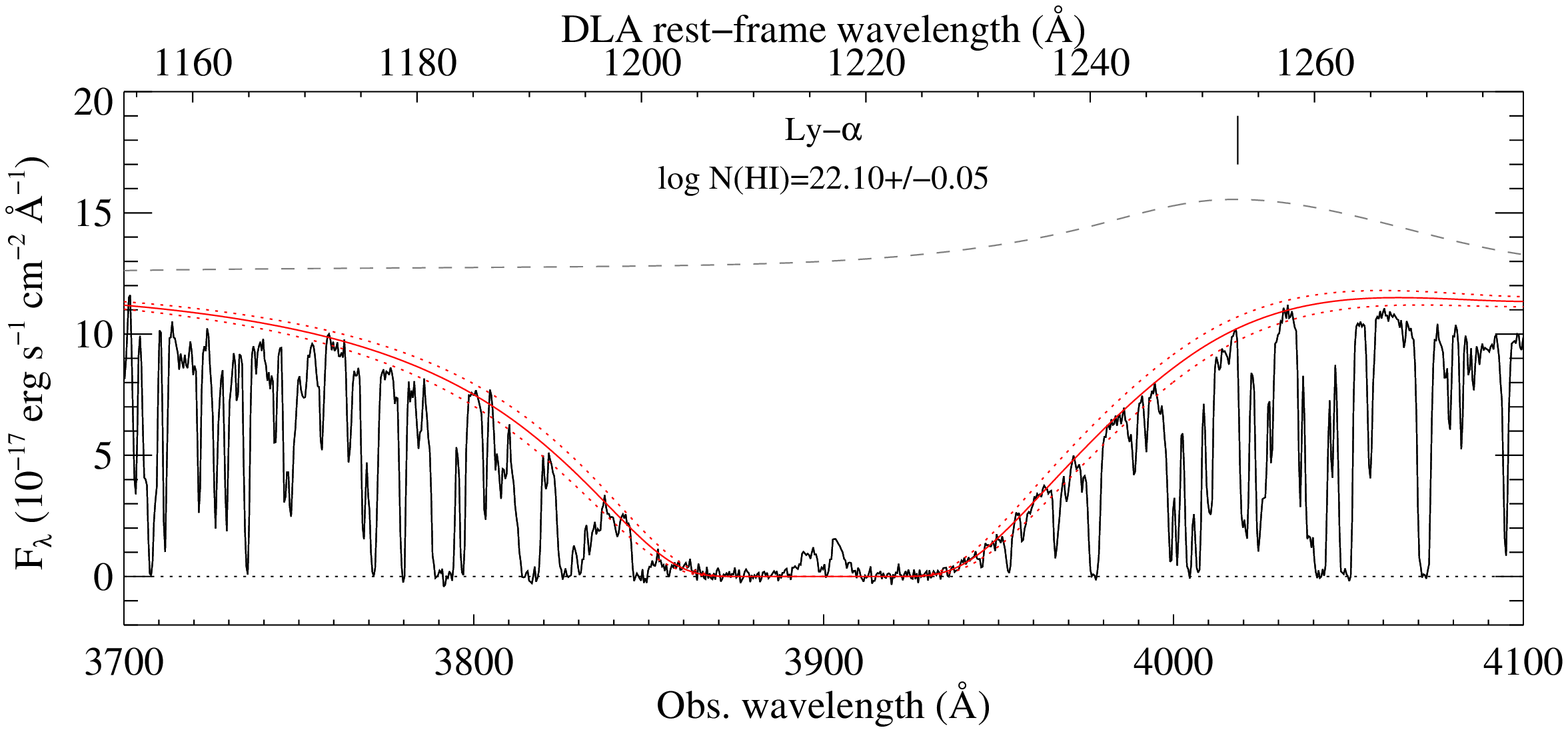}
\includegraphics[angle=90,bb=169 70 294 760,clip=,width=\hsize]{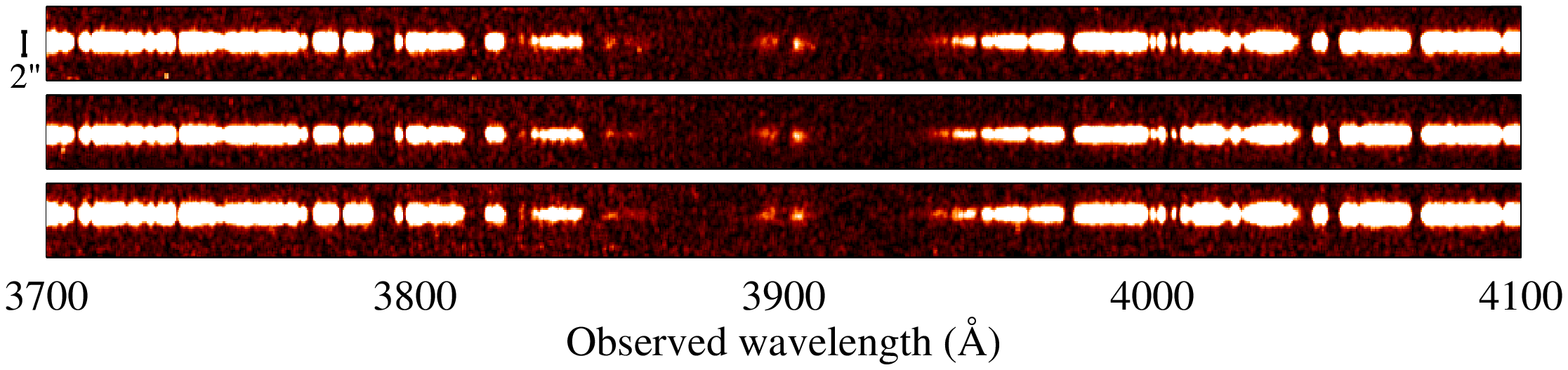}
\includegraphics[angle=90,bb=10 70 196 760,clip=,width=\hsize]{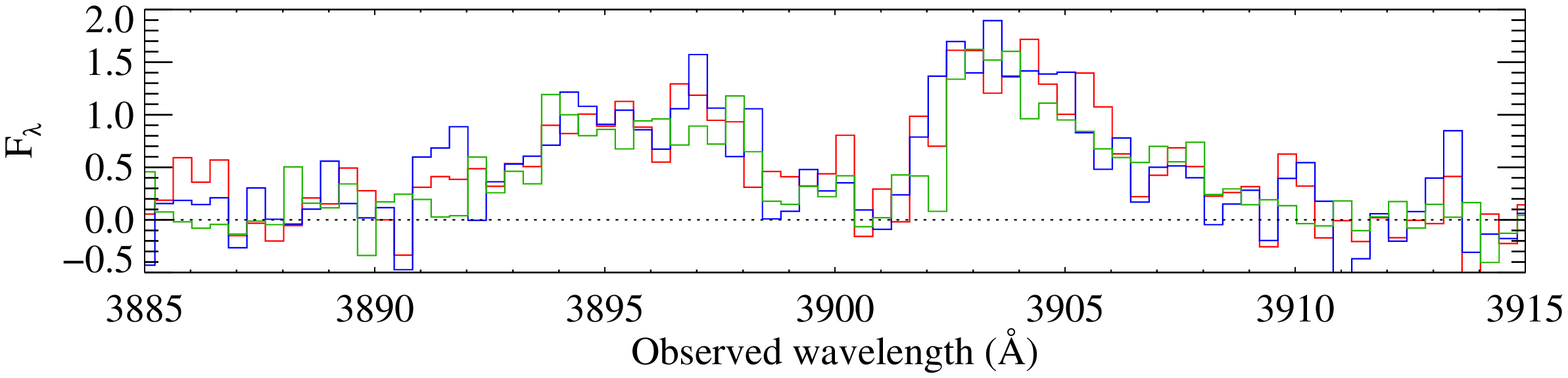}
\caption{{\sl Top}: One dimensional X-shooter spectrum around the DLA region (black). 
The best Voigt-profile fit of the DLA absorption is 
over-plotted in red together with the associated error (dotted 
red). The unabsorbed QSO continuum is represented by the dashed grey curve. Note the 
presence of the \lyb+\OVI\ emission line at $\lambda\sim4020$~{\AA} (indicated 
by a short vertical line).
{\sl Middle}: The corresponding three two-dimensional spectra.  
From top to bottom: PA=0\degr,60\degr and -60\degr. The short vertical segment indicates 
the spatial scale of the y-axis.
{\sl Bottom}: The three 1D spectra in the region around the double-peaked 
Ly-$\alpha$ emission (green: PA=0\degr; blue: PA=60\degr; red: PA=-60\degr).
\label{lya}}
\end{figure}

From fitting the damped Lyman-$\alpha$ absorption line, we measure the total column density of 
atomic hydrogen to be $\log N(\HI) (\cmsq)=22.10\pm0.05$ (see top panel of Fig.~\ref{lya}). Though 
the exact placement of the QSO continuum is somewhat complicated by the presence of the QSO \lyb+\OVI\ 
emission line, the \HI\ column density is very well constrained by the core of the DLA profile.
This is the largest \HI\ column density ever measured in a QSO absorber at high-$z$. 

\subsection{Metal and dust content \label{sec_met}}

\begin{figure}[!ht]
\centering
\renewcommand{\tabcolsep}{1pt}
\begin{tabular}{cc}
\includegraphics[bb=219 244 393 629, clip=, angle=90, width=0.45\hsize]{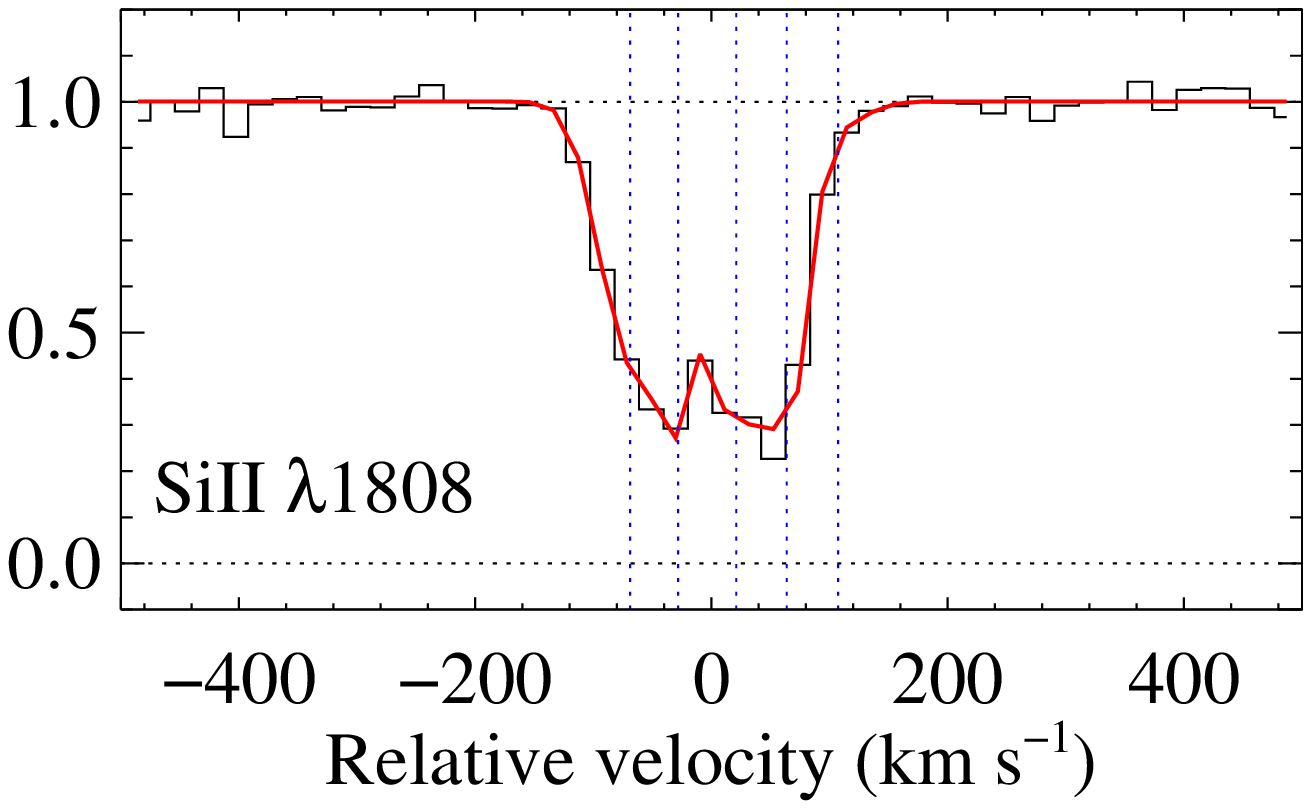}   &  
\includegraphics[bb=219 244 393 629, clip=, angle=90, width=0.45\hsize]{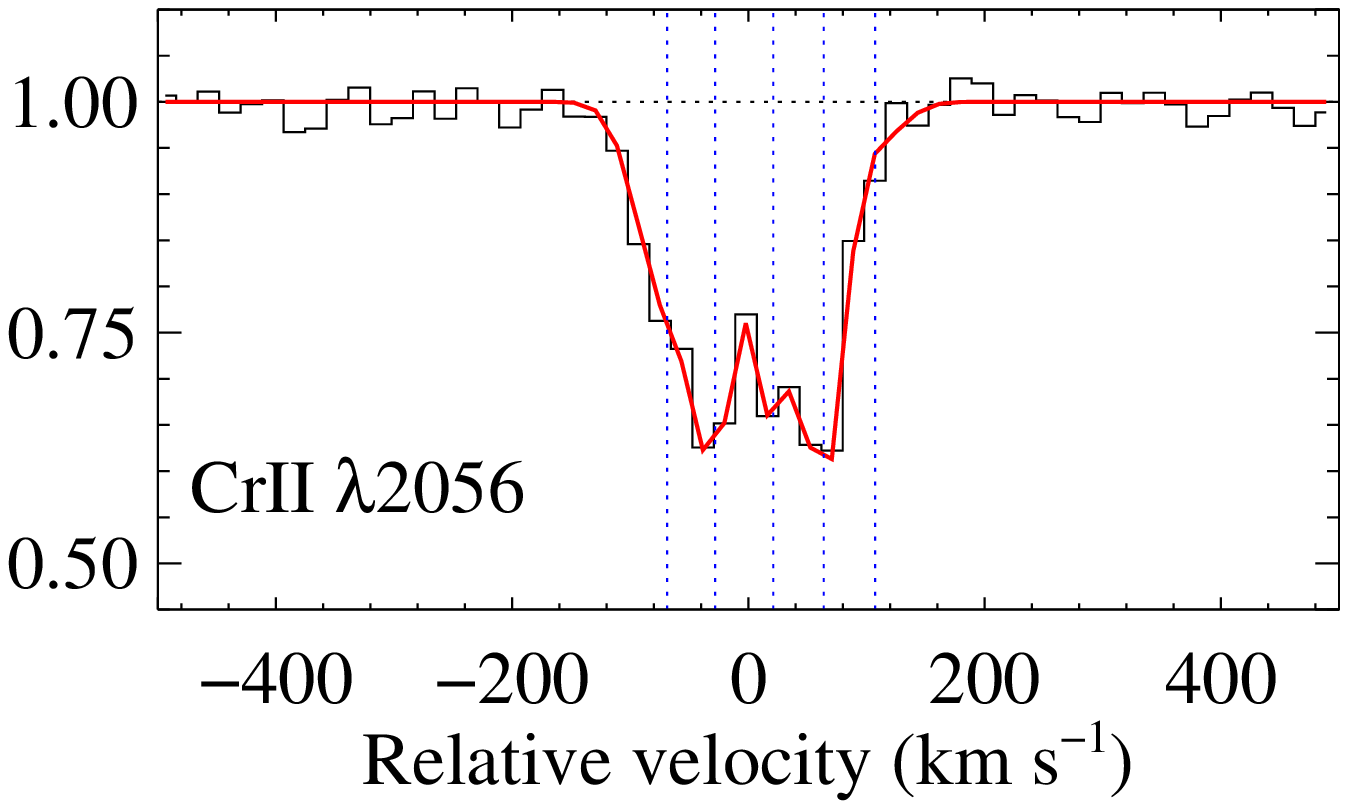}   \\ 
\includegraphics[bb=219 244 393 629, clip=, angle=90, width=0.45\hsize]{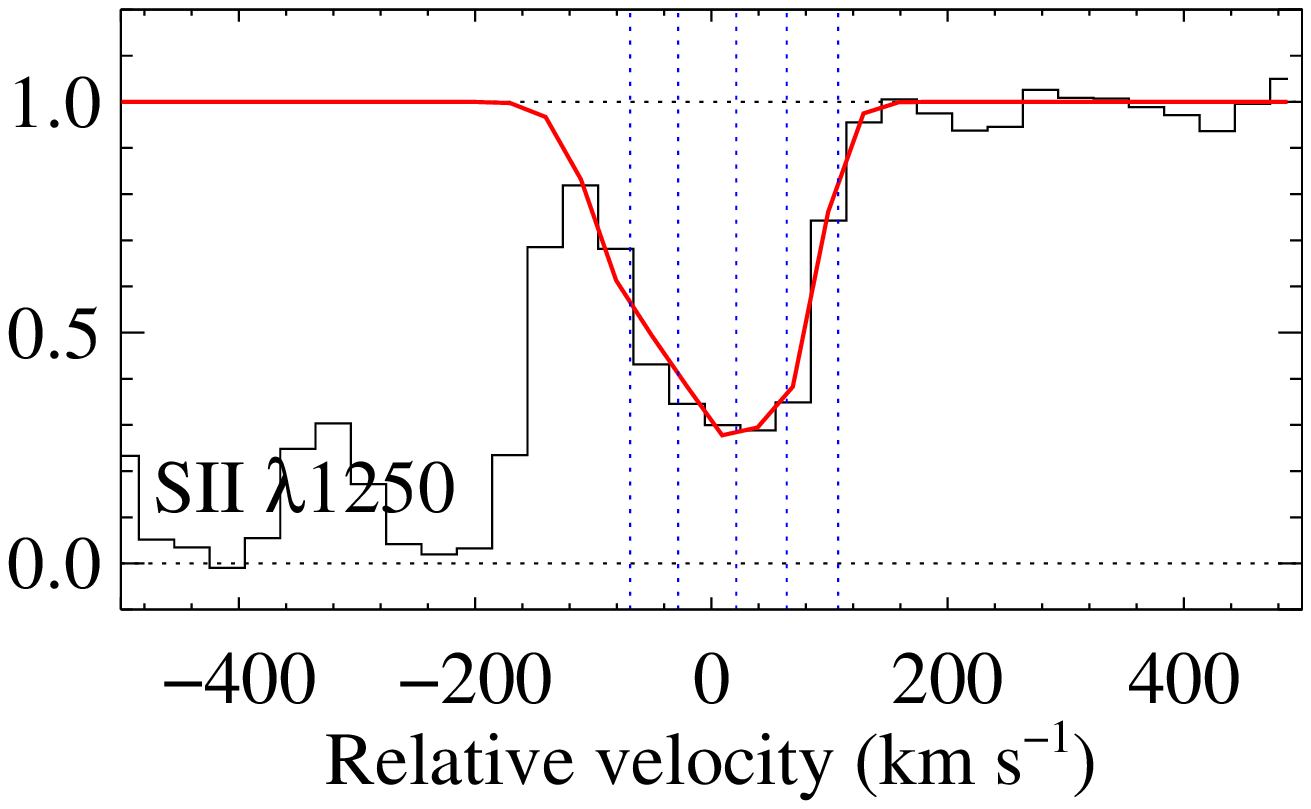}    &  
\includegraphics[bb=219 244 393 629, clip=, angle=90, width=0.45\hsize]{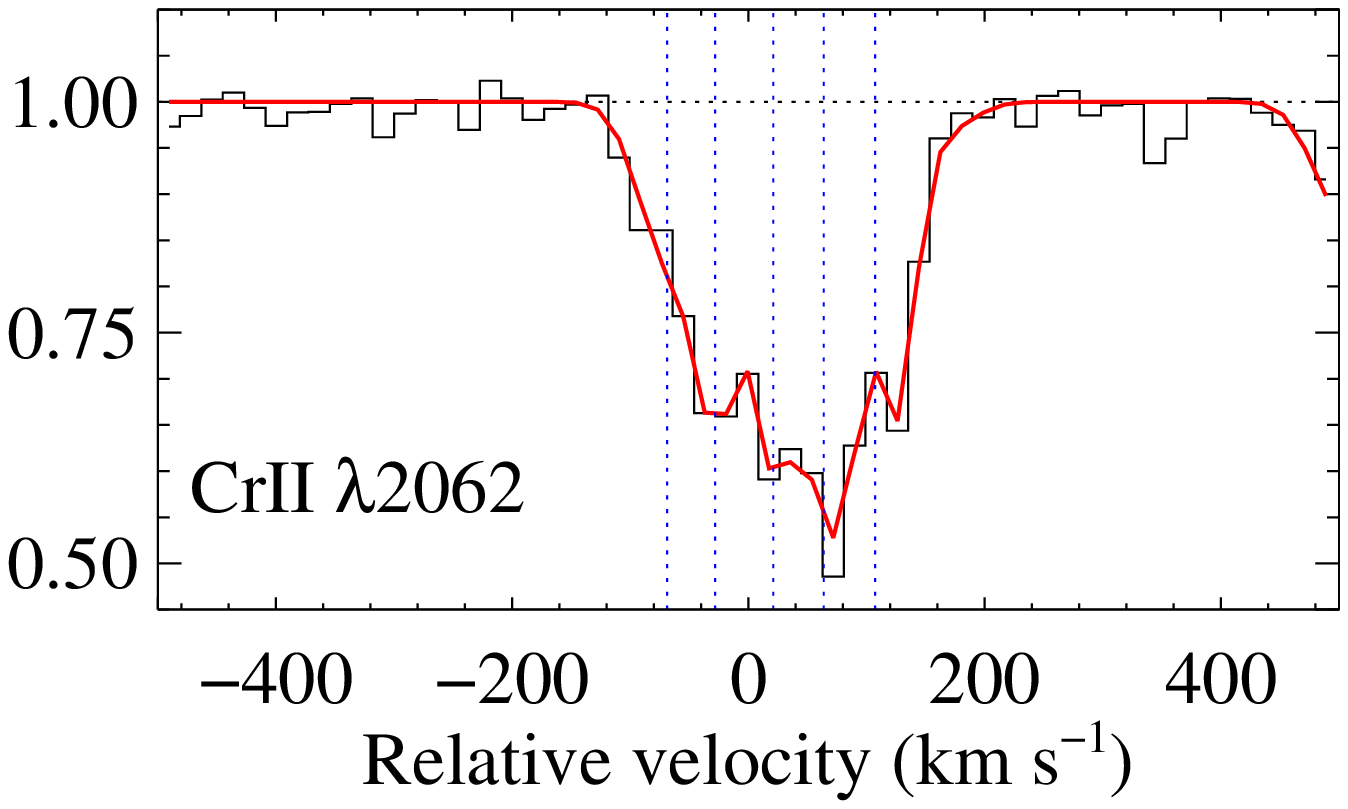}   \\ 
\includegraphics[bb=219 244 393 629, clip=, angle=90, width=0.45\hsize]{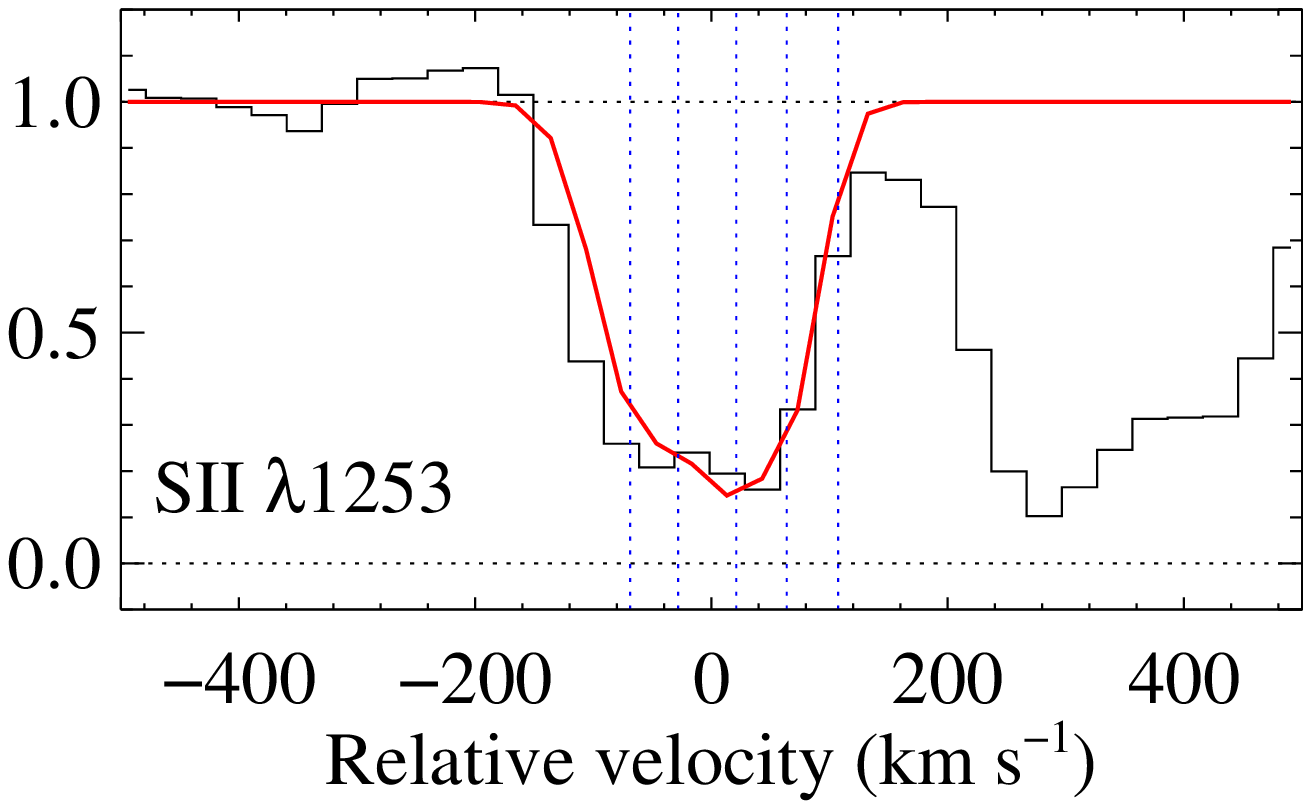}    &  
\includegraphics[bb=219 244 393 629, clip=, angle=90, width=0.45\hsize]{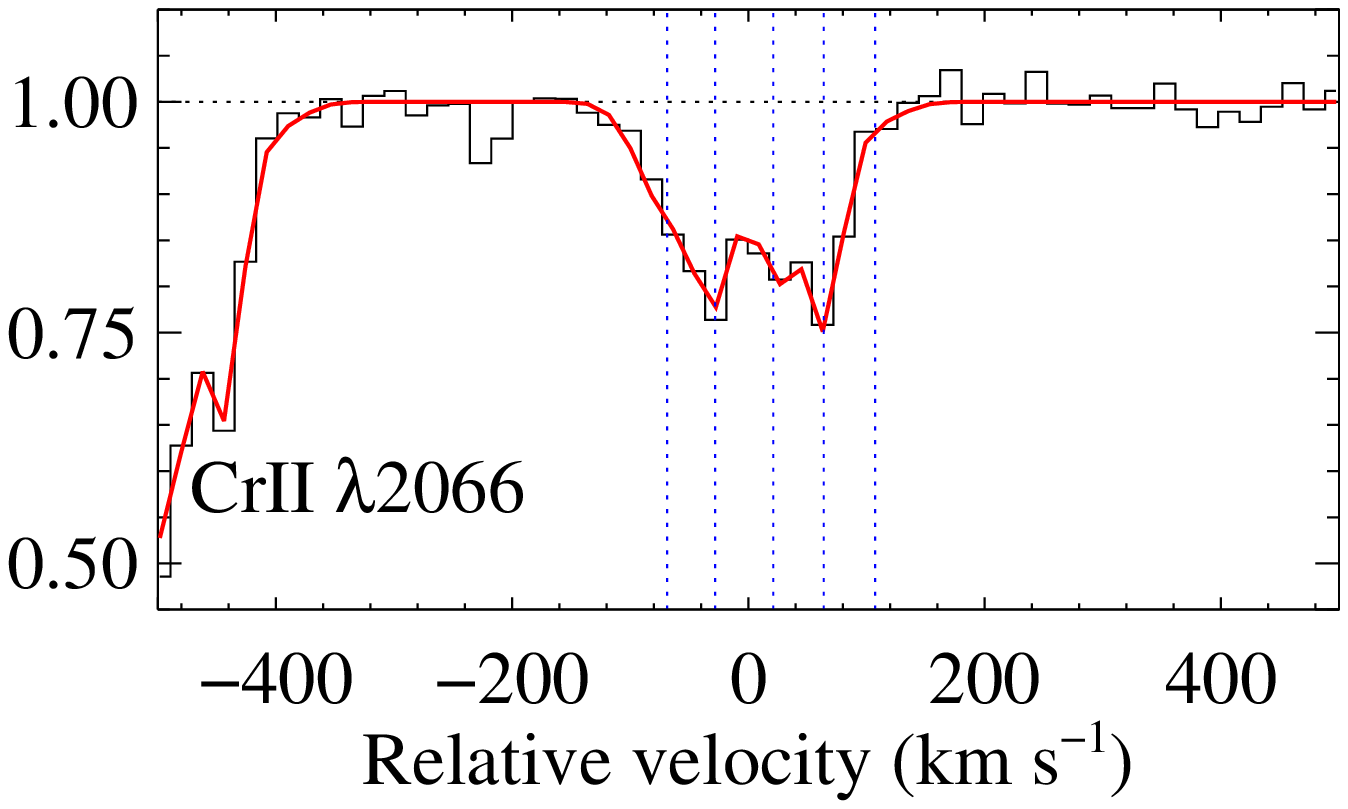}   \\
\includegraphics[bb=219 244 393 629, clip=, angle=90, width=0.45\hsize]{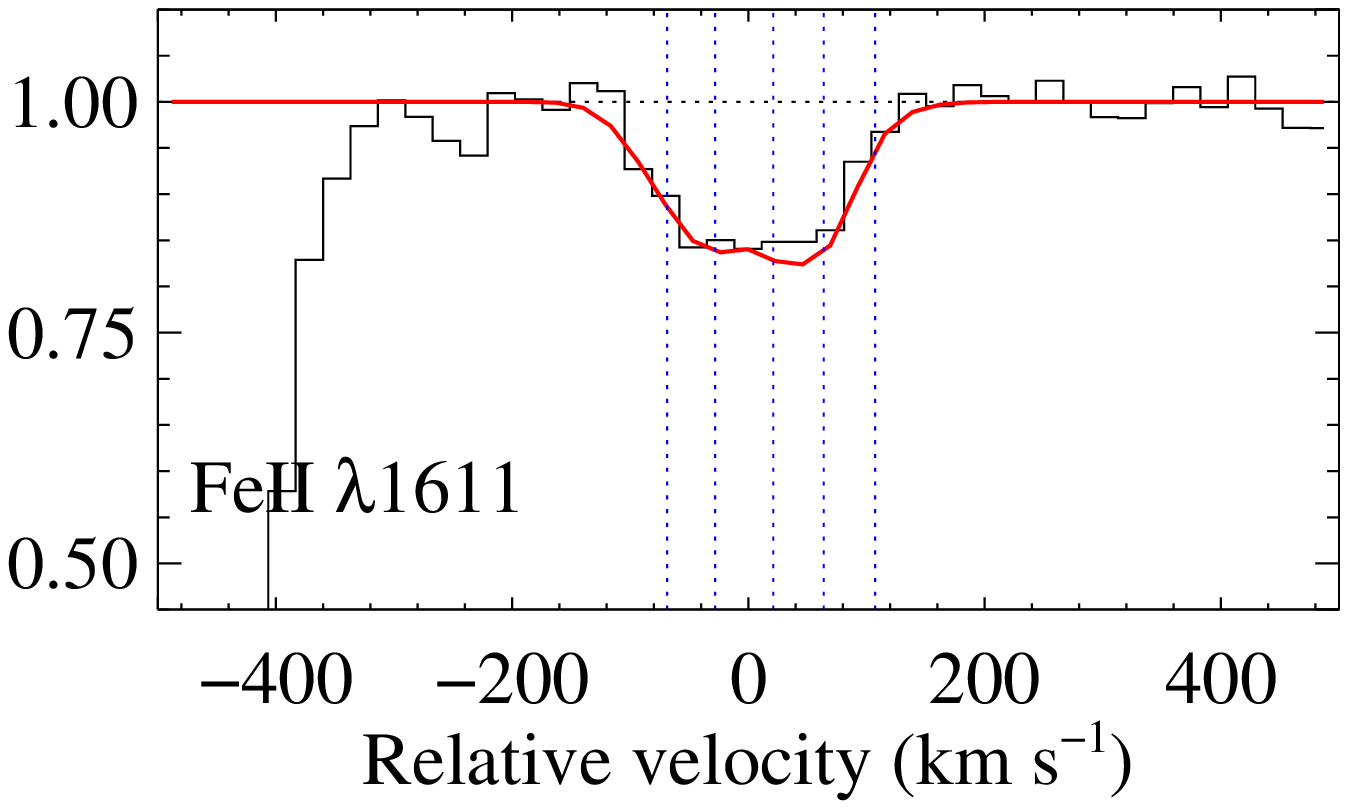} &
\includegraphics[bb=219 244 393 629, clip=, angle=90, width=0.45\hsize]{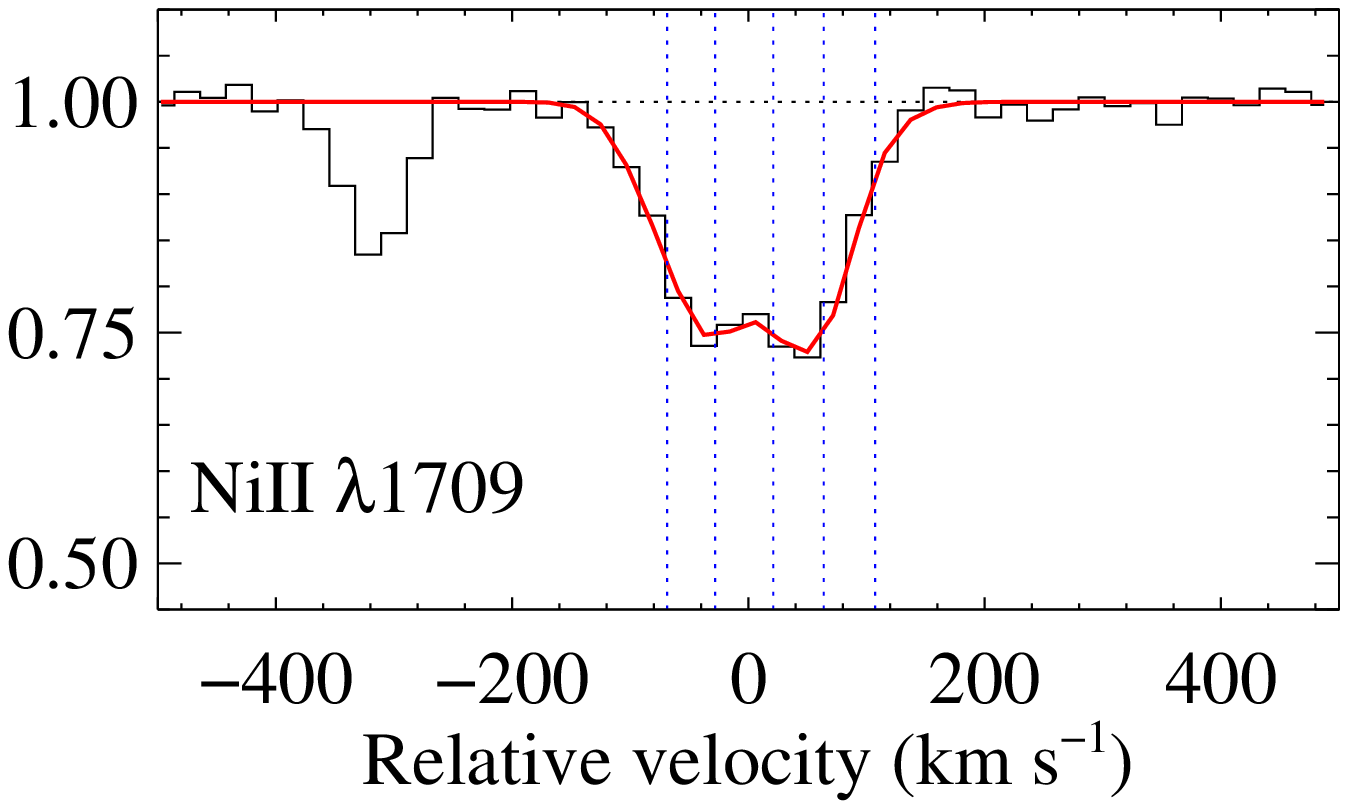} \\
\includegraphics[bb=219 244 393 629, clip=, angle=90, width=0.45\hsize]{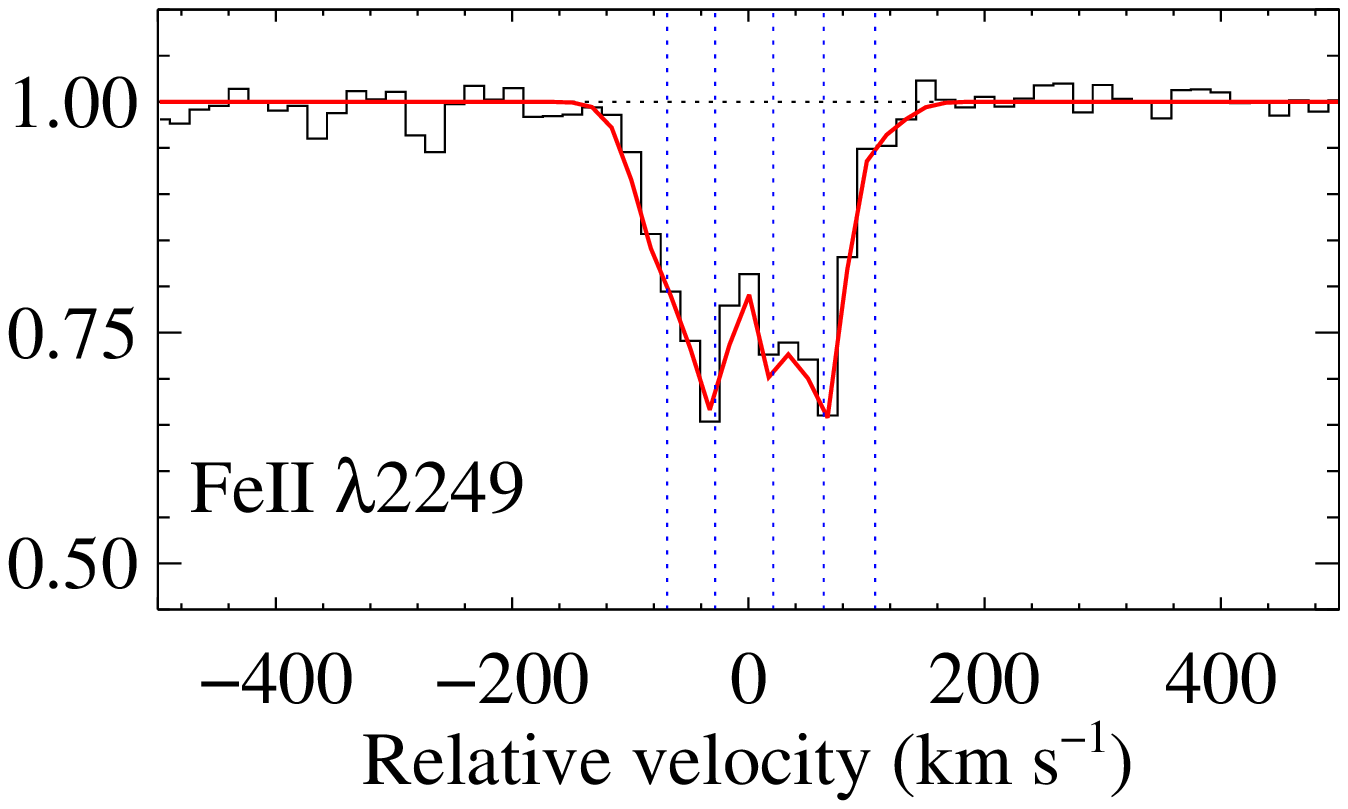} &
\includegraphics[bb=219 244 393 629, clip=, angle=90, width=0.45\hsize]{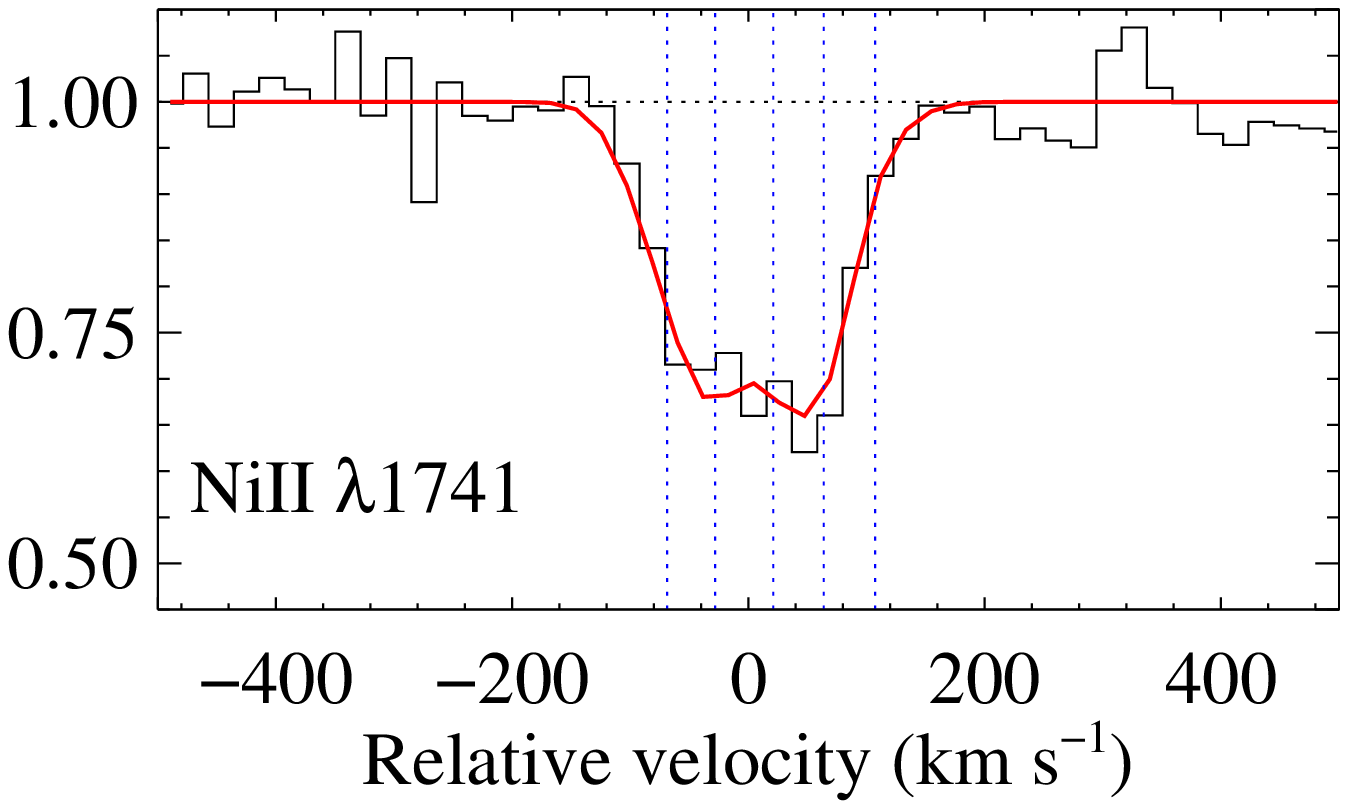} \\
\includegraphics[bb=219 244 393 629, clip=, angle=90, width=0.45\hsize]{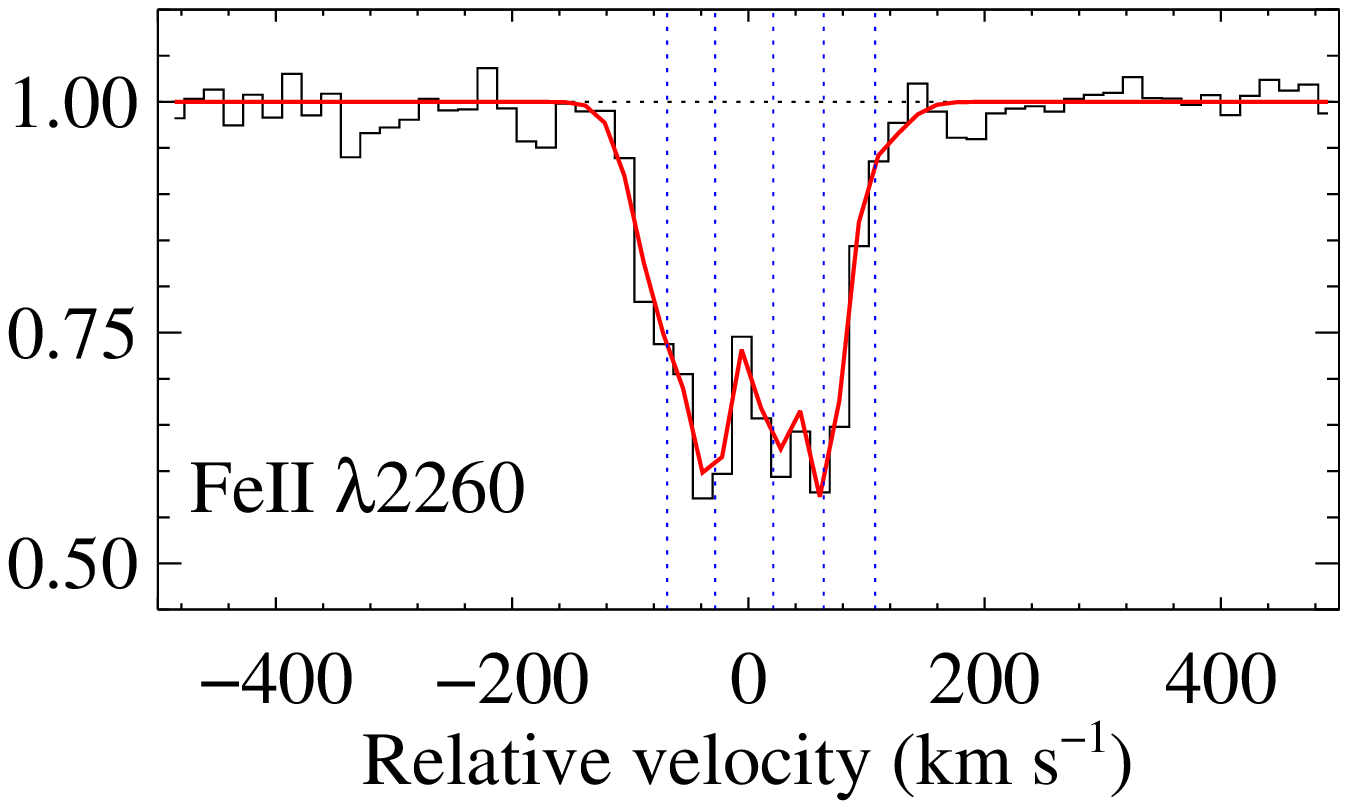} &
\includegraphics[bb=219 244 393 629, clip=, angle=90, width=0.45\hsize]{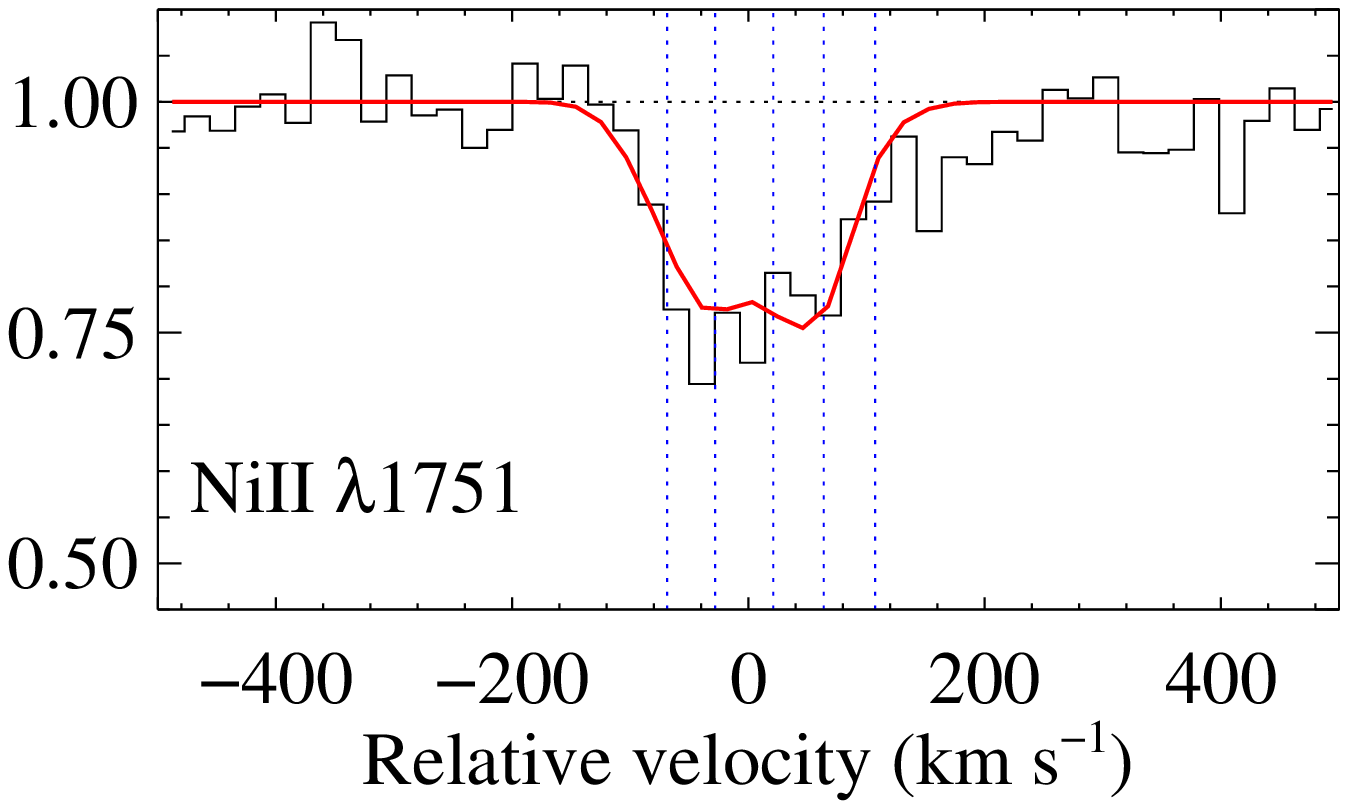} \\
\includegraphics[bb=219 244 393 629, clip=, angle=90, width=0.45\hsize]{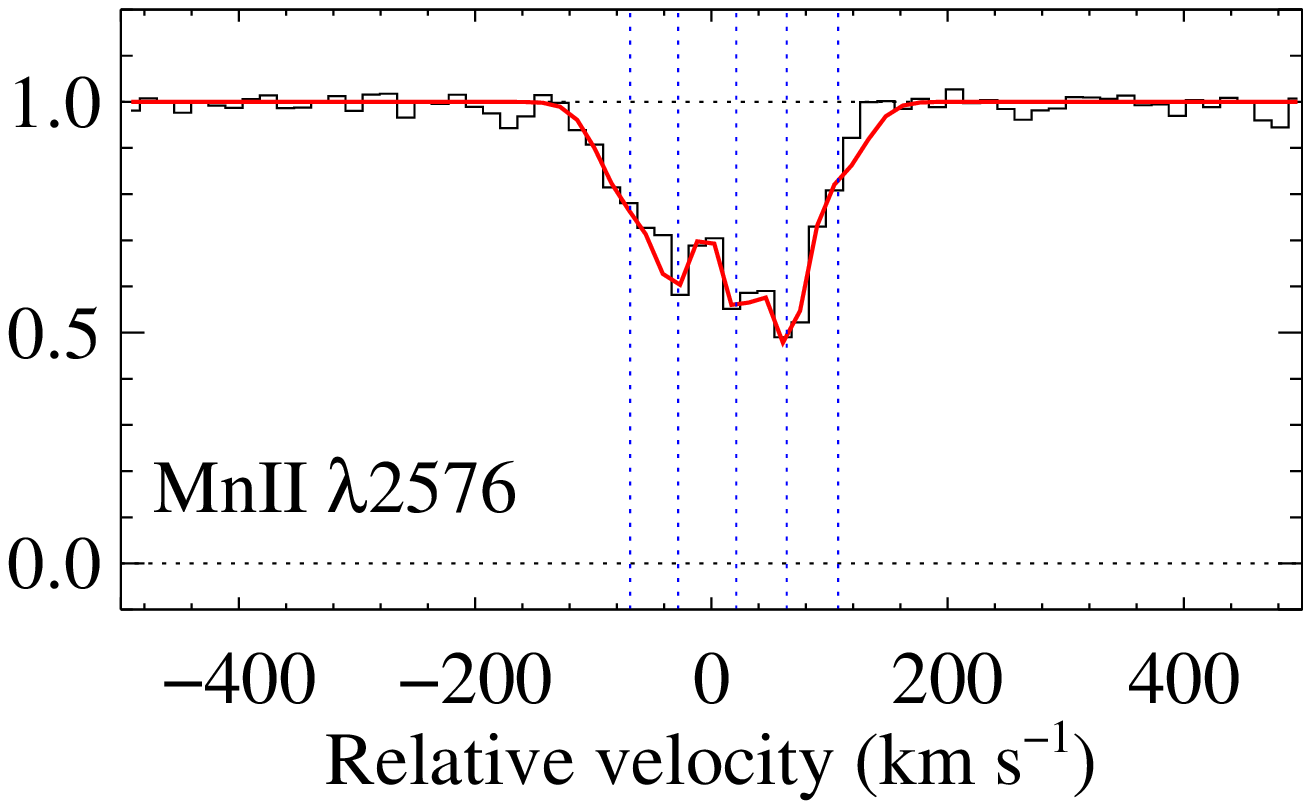}   &  
\includegraphics[bb=219 244 393 629, clip=, angle=90, width=0.45\hsize]{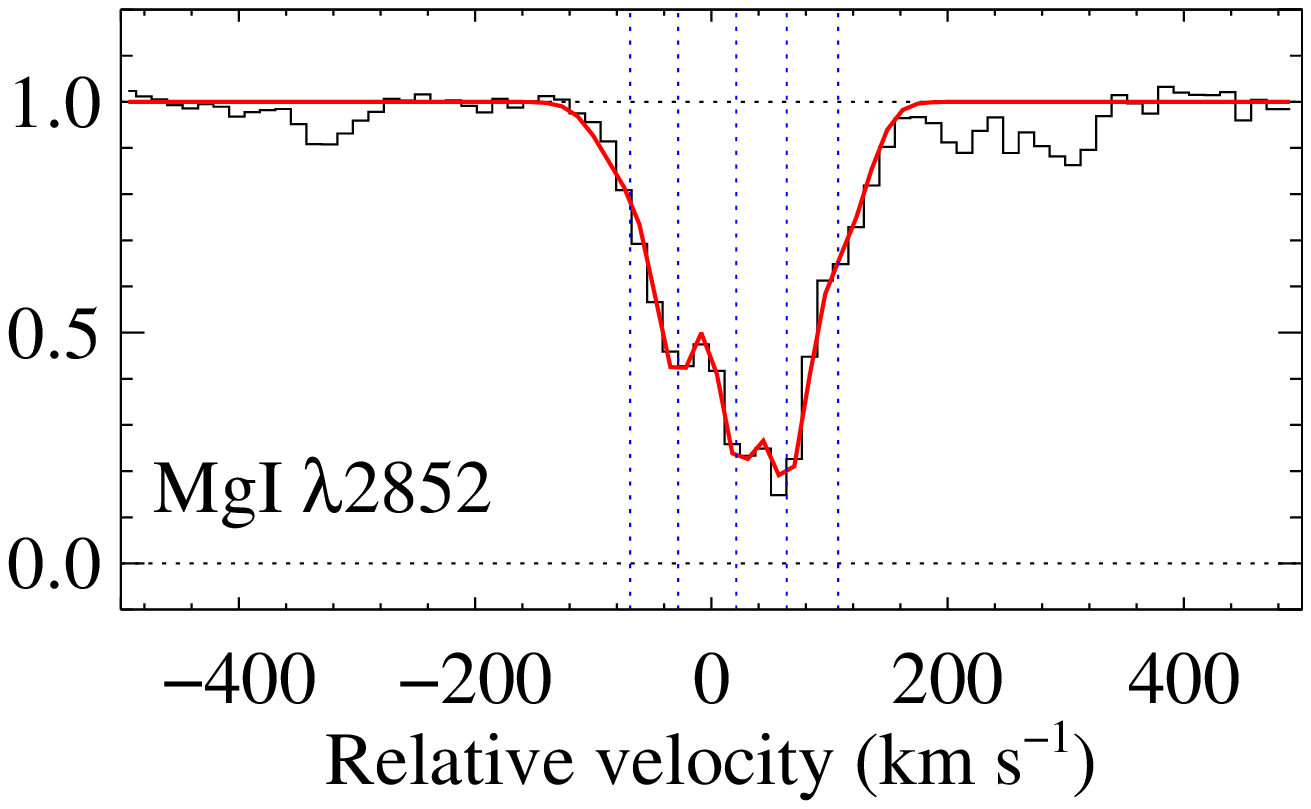}  \\
\includegraphics[bb=219 244 393 629, clip=, angle=90, width=0.45\hsize]{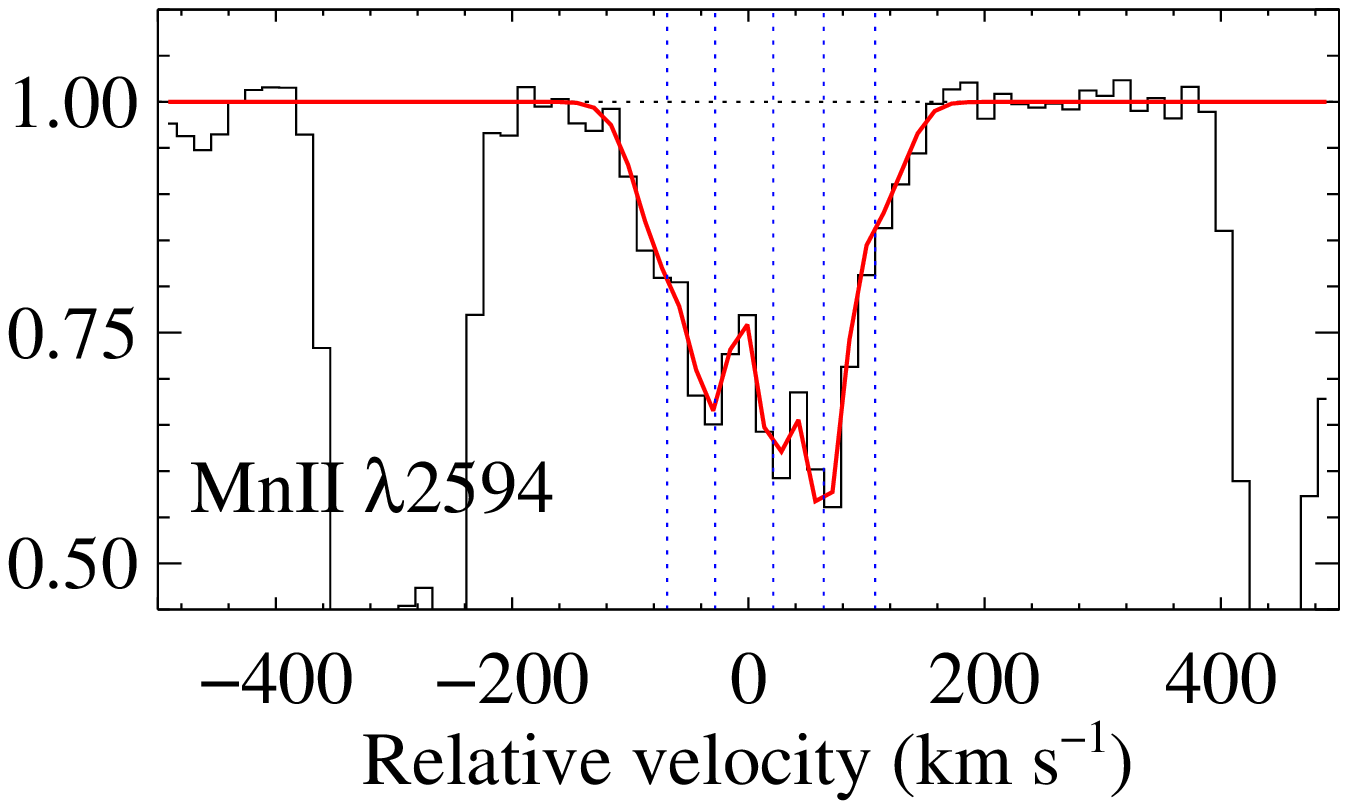}   &  
\includegraphics[bb=219 244 393 629, clip=, angle=90, width=0.45\hsize]{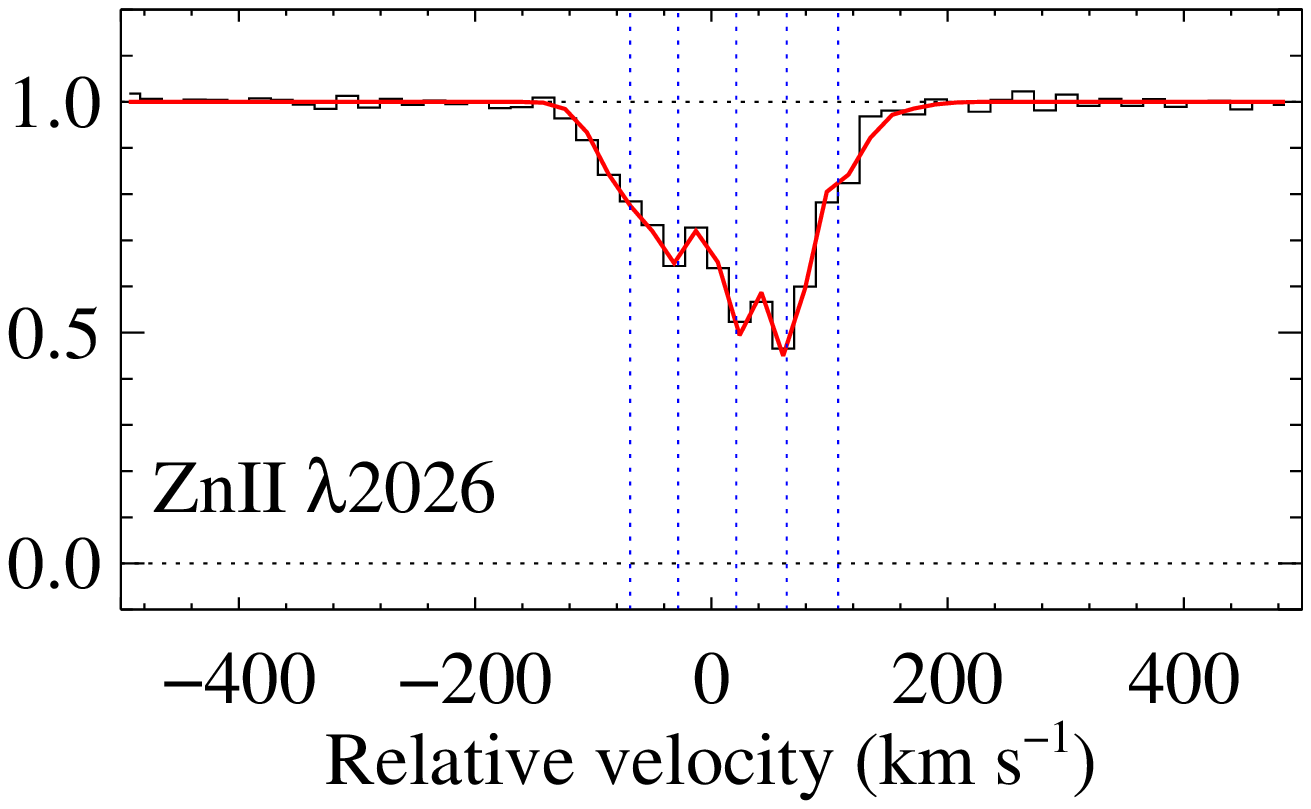} \\
\includegraphics[bb=165 244 393 629, clip=, angle=90, width=0.45\hsize]{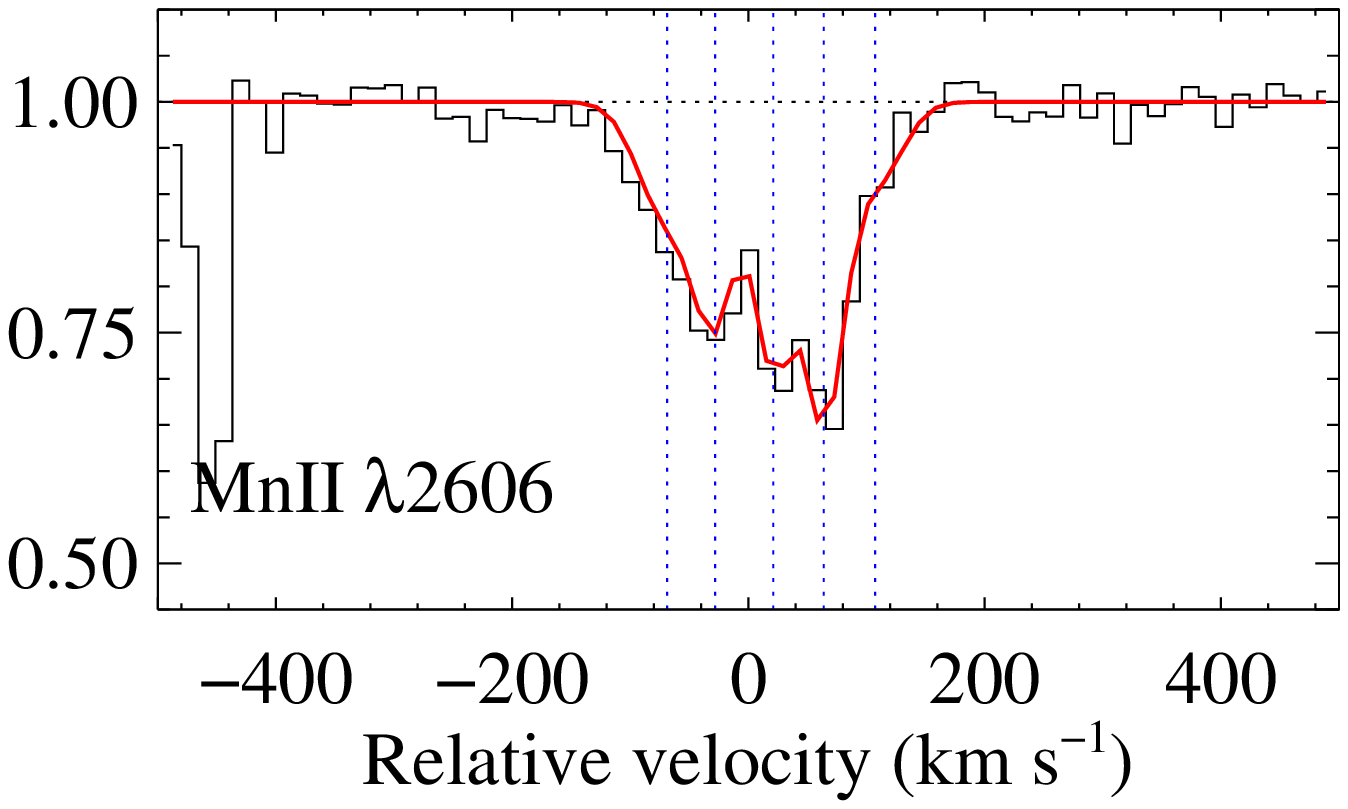}   &  
\includegraphics[bb=165 244 393 629, clip=, angle=90, width=0.45\hsize]{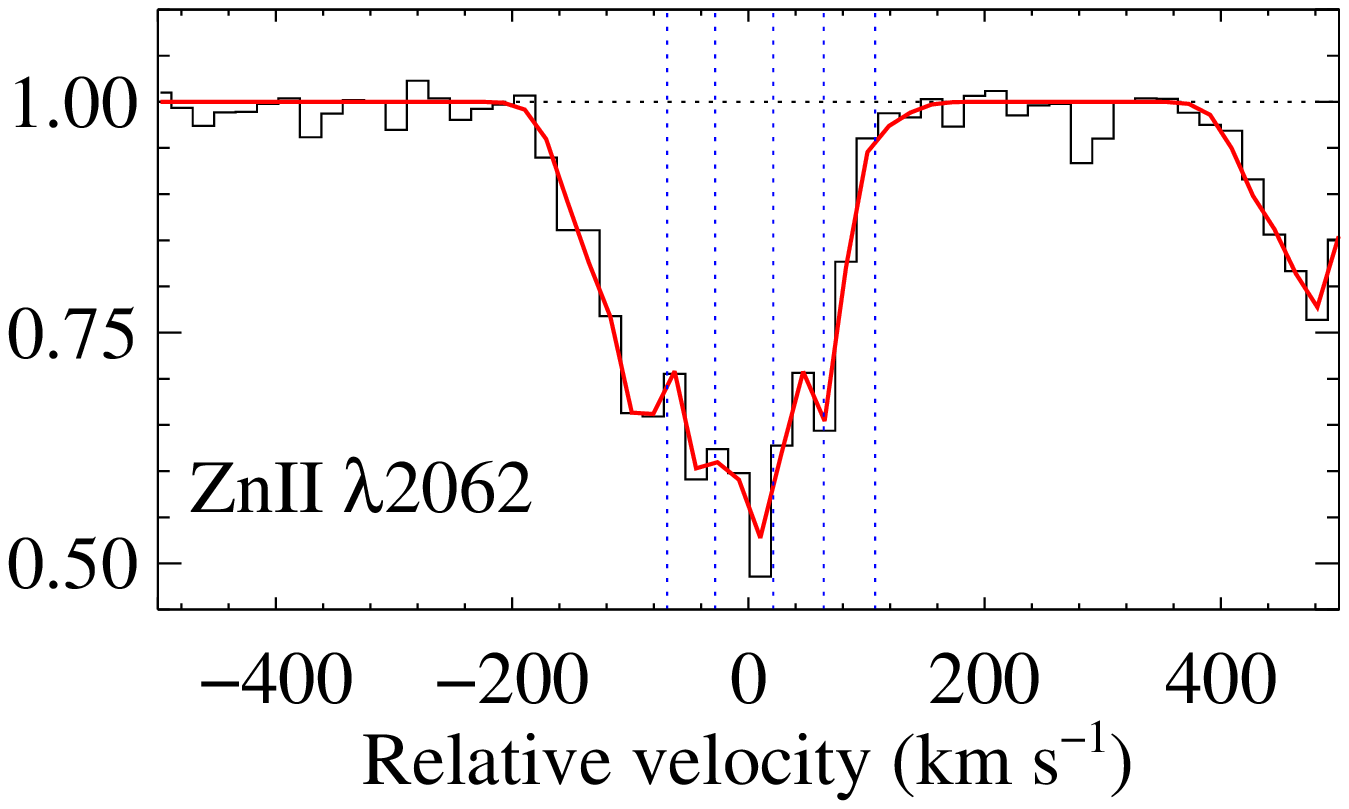} \\

\end{tabular}
\caption{Velocity plots of metal absorption lines detected in our X-shooter spectrum. Voigt profile 
fitting of the lines is represented by the red curve. 
The origin of the velocity scale is set to $z=2.2066$. Note that \ZnII$\lambda$2062 is blended with 
\CrII$\lambda$2062 and 
\ZnII$2026$ is blended with \MgI$\lambda$2026. \SII$\lambda$1253 is partially blended with an intervening 
absorption line from the \lya\ forest. \label{metals}}
\renewcommand{\tabcolsep}{6pt}
\end{figure}

Absorption lines from S$^+$, Si$^+$, Fe$^+$, Zn$^+$, Ni$^+$, Cr$^+$ and Mn$^+$ are detected spread 
over $\sim 180~\kms$ around $\zabs=2.207$ in the X-shooter spectrum. Five components were used to 
model the observed absorption lines. The continuum was fitted locally around 
each absorption line using low order splines over unabsorbed regions. The results from the 
simultaneous multiple Voigt-profile fitting are presented in Fig.~\ref{metals} and Table~\ref{metalst}. 
Note that, because of the lower spectral resolution in the UVB arm ($R\sim5200$) compared to the VIS 
arm ($R\sim9700$), the overall profile appears smoother for absorption lines in the UVB 
(\SII$\lambda\lambda$1250,1253, \FeII$\lambda$1611 and \NiII$\lambda\lambda\lambda$1709,1741,1751). 

\begin{figure}[!ht]
\centering
\renewcommand{\tabcolsep}{1pt}
\begin{tabular}{cc}
\includegraphics[bb=219 230 393 628, clip=, angle=90, width=0.48\hsize]{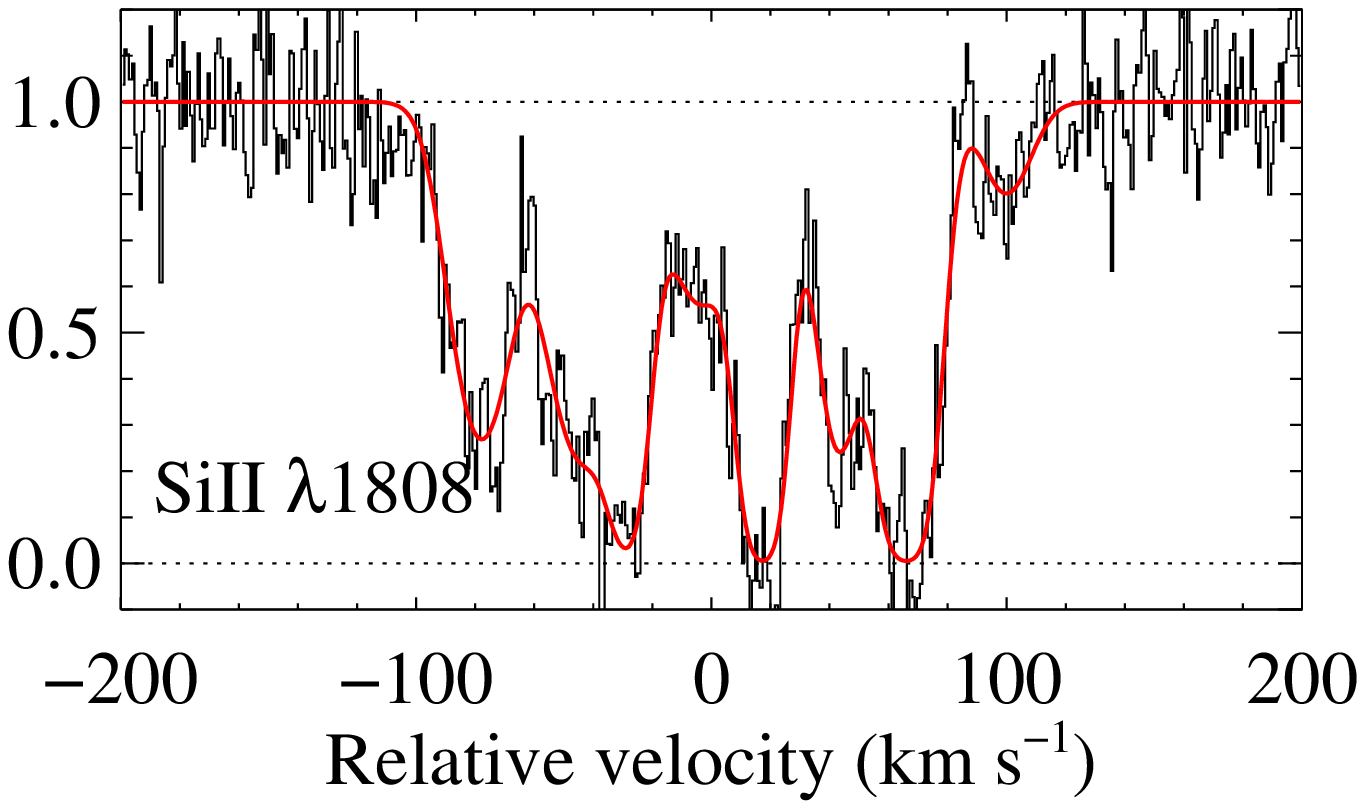} & 
\includegraphics[bb=219 230 393 628, clip=, angle=90, width=0.48\hsize]{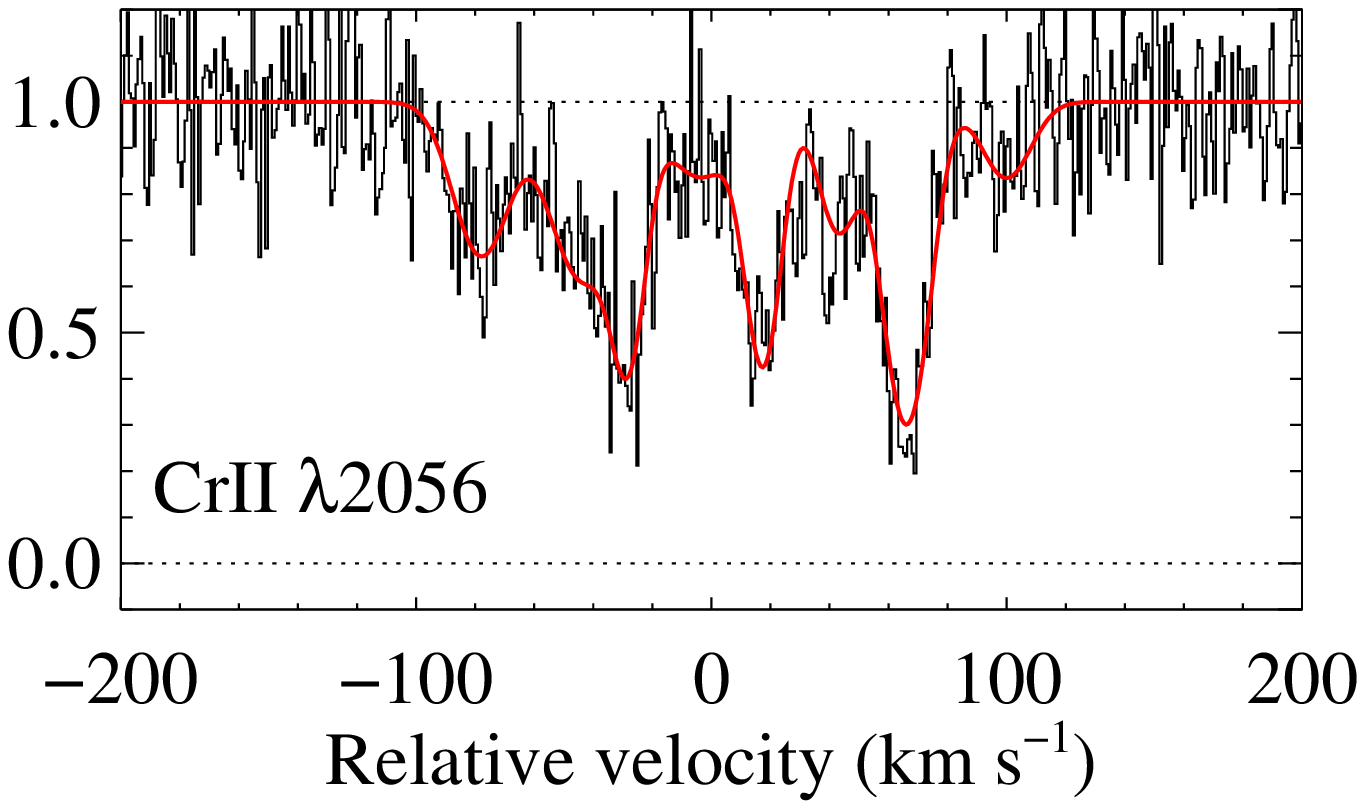} \\
\includegraphics[bb=219 230 393 628, clip=, angle=90, width=0.48\hsize]{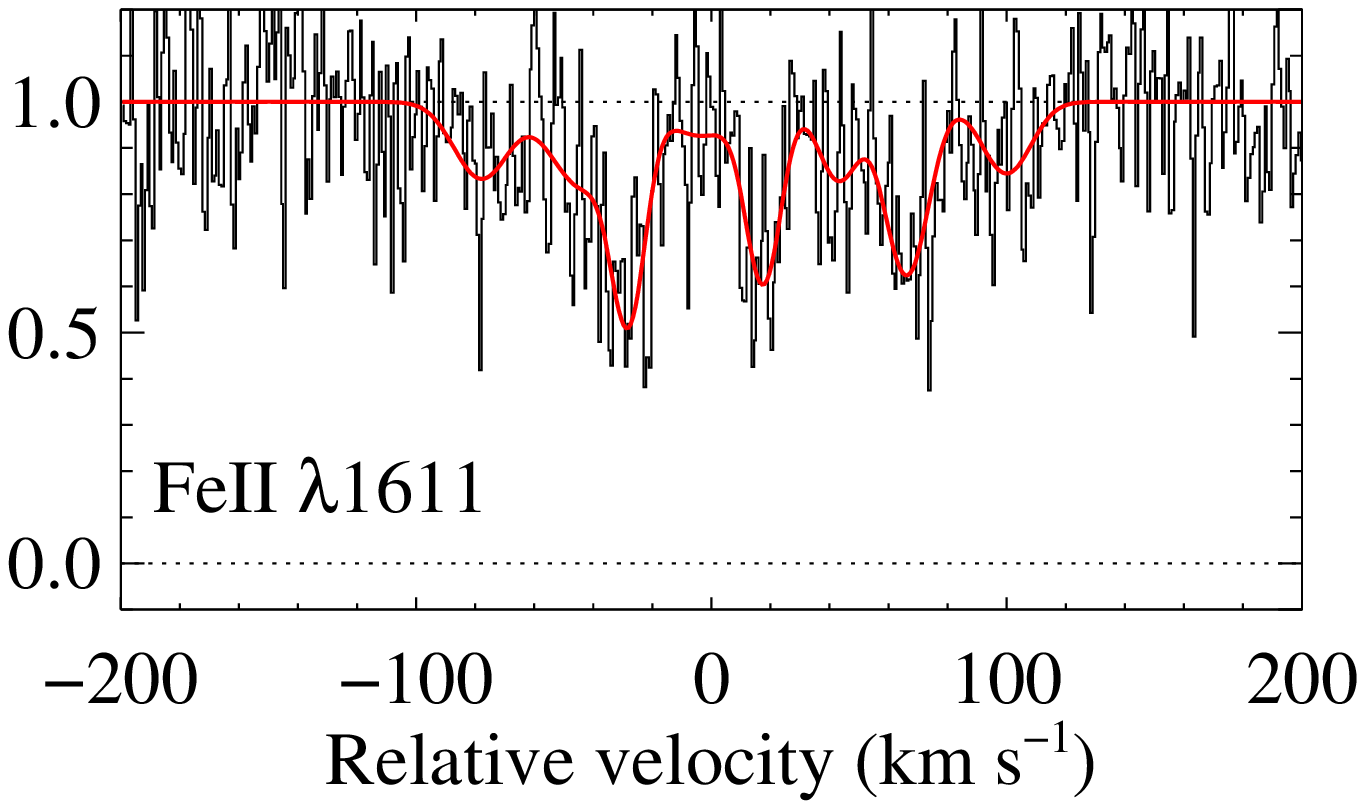} &
\includegraphics[bb=219 230 393 628, clip=, angle=90, width=0.48\hsize]{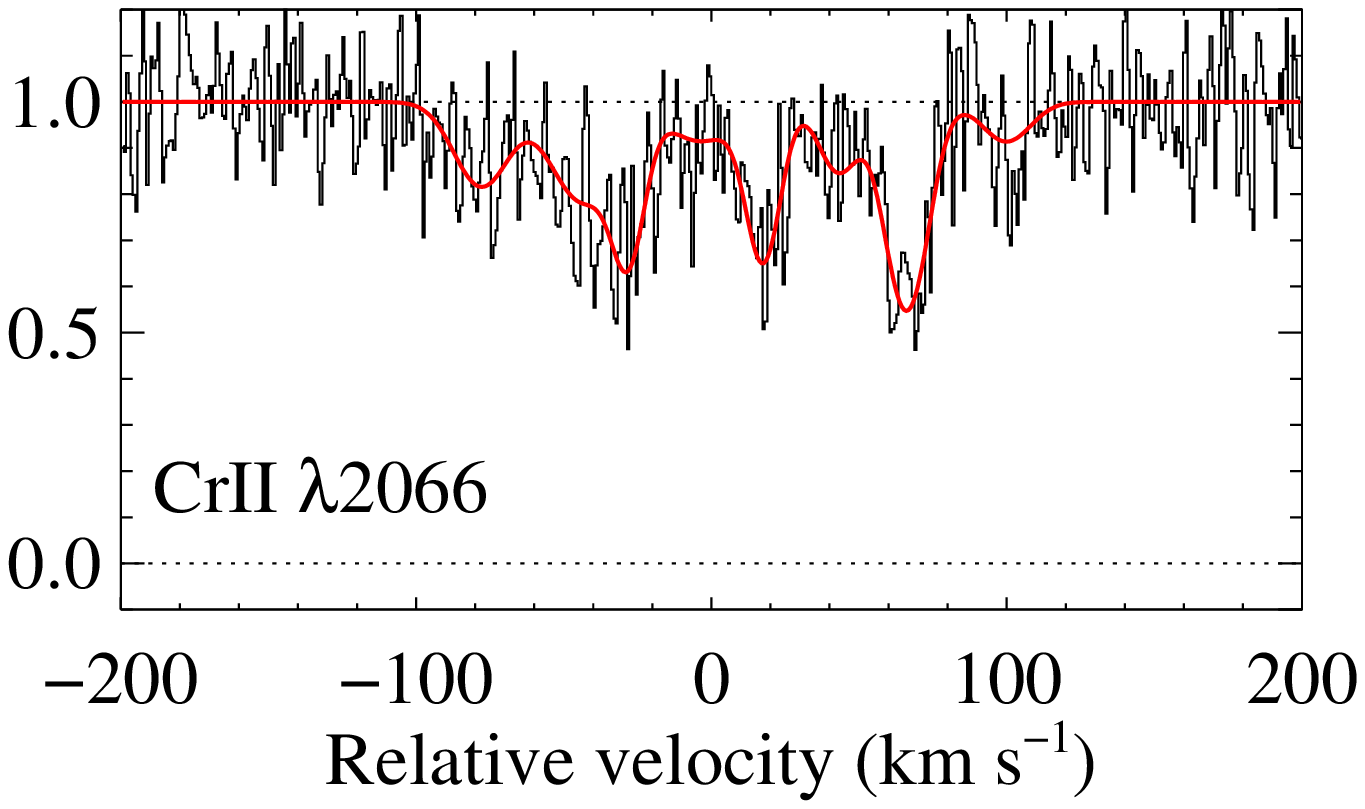} \\
\includegraphics[bb=219 230 393 628, clip=, angle=90, width=0.48\hsize]{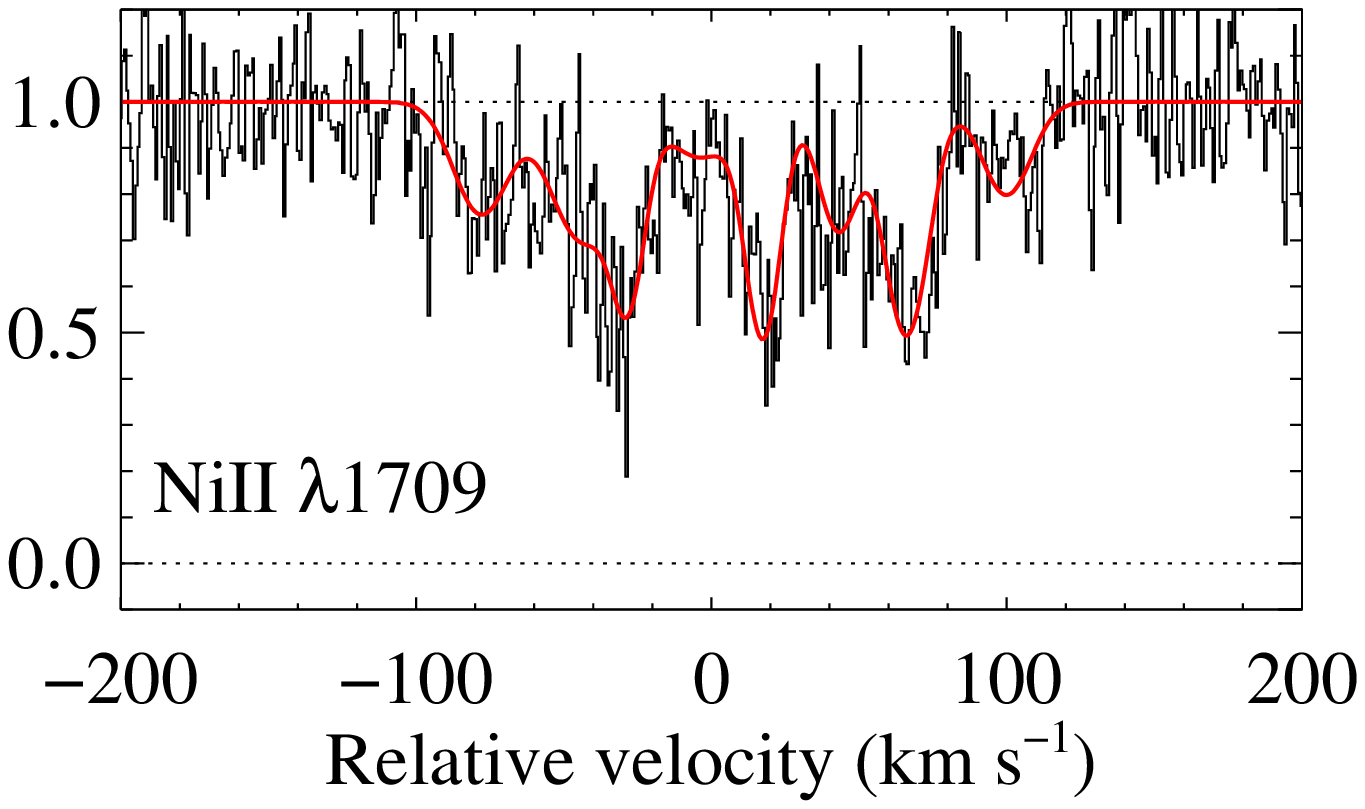} &
\includegraphics[bb=219 230 393 628, clip=, angle=90, width=0.48\hsize]{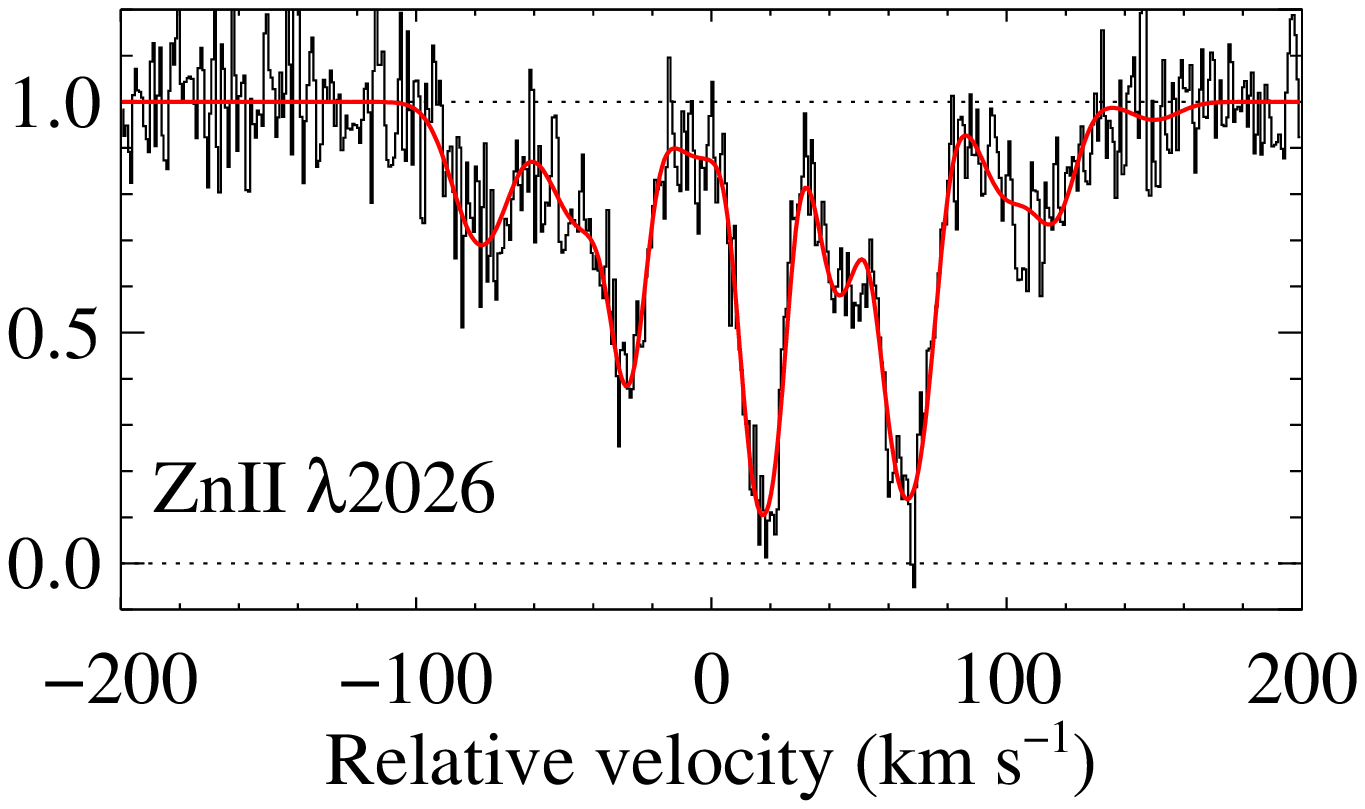} \\
\includegraphics[bb=165 230 393 628, clip=, angle=90, width=0.48\hsize]{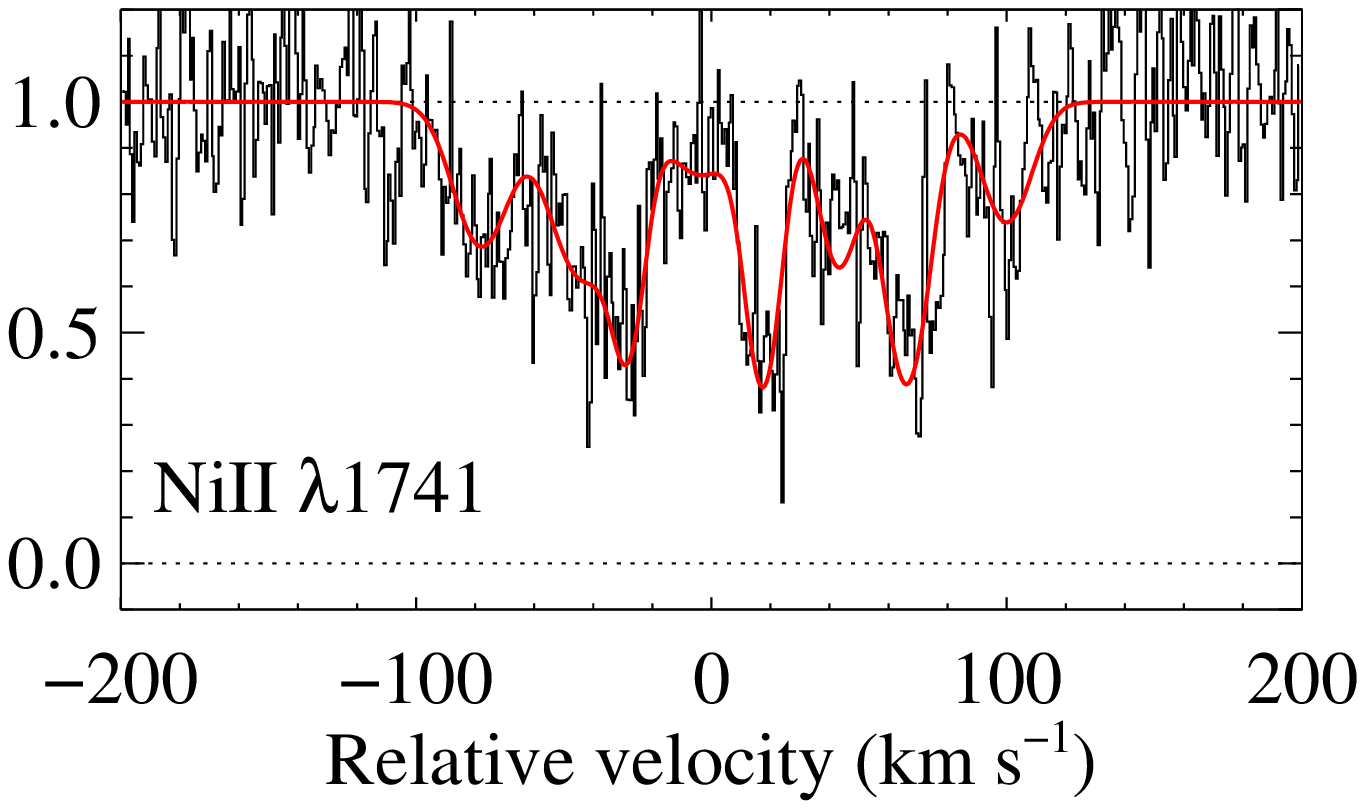} &
\includegraphics[bb=165 230 393 628, clip=, angle=90, width=0.48\hsize]{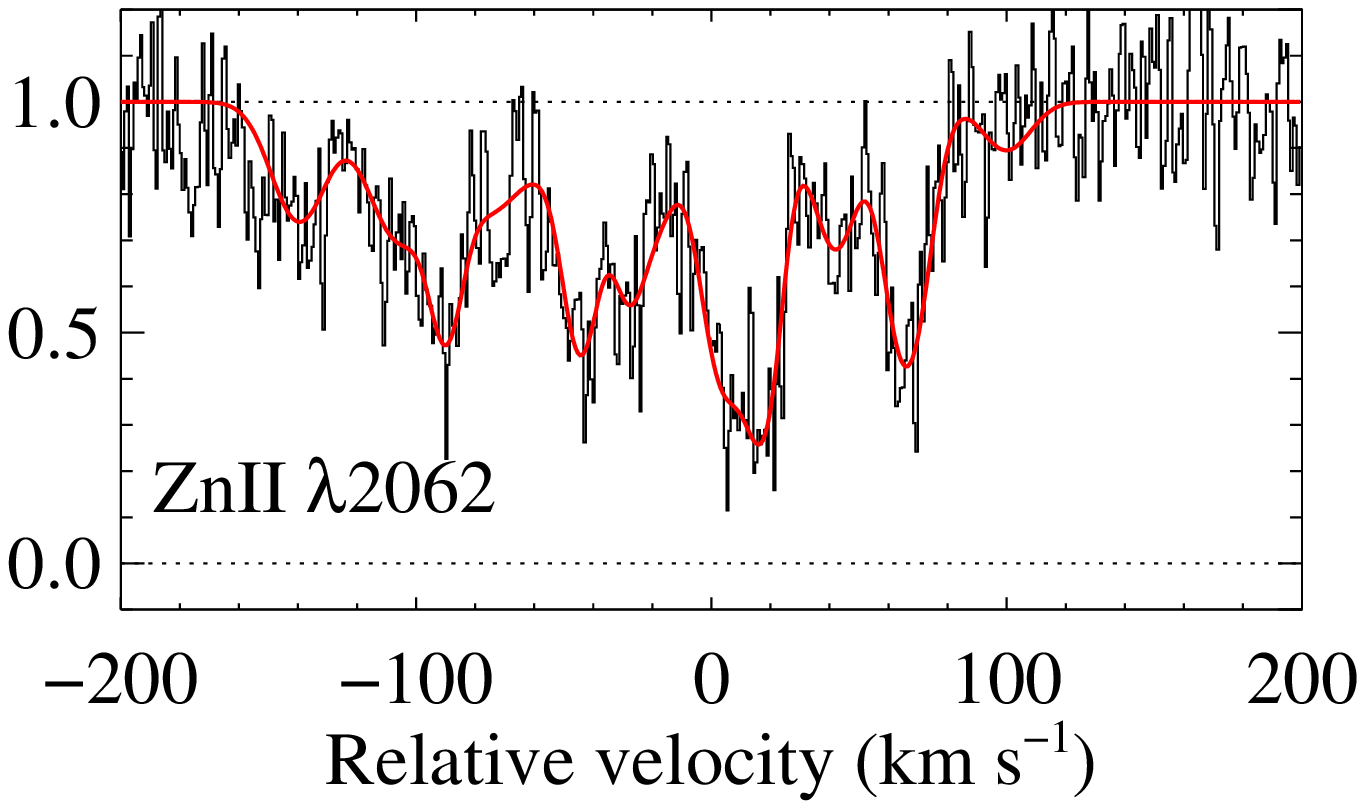} \\
\end{tabular}
\caption{Velocity plots of metal absorption lines (UVES). The vertical dashed lines indicate the position 
of the components as derived from the X-shooter spectrum. The zero velocity scale is defined as 
in Fig.~\ref{metals}.\label{metalsUVES}}
\renewcommand{\tabcolsep}{6pt}
\end{figure}

Measuring column densities using medium-resolution spectroscopy may lead one to underestimate 
the true values since saturation of 
the lines can be easily hidden and the velocity decomposition of the profile may be too simplistic. 
The availability of a high-resolution UVES spectrum in the ESO archive provides a good opportunity to check 
the values derived here. We performed an independent analysis of the UVES spectrum to measure 
the column densities of the metals. The high-resolution velocity profile (see Fig.~\ref{metalsUVES}) presents well 
separated clumps as inferred from the X-shooter spectrum (see Fig.~\ref{metals}). The total column 
densities obtained using UVES are given in the last column of Table~\ref{metalst}. The agreement between 
the X-shooter and UVES values is pretty good, the highest deviation being 0.19~dex for $N(\SiII)$ which 
is affected by saturation effects undetected in the X-shooter spectrum. The UVES-derived value could also be 
underestimated since saturation of \SiII$\lambda$1808 is also visible at high resolution.
For other species (except \SII\ which is also likely saturated), the better defined continuum 
in the X-shooter spectrum and the availability of several lines (e.g. 
\MgI$\lambda$2852, \FeII$\lambda\lambda$2249,2260, not covered by UVES) may compensate the lower 
spectral resolution. Therefore, while medium spectral resolution 
may generally lead to an underestimation of the column densities, here, X-shooter provides rather accurate values. 
\MgII\ lines are heavily saturated with a very large equivalent width ($W_{\rm r}^{\lambda2796}\simeq3.6$~{\AA}), making the  
system eligible as Ultra-Strong \MgII\ system \citep[US\MgII;][]{Nestor07}.

The overall DLA metallicity is found to be [Zn/H]~$\approx -1.1$, which is typical of the overall population 
of DLAs \citep{Prochaska02}. Interestingly, the zinc and hydrogen column densities measured here place the DLA 
close to the observational limit proposed by \citet{Prantzos00} beyond which the quasar would be
too attenuated to be observable \citep[see also][]{Boisse98,Vladilo05}.

\begin{table*}
\centering
\caption{Column densities in individual components and mean abundances of the DLA \label{metalst}}
\begin{tabular}{c c c c c c c c}
\hline
\hline
{\large \strut}Species  & \multicolumn{7}{c}{$\log N$ (\cmsq)\tablefootmark{a}} \\
         & \multicolumn{5}{c}{X-shooter} & X-shooter & UVES \\
         & $\zabs$=2.2059 & $\zabs$=2.2063 & $\zabs$=2.2068 & $\zabs$=2.2073 & $\zabs$=2.2077 & Total & Total \\
\hline
{\large \strut}
Mg$^0$   &11.79$\pm$0.06&12.49$\pm$0.01&12.81$\pm$0.01&12.87$\pm$0.04&12.21$\pm$0.04&13.28$\pm$0.02& 13.22$\pm$0.19\\
Si$^+$   &15.68$\pm$0.04&15.73$\pm$0.05&15.74$\pm$0.03&15.76$\pm$0.03&14.48$\pm$0.17&16.34$\pm$0.02\tablefootmark{b}& 16.53$\pm$0.04\tablefootmark{b}\\
S$^+$    &15.35$\pm$0.10&15.18$\pm$0.14&15.82$\pm$0.12&15.70$\pm$0.16&              &16.19$\pm$0.08\tablefootmark{b}&               \\
Cr$^+$   &13.42$\pm$0.04&13.54$\pm$0.03&13.48$\pm$0.02&13.61$\pm$0.02&12.65$\pm$0.11&14.13$\pm$0.01& 14.14$\pm$0.01\\
Mn$^+$   &12.77$\pm$0.04&12.92$\pm$0.03&13.01$\pm$0.01&13.12$\pm$0.02&12.58$\pm$0.05&13.62$\pm$0.01&               \\
Fe$^+$   &14.98$\pm$0.04&15.23$\pm$0.02&15.12$\pm$0.01&15.22$\pm$0.01&14.29$\pm$0.09&15.76$\pm$0.01& 15.84$\pm$0.03\\
Ni$^+$   &13.83$\pm$0.05&14.01$\pm$0.04&13.94$\pm$0.03&14.06$\pm$0.03&13.11$\pm$0.16&14.59$\pm$0.02& 14.59$\pm$0.05\\
Zn$^+$   &12.71$\pm$0.04&12.77$\pm$0.04&13.02$\pm$0.02&13.09$\pm$0.02&12.08$\pm$0.12&13.54$\pm$0.01& 13.60$\pm$0.01\\
\hline
\end{tabular}
\tablefoot{
\tablefoottext{a}{Quoted errors are rms errors from fitting the Voigt profiles.}
\tablefoottext{b}{These values are likely underestimated due to saturation effects.}
}
\end{table*}

The presence of dust can be inferred from the depletion factors of iron, manganese, nickel and chromium
compared to zinc: 
[Zn/Fe]~$=0.72$, [Zn/Cr]~$=0.49$, [Zn/Mn]~$=0.79$ and [Zn/Ni]~$=0.61$. These values are marginally higher
than the typical values found in DLAs \citep[see e.g.][]{Noterdaeme08} and in-between the values found in warm 
neutral gas in the Galactic disk and the Halo \citep{Welty99}. Similar depletion pattern have 
been observed along gas-rich lines of sight in the SMC \citep{Welty01}. 
Interestingly, this depletion pattern is also very similar to that found in the 
strong DLAs at $\zabs=2.4$ towards HE\,0027$-$1836 \citep{Noterdaeme07lf} and at $\zabs=3.3$ towards 
SDSS\,J081634$+$144612 (Guimar\~aes et al., submitted), both of which have very large \HI\ column densities 
($\log N(\HI)=21.75$ and 22.0, respectively).  
Excess of UV flux from local star formation activity was inferred in the case of HE\,0027$-$1836 
from the excitation of H$_2$ lines. SDSS\,J081634$+$144612 is also an H$_2$-bearing DLA. 
Unfortunately, in the case of \qso, H$_2$ lines fall 
bluewards of the Lyman-break from a Lyman-limit system at $\zabs=2.878$. 
In addition, CO lines are not detected. 

\subsection{Dust reddening}
\begin{figure}
\centering
\includegraphics[bb= 85 175 495 400, width=0.9\hsize]{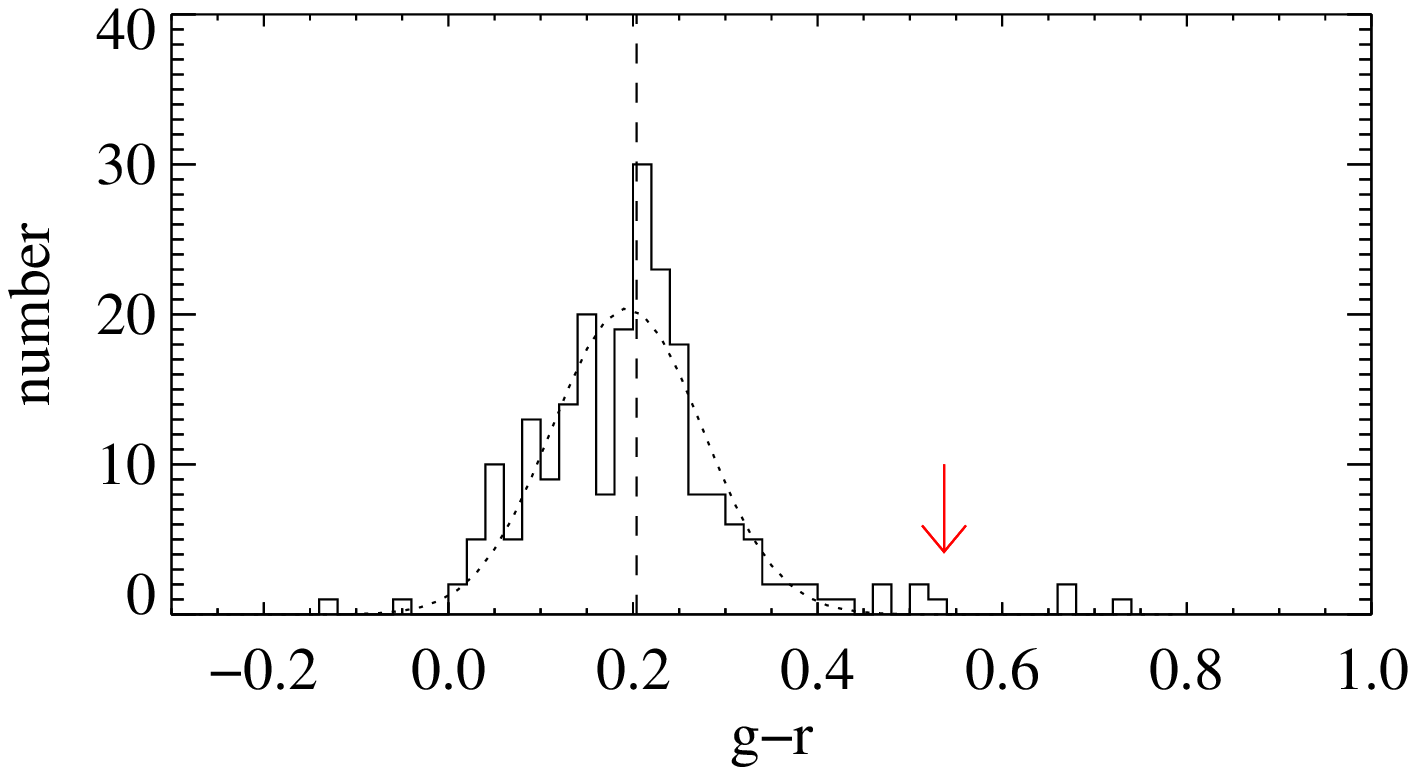}
\caption{SDSS $g-r$ colour distribution for a control sample of 222 non-BAL QSOs at redshift close 
to that of \qso\ and $i<19.5$. The arrow indicates the location of \qso. The vertical dashed line 
marks the median $g-r$ value. The dotted line gives the best fitted Gaussian  distribution.\label{gr}}
\end{figure}

\begin{figure*}
\centering
\includegraphics[bb=200 90 385 740,angle=90,width=0.98\hsize]{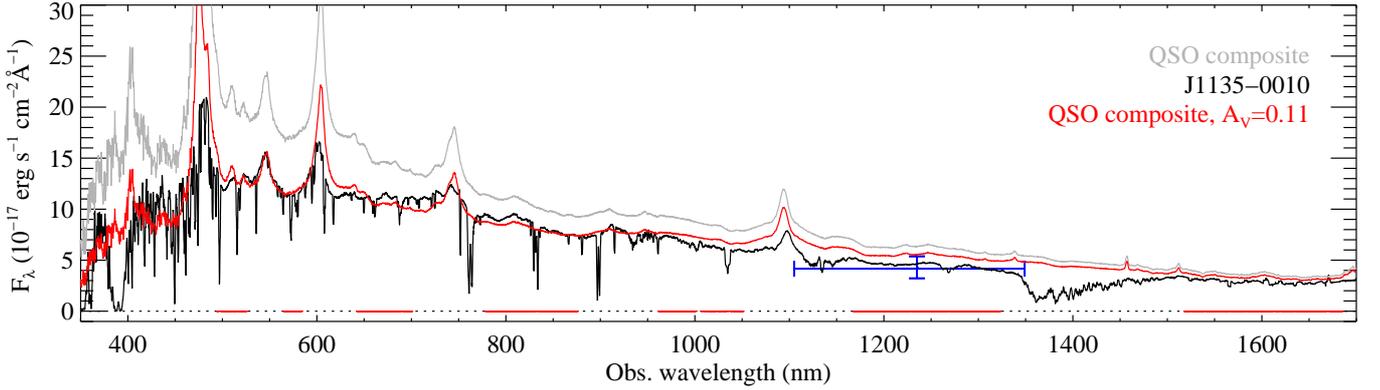}
\caption{The X-shooter spectrum of \qso\ (black, corrected for Galactic extinction) is plotted 
along with the QSO HST composite from \citet{Telfer02} (grey). The reddened composite, using 
$A_{\rm V}=0.11$ and SMC extinction curve is shown in red. The red horizontal bars at $F_{\lambda}=0$ 
indicate the wavelength ranges used to constrain the fit. The X-shooter spectrum has been smoothed 
with boxcar 30 pixels for presentation purpose only. The blue error bar corresponds to J-band 
magnitude extracted from 2MASS images by \citet{Schneider10}. \label{sed}}
\end{figure*}

Having established that some dust is present in the system --with a {\sl relative abundance} quantified by 
the depletion factor-- we can now study the {\sl integrated} effect of this dust along the line of sight. 
The median $g-r$ colour for a control sample of 222 non-BAL QSOs with $i<19.5$ and similar redshift 
(within $\Delta z = \pm 0.03$ around $\zem=2.891$) is $\avg{g-r}=0.20$ with a dispersion 
of 0.08~mag (see Fig.~\ref{gr}). \qso\ is among the 2\% redder QSOs with $g-r=0.54$. 
This means that the significance 
of \qso\ being redder than the average is between 2.4\,$\sigma$ (face value, assuming the red tail 
of the colour distribution is only due to intrinsic spectral shape variation) and 4.2\,$\sigma$ (considering 
the dispersion of 0.08~mag
around the median value, i.e. that other QSOs in the red tail are also reddened by intervening 
systems\footnote{Note that errors from the SDSS photometric calibrations are less than 0.02~mag.}). 

We can also measure the reddening of the QSO light by the intervening DLA following the technique described in 
\citet{Srianand08bump} and \citet{Noterdaeme10co}: The observed X-shooter spectrum is reasonably well 
matched with a QSO composite \citep{Telfer02}, reddened by an SMC extinction law \citep{Gordon03} with 
A$_{\rm V}=0.11$~mag or $E(B-V) \sim 0.04$. This is higher than the {\sl mean} value for the overall DLA population  
\citep{Frank10,Khare11}. However, because of the large column density, the extinction-to-gas ratio 
$A_{\rm V}/N(\HI)\approx9\times10^{-24}$~mag\,cm$^2$ is lower than the mean value for DLAs 
\citep[$A_{\rm V}/N(\HI)\approx2-4\times10^{-23}$~mag\,cm$^2$;][]{Vladilo08} whereas the depletion factors are only slightly 
above typical DLA values. This could be the consequence of a non-uniform distribution of dust in the \HI\ gas. 
It seems that in the present system, the relative abundance of dust is quite typical of the overall DLA population 
(i.e. low), but its large column density produces a detectable reddening of the background QSO. 

Similarly, the quasar Q\,0918$+$1636 is reddened by a foreground DLA galaxy \citep{Fynbo11}. In that case 
molecular hydrogen is detected --a clear indication for the presence of dust \citep{Ledoux03,Noterdaeme08}. 
These examples of dust attenuation of the background quasar reinforce the view that optically-selected QSO samples may be 
biased against lines of sight passing through the inner (gas-, molecular- and metal-rich) regions of foreground 
galaxies (e.g. \citealt{Noterdaeme09co, Noterdaeme10co}).

In conclusion, the large amount of neutral gas, the metallicity and the dust depletion 
pattern in the DLA towards \qso\ are all similar to what is seen along SMC lines of sight 
\citep{Tumlinson02,Welty01}. Interestingly, the latter is a vigorous star-forming dwarf galaxy. 

\section{Properties of the DLA galaxy \label{em}}

\subsection{Impact parameter and star formation rate \label{impact_sfr}}

\begin{figure}
\centering
\includegraphics[angle=90,bb=70 70 350 750,clip=,width=\hsize]{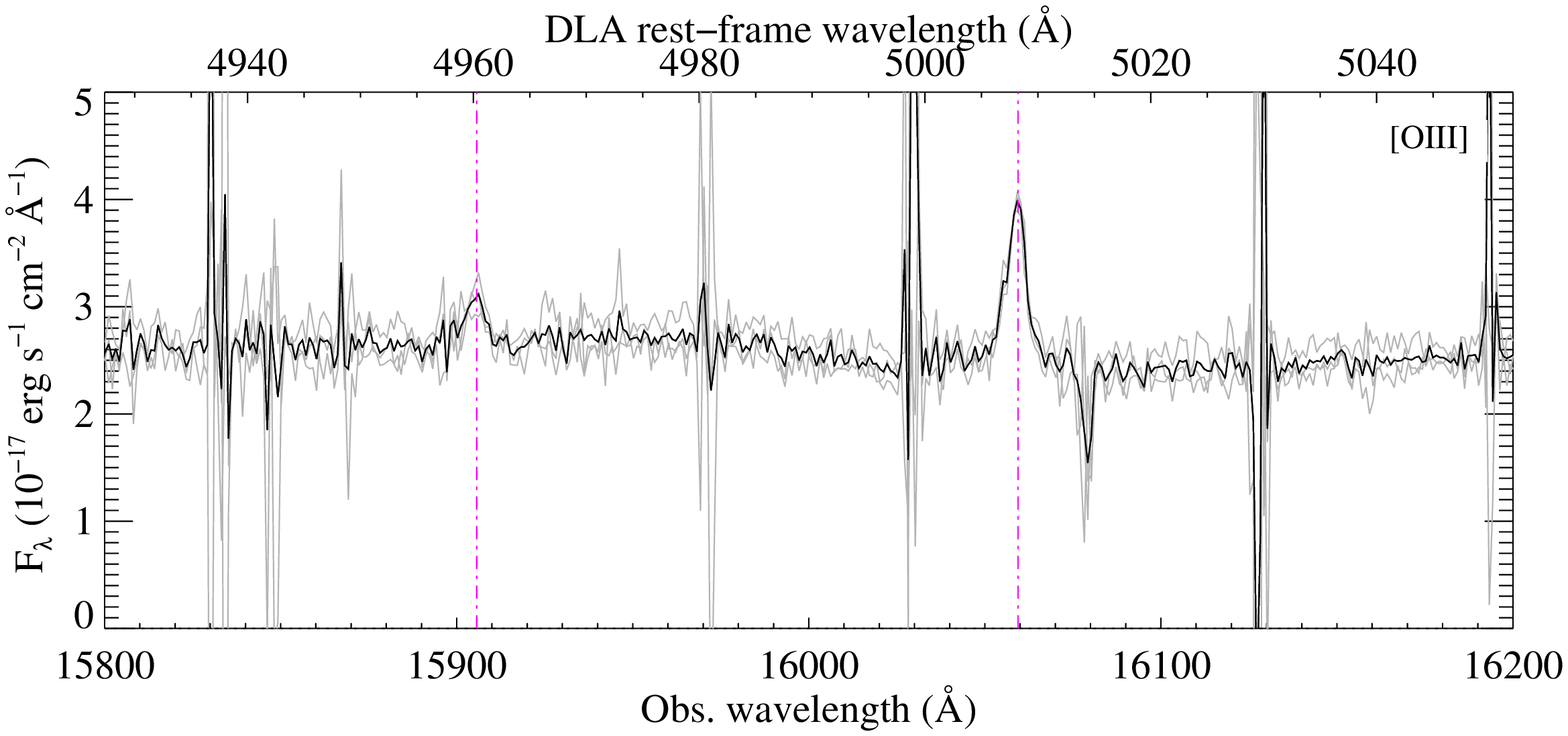}
\includegraphics[angle=90,bb=140 70 295 750,clip=,width=\hsize]{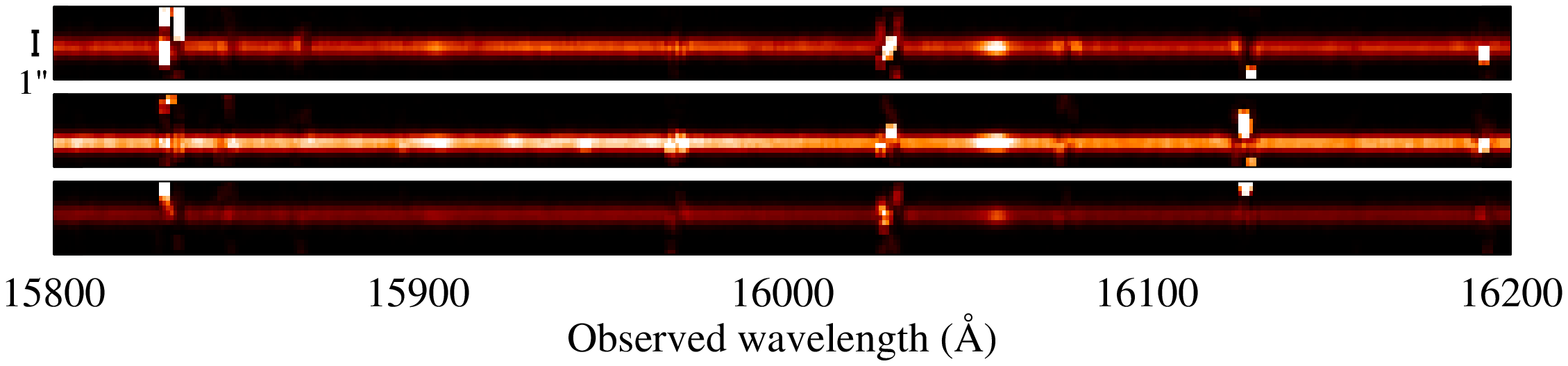}
\caption{Same as Fig.~\ref{lya} for the region around [\OIII]$\lambda$4959 and [\OIII]$\lambda$5007 (marked by 
dash-dotted vertical lines). Individual spectra for each PA are shown in grey. \label{oiii}}
\end{figure}

\begin{figure}
\centering
\includegraphics[angle=90,bb=70 70 350 750,clip=,width=\hsize]{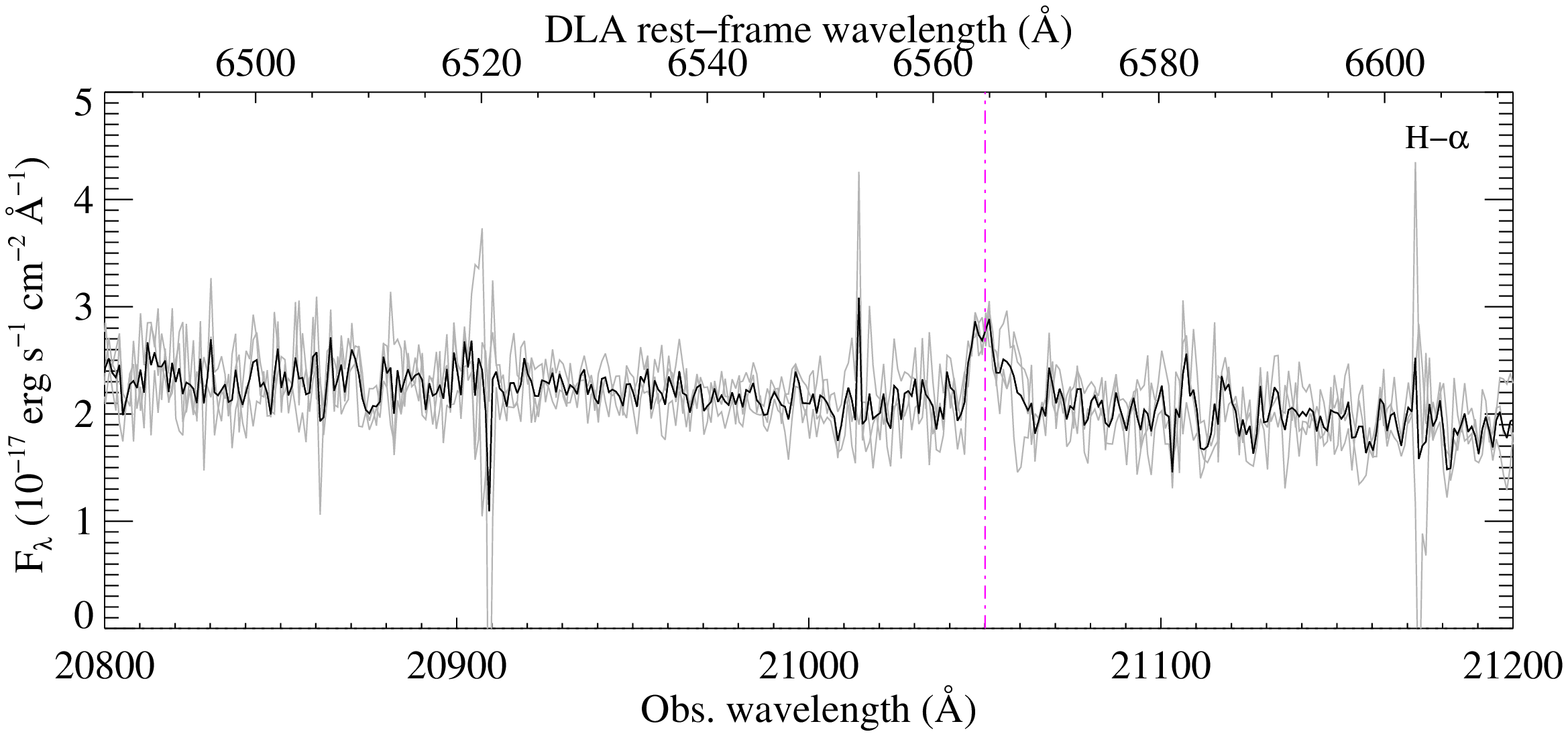}
\includegraphics[angle=90,bb=140 70 295 750,clip=,width=\hsize]{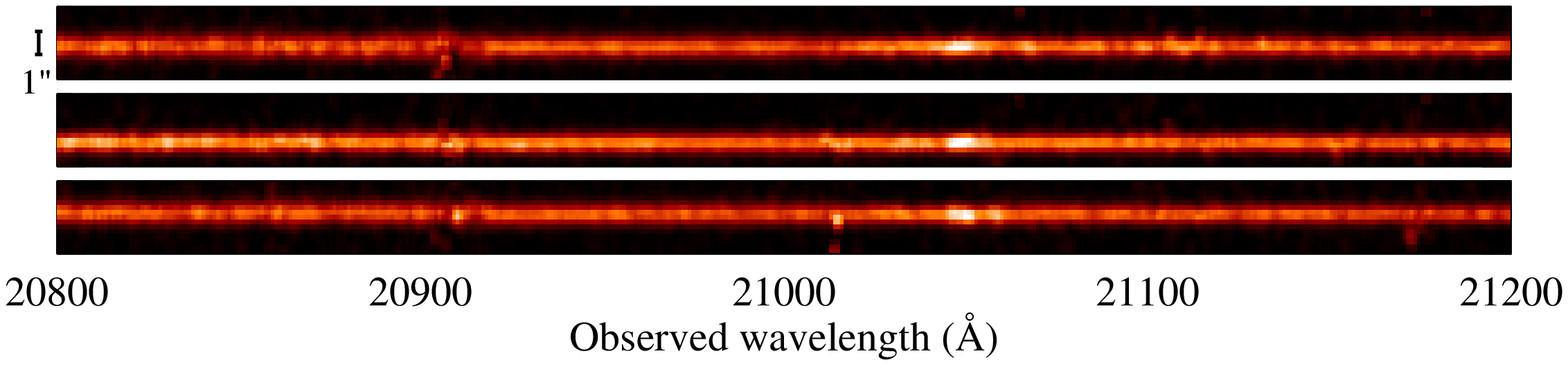}
\caption{Same as Fig.~\ref{oiii} for the region around \Ha. \label{ha}}
\end{figure}

In addition to the double-peaked \lya\ emission (see Figs.~\ref{lya_mage} and \ref{lya}), we 
detect [\OIII]$\lambda\lambda$4959,5007 (Fig.~\ref{oiii}) and H-$\alpha$ (Fig.~\ref{ha}) 
at the three position angles. Unfortunately, because the H-$\beta$ line falls on a sky emission line, 
we could not measure its strength. The [\OII]$\lambda$3727 doublet is not detected 
but a stringent upper limit on the corresponding flux can be derived from the data (see Table~\ref{line_fluxes}).

\begin{figure}
\centering
\includegraphics[bb=70 180 490 570,clip=,width=\hsize]{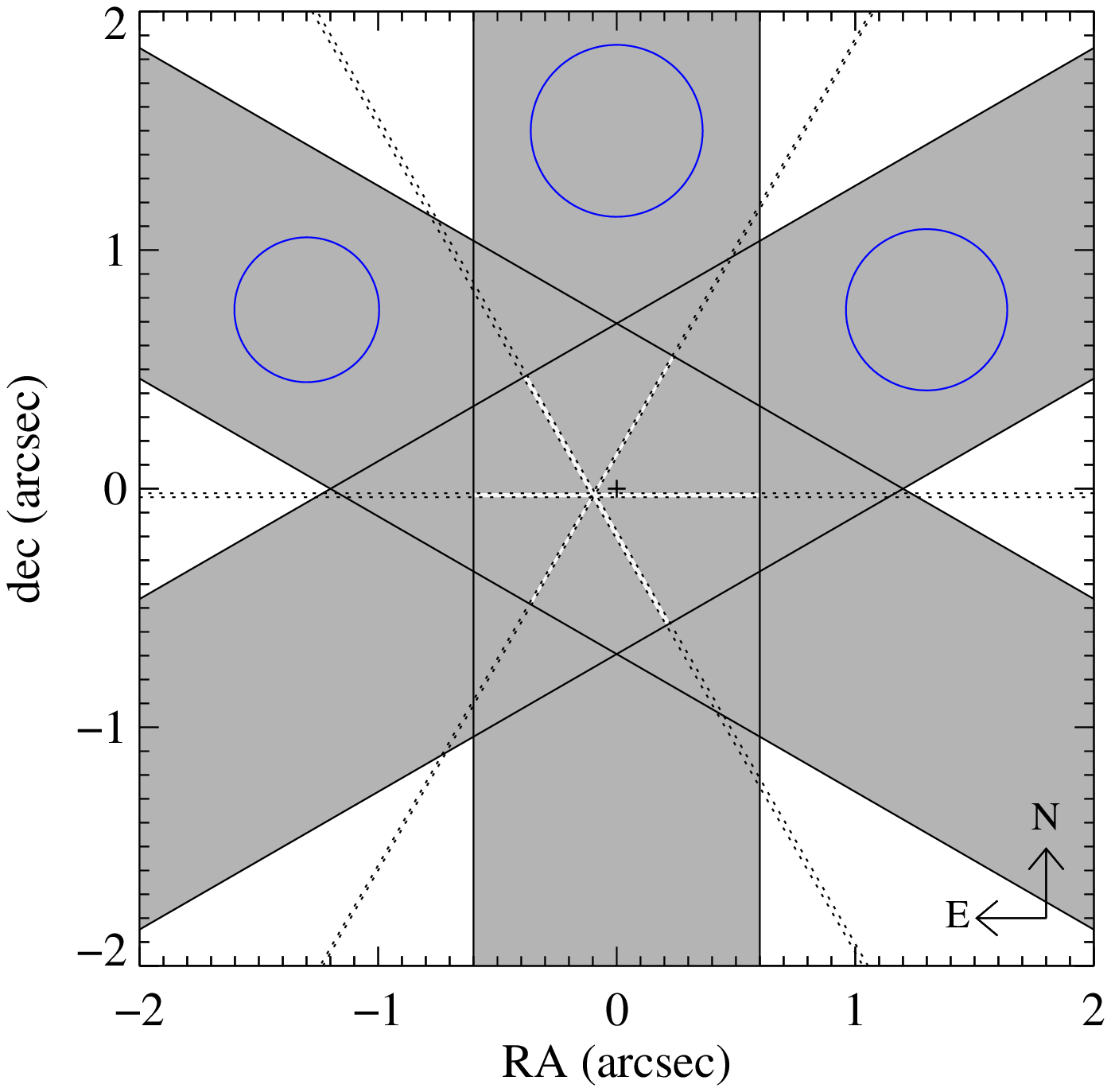}
\caption{Layout of the X-shooter slits. The position of the QSO is marked by the cross. 
The blue circles illustrate the seeing for each observation, measured from the 
QSO trace at $1.6\,{\rm \mu m}$.
The dotted lines mark the 2\,$\sigma$ regions for the measured impact parameter 
of [\OIII]$\lambda$5007 for each PA. The triangulation gives 
an impact parameter $b\simeq0.1$\arcsec\ East from the QSO. \label{impact_p}}
\end{figure}

By triangulating the position of the [\OIII]$\lambda$5007 emission with respect to that of the QSO trace 
(as done by \citealt{Moller04} and \citealt{Fynbo10}), we measure precisely the impact parameter between 
the QSO and the galaxy (see Fig.~\ref{impact_p}). We 
note that while only two slit angles would suffice to determine the position of the galaxy emission, 
the use of three different position angles nicely confirms the location, in spite of the impact parameter being 
much smaller than the seeing. The small impact parameter ($b\approx0.1\arcsec$, corresponding to 0.9~kpc 
at $z=2.207$) explains the almost equal emission line strengths measured at different PA 
(hence no differential slit losses) and the absence of detectable velocity shift between [\OIII] emission line 
measurements at different PAs (that would result from different locations with respect to the slit axis).
Finally, the spatial extent of the [\OIII] emission equals that of the QSO trace within errors for PA=0 
and $+$60\degr, and is marginally larger than the QSO trace for PA=$-60$\degr. This indicates that 
the [\OIII] emission is compact, with some spatial elongation of $FWHM\sim0.3$~arcsec in the $-60$\degr\ 
direction. The \Ha\ emission is also spatially unresolved and consistent with $FWHM \la 0.3$~arcsec.
From the unresolved (or only barely resolved) [\OIII] and \Ha\ emission, the very small impact 
parameter and the good seeing (see Fig.~\ref{impact_p}), it is also clear that slit losses are negligible.

Table~\ref{line_fluxes} presents the line fluxes measurements. 
From the luminosity of the \Ha\ line, we derive a significant star formation rate of about 25~M$_\odot$\,yr$^{-1}$ 
(uncorrected for dust), 
adopting the SFR-luminosity calibration from \citet{Kennicutt98}.

Because the most sensitive key emission line diagnostics (e.g. $R_{23}$; [\NII]/\Ha) cannot be used here, 
there is some degeneracy between the measurement of the ionisation parameter 
and the metallicity of the emitting region. However, the strict lower limit on the [\OIII] to [\OII] 
ratio indicates a high ionisation parameter $q> 5\times 10^{7}$~cm\,s$^{-1}$ \citep[fig.~5 of ][]{Kobulnicky04} and  
the low \Ha\ to [\OIII] ratio therefore indicates 
a metallicity in the range $0.1-0.5 Z_{\odot}$ \citep[see fig.~9 of][]{Weatherley05}. 
The metallicity in the absorbing gas is found at the low end of this range, but the 
impact parameter is too small to interpret this as a gradient \citep[see e.g.][]{Peroux11}.
We note however that the two metallicities do not apply to the same phase (ionised versus 
neutral) and that \HI\ gas arise in several clouds, some of which (possibly infalling) are 
likely to have a metallicity smaller than in the star-forming region. 
Interestingly, the high $q$-value agrees well with the typical value for LBGs \citep[e.g.][]{Pettini01a} and 
active compact starburst galaxies \citep{Martin97}. The high excitation of [\OII] places indeed the DLA galaxy 
in a region of the [\OIII]/[\OII] vs [\NII]/[\OII] diagram mostly populated by blue compact galaxies, 
metal poor galaxies and GRB host galaxies \citep[fig.~5 of][]{Levesque10}.

\begin{table}
\centering
\renewcommand{\tabcolsep}{2pt}
\caption{Emission line properties \label{line_fluxes}}
\begin{tabular}{c c c c}
\hline
\hline
{\large \strut}Line                   & Flux                         & Luminosity        & Derived               \\
                       & ($10^{-17}$~erg/s/cm$^{2}$)    & ($10^{42}$~erg/s)  &  quantities\\
\hline
{\large \strut}\lya\                  & 14.3$\pm$0.5                 &  5.6$\pm$0.2      & $f_{esc} = 0.2$\tablefootmark{(a)}          \\
$[\OII]\lambda$3727    & $<3.3$ (3\,$\sigma$)         &  $<1.3$           & \\
$[\OIII]\lambda$4959   & 3.0$\pm$0.7                  &  1.2$\pm$0.3      &                         \\
$[\OIII]\lambda$5007   & 12.3$\pm$0.9                 &  4.8$\pm$0.4      & $q$ (cm\,s$^{-1}$) $\sim 10^8$\\
\Ha\                   & 8.2$\pm$2.1                  &  3.2$\pm$0.8      & SFR (M$_{\odot}$\,yr$^{-1}$) = 25$\pm$6\tablefootmark{(b)}           \\

\hline
\end{tabular}
\tablefoot{
Quoted errors are measurement errors and do not include systematic uncertainties from the flux calibration.
\tablefoottext{a}{Escape fraction assuming case B recombination, i.e. $L($Ly-$\alpha)=8.7 L($H-$\alpha)$.}
\tablefoottext{b}{SFR is not dust-corrected and derived using the calibration from \citet{Kennicutt98}.}}
\renewcommand{\tabcolsep}{6pt}
\end{table}

\subsection{The double-peaked \lya\ emission \label{dplya}}

The \lya\ emission is characterised by an almost-symmetric double-peak profile 
which is detected in the DLA trough in each UVB spectrum (all three position angles, see Fig.~\ref{lya}).
Several physical processes can produce \lya\ radiation at comparable levels: \emph{i)} 
star formation activity, \emph{ii)} Lyman-$\alpha$ fluorescence induced by the UV background or the QSO 
\citep{Spitzer78}, \emph{iii)} cooling radiation \citep[e.g.][]{Dijkstra06} due to the release 
of part of the gravitational energy from infalling gas into \lya\ radiation (this is more likely 
to take place if the galaxy is massive).
When only \lya\ is detected, it is not straightforward to identify the astrophysical 
origin of the emission \citep[see discussion in][]{Rauch08}. Here, since we detect other 
lines --from which we derive a high star formation rate-- most \lya\ photons seemingly 
originate from star formation activity.

\begin{figure}
\centering
\includegraphics[bb=216 20 396 810, clip=, angle=90, width=\hsize]{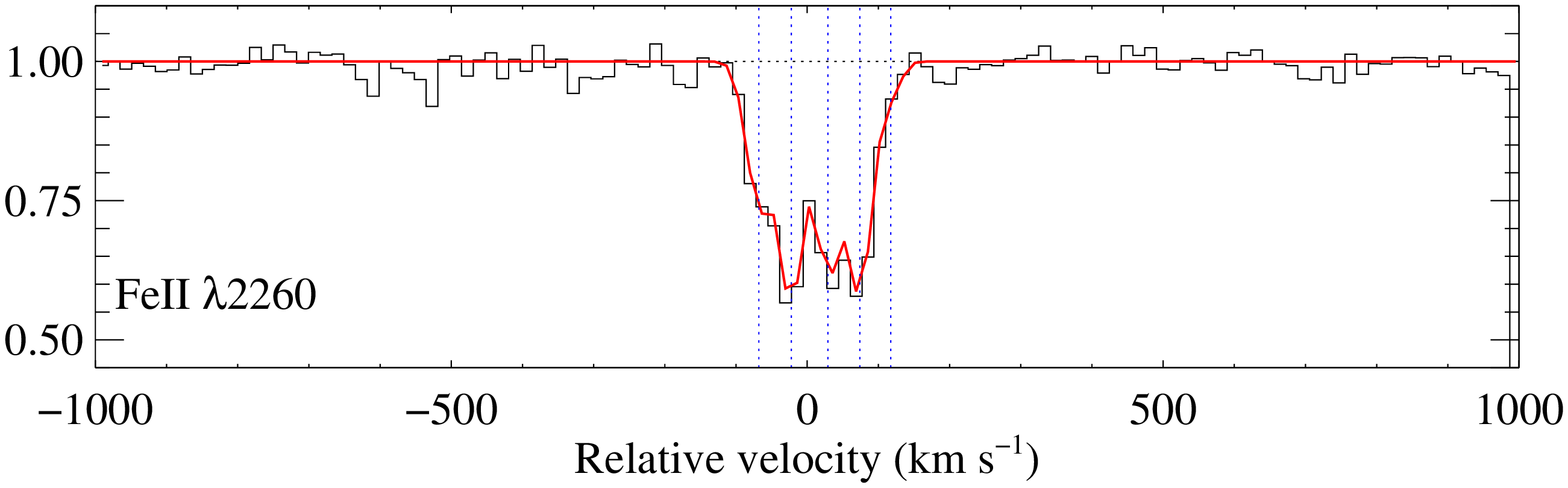}\\
\includegraphics[bb=216 20 396 810, clip=, angle=90, width=\hsize]{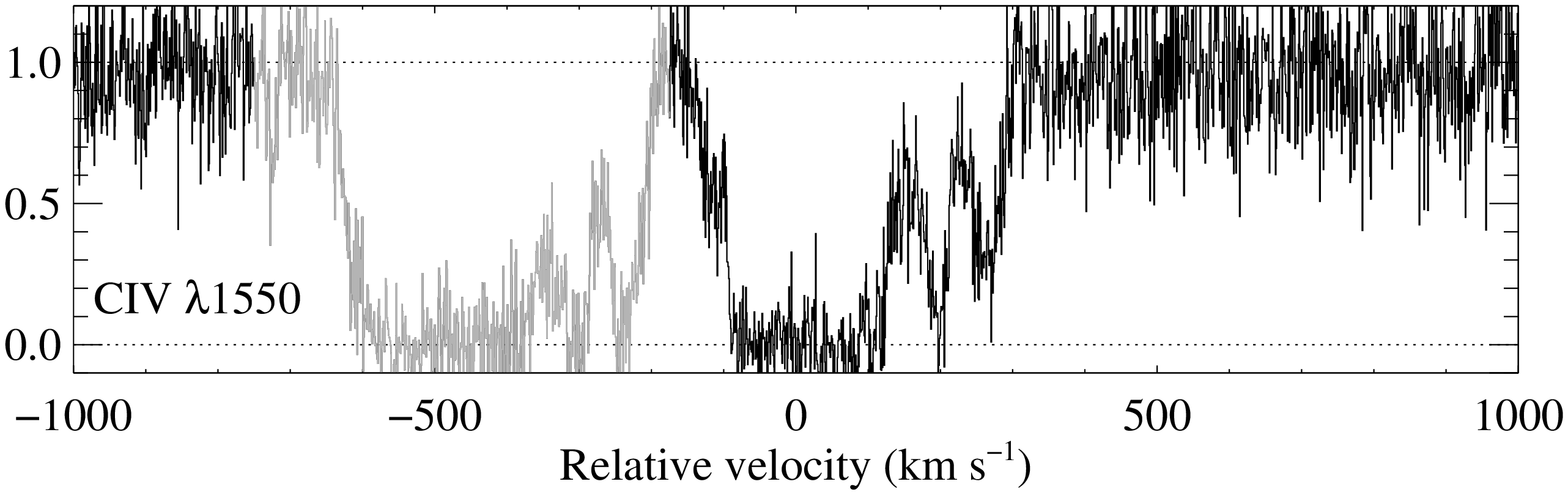}\\
\includegraphics[bb=216 20 396 810, clip=, angle=90, width=\hsize]{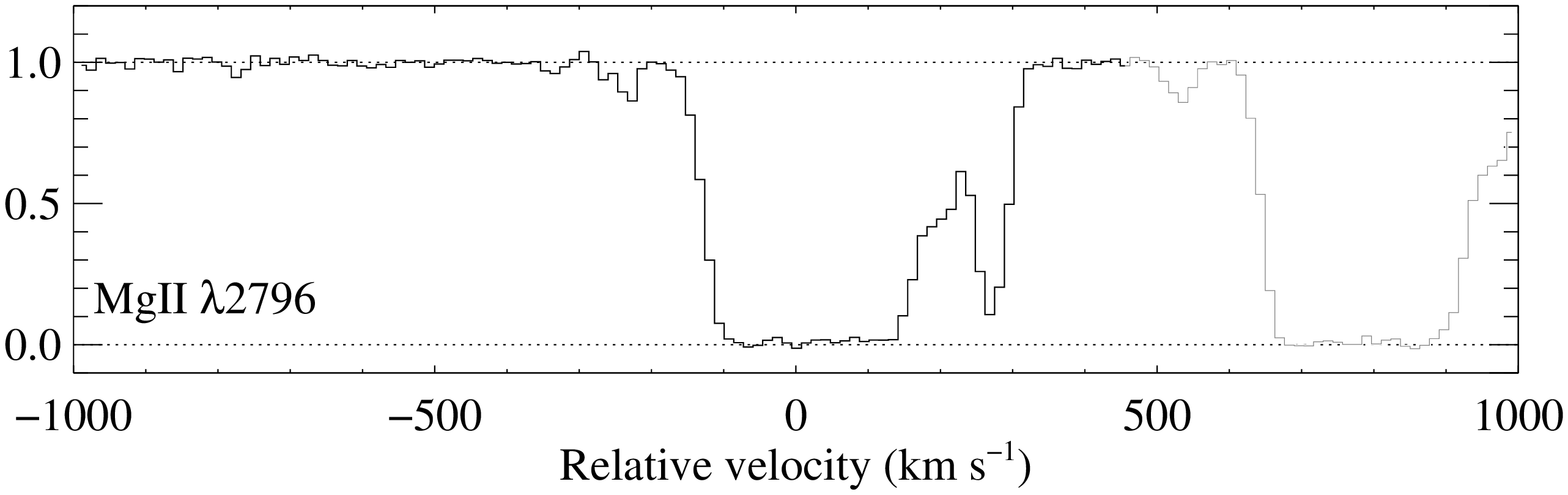}\\
\includegraphics[bb=216 20 510 810, clip=, angle=90, width=\hsize]{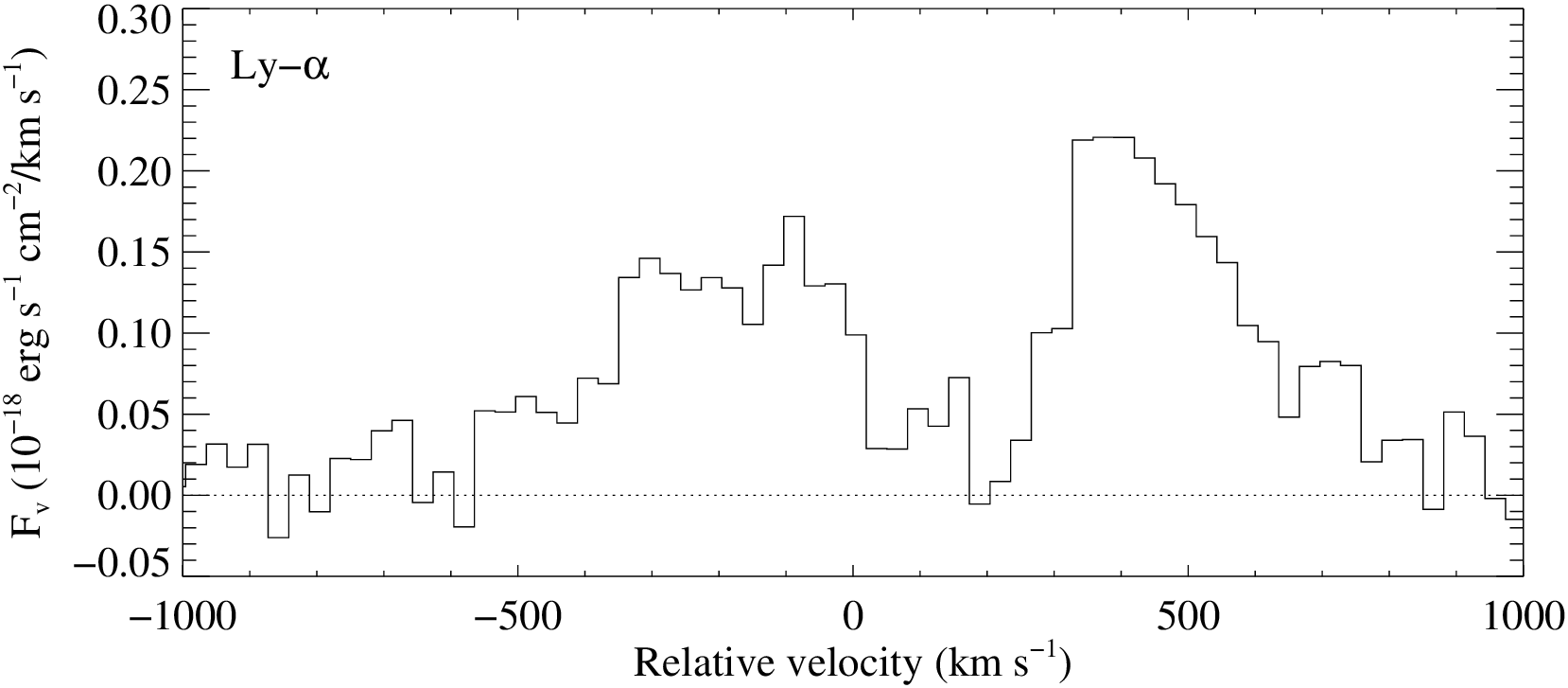}\\
\includegraphics[bb=216 20 510 810, clip=, angle=90, width=\hsize]{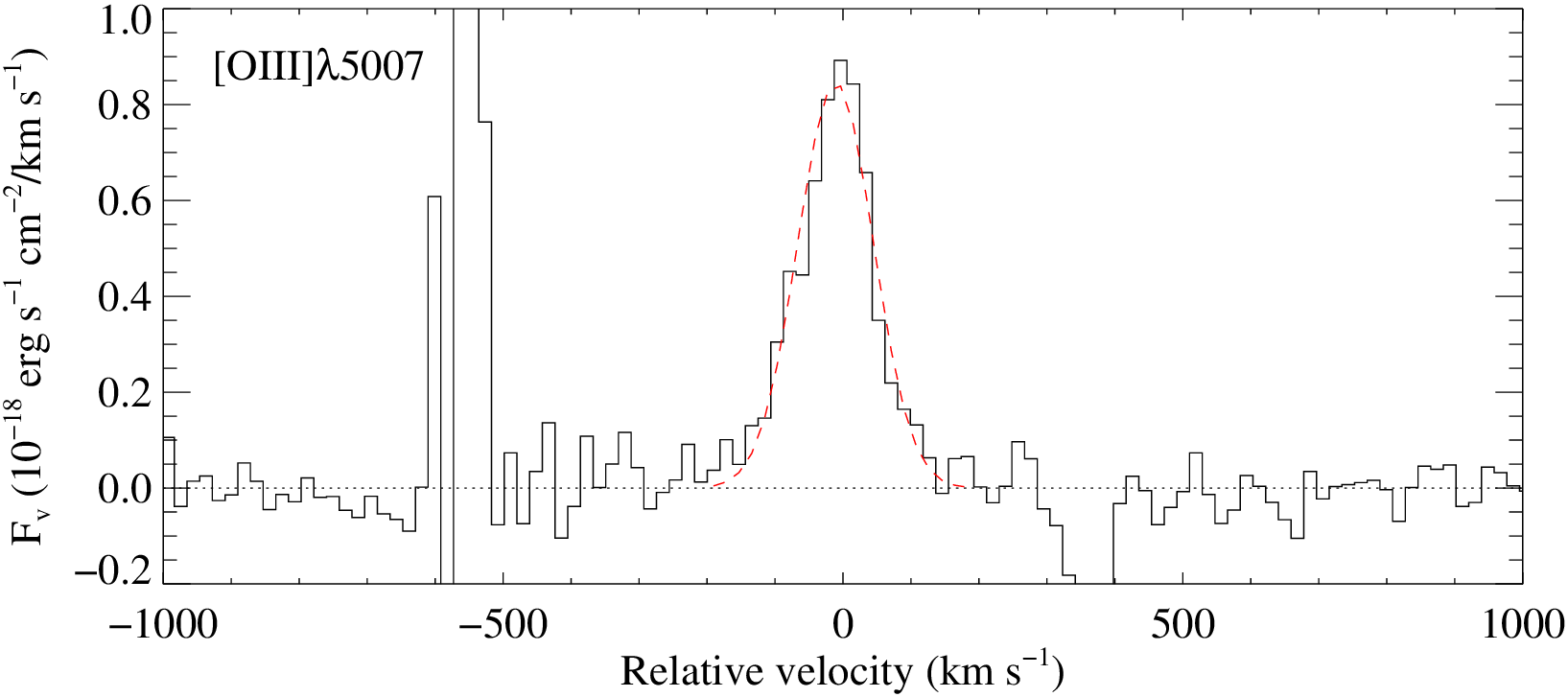}\\
\includegraphics[bb=216 20 510 810, clip=, angle=90, width=\hsize]{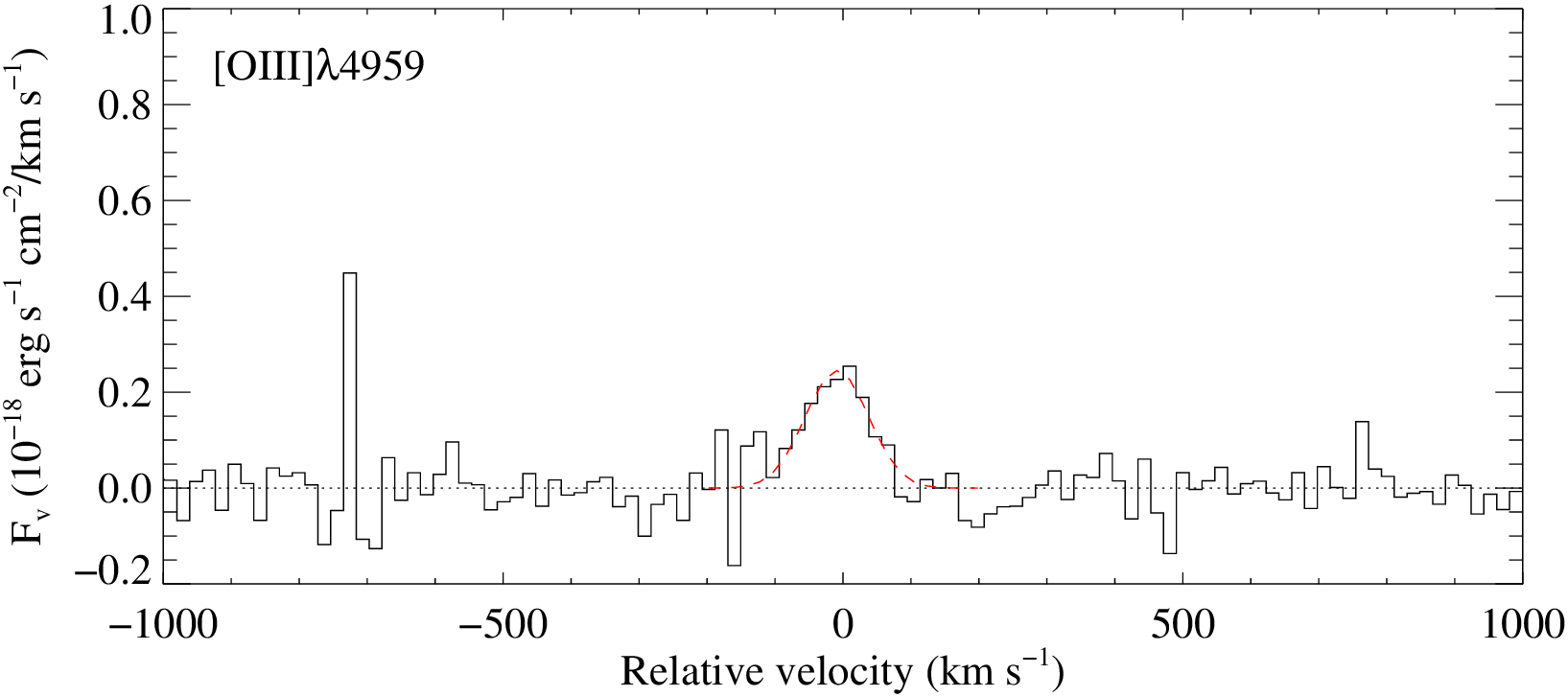}\\
\includegraphics[bb=165 20 510 810, clip=, angle=90, width=\hsize]{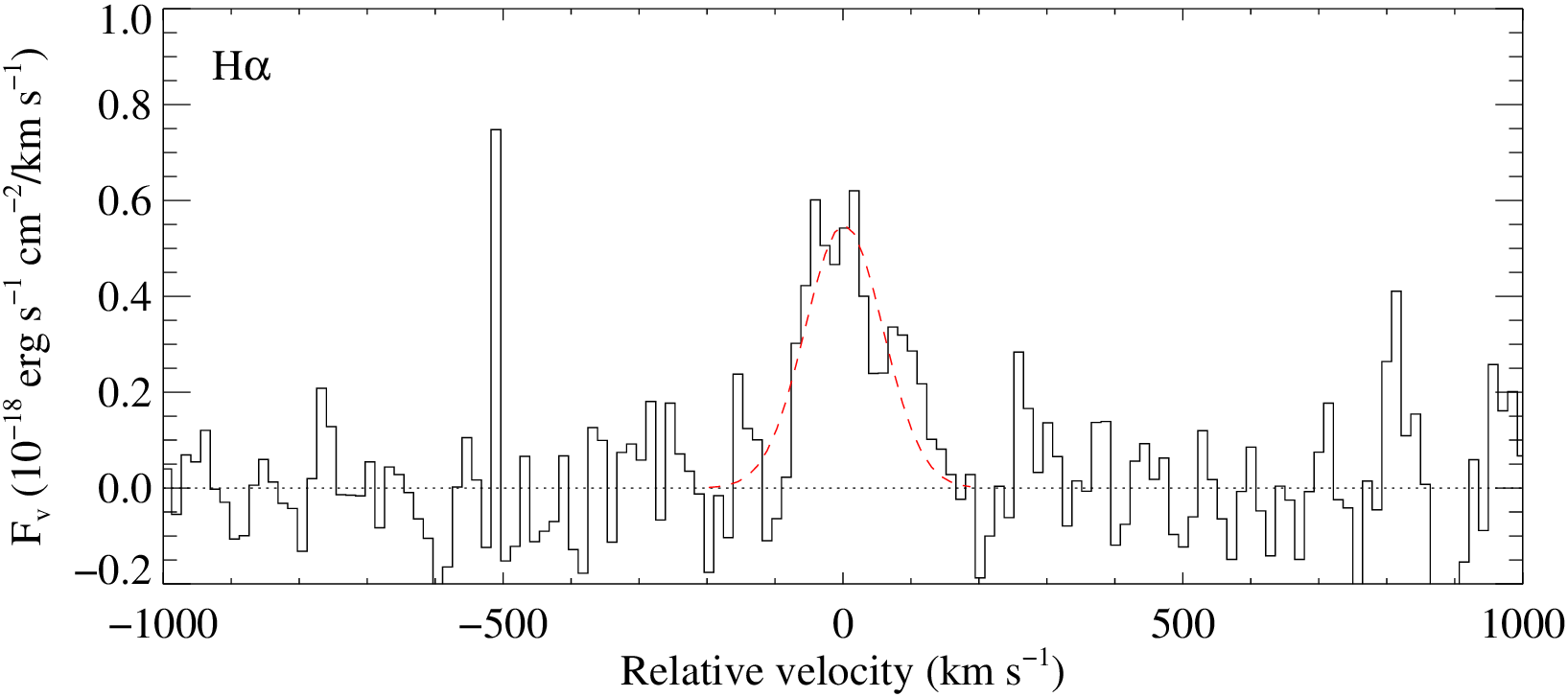}\\
\caption{Velocity plots of absorption lines (normalised spectra, top three panels) compared to that of emission 
lines (continuum-subtracted spectra, bottom four panels). 
The quasar continuum has been subtracted to isolate the emission lines. The grey absorption features at 
$v<-200$~\kms\ (resp. $v>450$~\kms) in the \CIV$\lambda$1550 panel (resp. \MgII$\lambda$2796) 
are due to \CIV$\lambda$1548 (resp. \MgII$\lambda$2803). 
The \CIV$\lambda$1550 panel shows the UVES spectrum smoothed by 4 pixels. Other panels present X-shooter data.
The features at $v=-550$ and +350~\kms in the [\OIII]$\lambda$5007 panel are skyline subtraction residuals. 
\label{vplot}}
\end{figure}

The velocity spread of the \lya\ line is much larger than that of \Ha\ and [\OIII] (Fig.~\ref{vplot}). This 
can be understood as the natural consequence of radiative transfer through optically thick \HI\ medium 
\citep[e.g.][]{Zheng02a, Dijkstra06a}: because \lya\ 
photons need to scatter out of the line centre in order to escape their emission region, we can expect an 
extended double-peaked profile. However, in practice, the resulting profile is a complex convolution of the input 
spectrum, the velocity and density fields as well as the turbulent and thermal motions.
Qualitatively, in case of an homogeneous, isotropic expanding shell around the source, the blue peak is 
severely diminished because \lya\ photons 
with higher frequencies are seen in the resonant frequency by the outflowing gas in the observer's 
direction. \lya\ photons can nevertheless escape being redshifted by back-scattering, producing a red peak. 
This is probably the case for the \lya\ emission associated to the $\zabs=2.35$ DLA towards Q\,2222$-$0946 
\citep{Fynbo10}. Conversely, the red peak is diminished in case of infalling gas. 

Here, the double-peaked profile 
could be more representative of a quasi-static medium around the emitting source \citep{Verhamme06}. 
However, in this scenario, we would expect that \lya\ photons with different frequencies are seen coming 
from the same projected region, i.e. that the spatial locations of the red and blue peaks coincide. We will 
see further (Sect.~\ref{sec_spa_lya}) that this is not the case.
Another possibility is the presence of scattering gas with high velocities in an anisotropic configuration. 
This can be the case for collimated winds or streams of infalling gas. 
Considering the spectral shape of \lya\ emission in a sample of high-$z$ UV-selected star-forming 
galaxies (with properties very similar to that studied here), \citet{Kulas12} also conclude that the \lya\ 
emission reveals complex velocity structures that cannot be reproduced with simple shell models. 
It is therefore important to study the kinematics of the 
emitting and scattering regions in order to understand the shape of the \lya\ profile as well as its 
spatial and velocity spread.

The velocity dispersion of the emitting galaxy can be inferred from the spectral width of the 
[\OIII] and \Ha\ emission lines\footnote{These lines are resolved at the spectral resolution of about 40~\kms\ in the NIR.} 
and is about FWHM$\sim$120~\kms. 
This provides an estimate of the projected rotational/dispersion velocity of the galaxy, though we note 
that turbulent motion in the star formation region can contribute to the width of the line. Nevertheless, the 
measured [\OIII] velocity width is also consistent with the velocity spread of metal absorption lines (see Fig.~\ref{vplot}). 
Using X-ray data, \citet{Martin99} have estimated that gas will escape through starburst-driven 
galactic winds from galaxies with $v_{\rm rot}<130$~\kms. 
The velocity measured here would therefore favour a wind scenario. 
In that case, as discussed above, the outflowing gas must be ejected anisotropically to preserve the blue and red peaks. 
Interestingly, from the UVES spectrum, two \CIV\ components are seen detached from the main profile at $v\sim 250$~\kms. These 
detached high-ionisation components likely arise from outflowing gas \citep[see e.g.][]{Fox07a,Fox07b}.
 
Another observational constraint comes from the \lya\ escape fraction. 
From the Ly$\alpha$/H$\alpha$ ratio, we observe $f_{\rm esc}\sim20\%$, considering case B recombination theory 
\citep{Brocklehurst71}. Such a high escape 
fraction is intriguing given the high \HI-column density measured close to the star-forming region.
This suggests that \lya\ photons escape efficiently along particular directions with low column densities 
\citep[e.g.][]{Tenorio-Tagle99,Mas-Hesse03}, high velocity gradients \citep{Kunth98,Ostlin09,Atek08} and/or bounce 
off of the surface of dense clouds in a clumpy medium \citep[e.g.][]{Neufeld91, Atek09}. 
Finally, we note that absorption by the intergalactic medium (IGM) could in principle still affect the blue peak 
of the \lya\ emission. However, \citet{Laursen11} showed that the consequences of intergalactic \lya\ transmission 
are important only for $z>3.5$, where the \lya\ forest is denser. 

\subsection{Spatial configuration and mass \label{sec_spa_lya}}

\begin{figure}
\centering
\includegraphics[bb = 95 220 485 398,clip=,width=0.9\hsize]{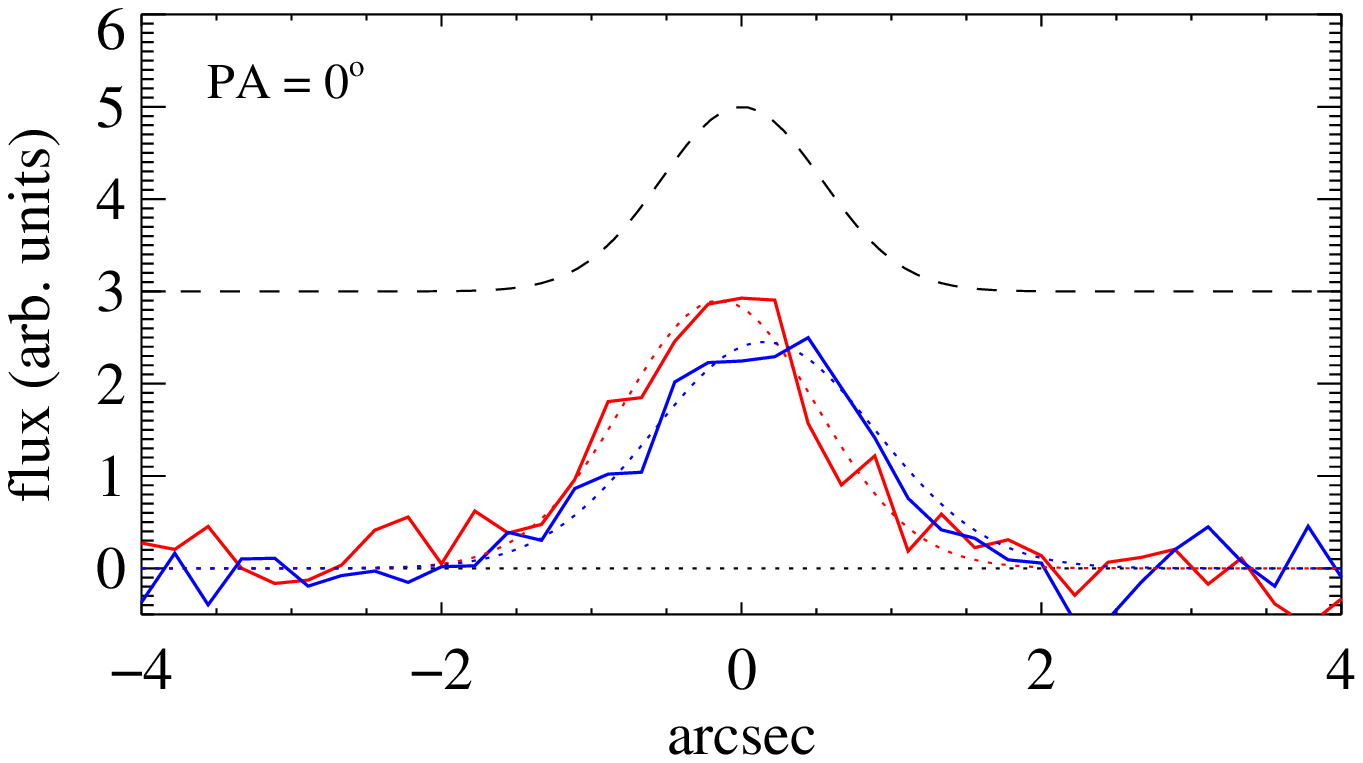}
\includegraphics[bb = 95 220 485 398,clip=,width=0.9\hsize]{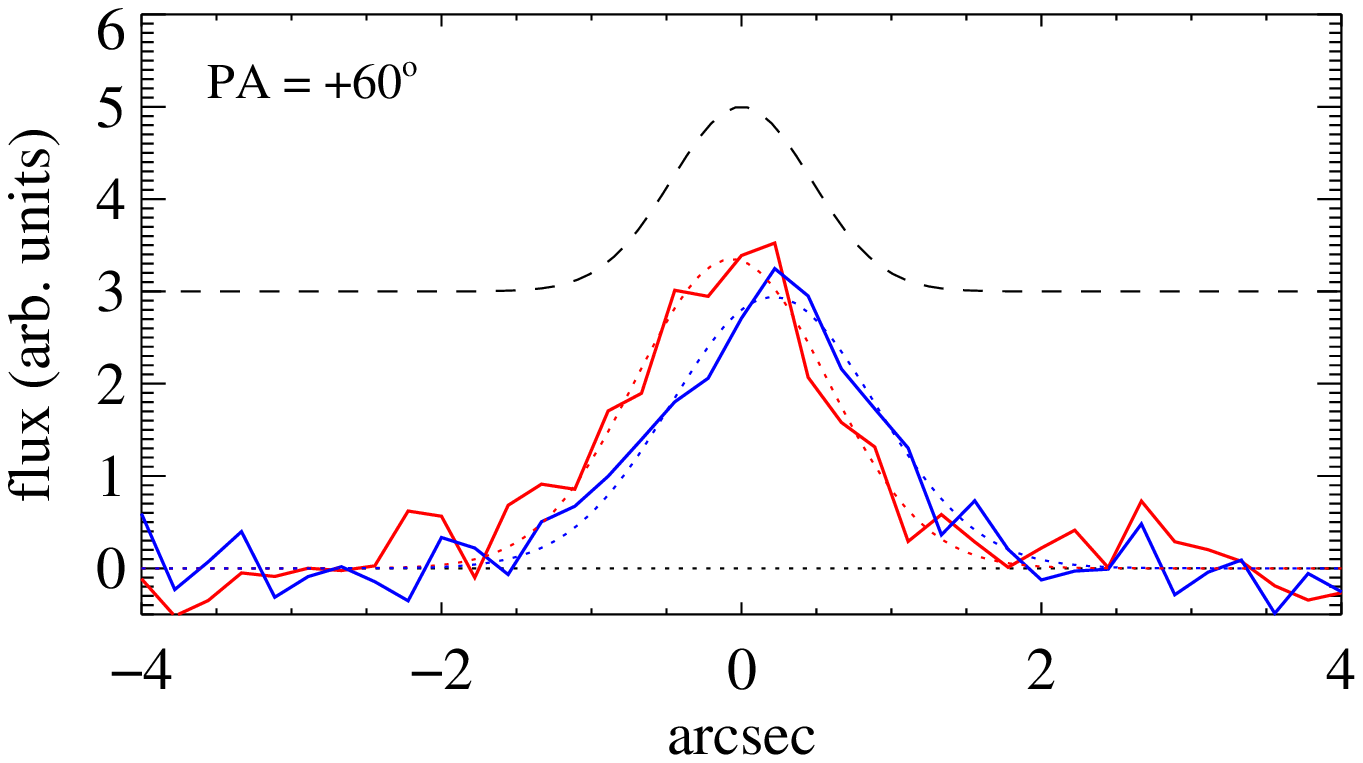}
\includegraphics[bb = 95 180 485 398,clip=,width=0.9\hsize]{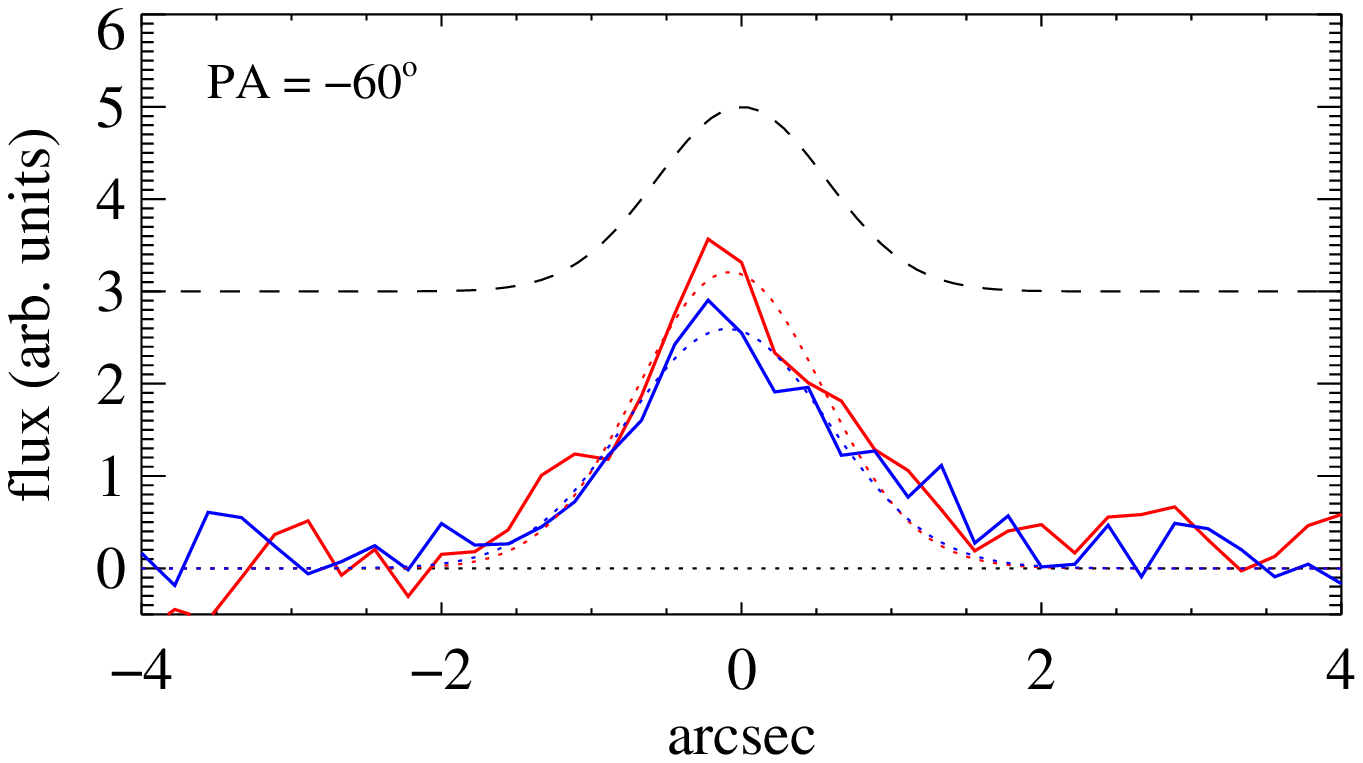}
\caption{Observed spatial profiles of the \lya\ emission. The red and blue profiles correspond 
to integrations of the emission over the velocity ranges [-500,0] and [200,700]~\kms~(see Fig.~\ref{vplot}), 
respectively. The dotted curves are Gaussian fits to the observed profiles. The dashed profile illustrates the 
spectral PSF, obtained from fitting the QSO trace. \label{spa_lya}}
\end{figure}

\begin{figure}
\centering
\includegraphics[bb=70 180 520 610,clip=,width=\hsize]{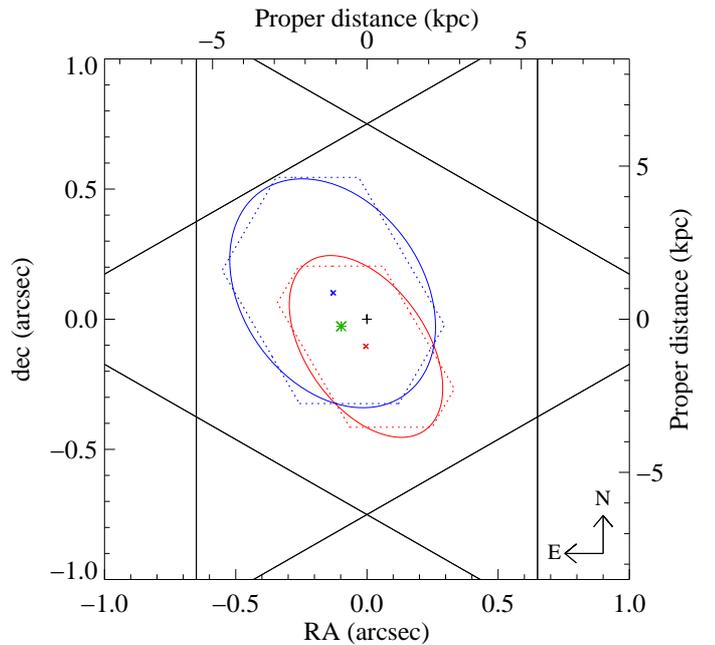}
\caption{Summary of the spatial configuration. The QSO location is marked by the plus sign ($+$). 
The location of the [\OIII] emission is marked by a green asterisk. The red and blue dotted polygons 
illustrate the spatial FWHM for the red and blue peaks (see Fig.~\ref{vplot}) as shown in Fig.~\ref{spa_lya}. 
The spatial FWHM extent of \lya\ peaks is then fitted with ellipses.
Triangulating the impact parameters for each PA provides the location of the centroids ($\times$). 
Solid black lines represent the layout of the three slits. Top and right axis provide the physical 
projected distance in proper units at $z=2.207$. \label{im_lya}}
\end{figure}

In Fig.~\ref{spa_lya}, we compare the location and extension of the \lya\ emission along the slit with 
the spectral PSF, obtained from fitting the QSO trace in the same wavelength region. We measure the 
positions of the blue and red peaks along each slit (for each PA) by fitting Gaussian profiles to
the corresponding Ly$\alpha$ emissions. 
The triangulation strongly constrains the location of the centroid of the red and 
blue emitting regions (see Fig.~\ref{im_lya}). As for [\OIII], the three constraints from the different PAs 
intersect well in a single small region. Interestingly, the centroid of the blue and red emissions do 
not coincide and are found on each side of the location of the [\OIII] emission.

From Fig.~\ref{spa_lya} it is also clear that the \lya\ emission is wider than the seeing. While the 
observed spatial profile is the convolution of the spectral PSF and the actual brightness profile, we 
found that this can be well reproduced by a simple Gaussian model. We then deconvolved 
the observed profile and obtained the spatial extend of the emission for each PA. The results are shown in 
Fig.~\ref{im_lya} as dotted polygons which are the intersection of the FWHMs along each PA. These regions 
where \lya\ photons are detected are then well fitted by ellipses.

The observed configuration
can be explained if the QSO line of sight passes through the inner gas-rich parts 
(possibly disc) of a compact galaxy with a radius of the order of $\sim3$~kpc (as constrained by the spatial extent of 
the [\OIII] emission). The expansion of a starburst-driven super-bubble is 
quickly stopped in the dense gas region, and naturally shape bipolar outflows \citep[e.g.][and references therein]{Veilleux05}. 
\lya\ photons arising from the vigorous star formation region escape preferentially in the wind directions where the column 
densities are lower \citep{Heckman01} and scatter off collimated outflowing gas (with velocities of the order 
of at least several 100~\kms, as inferred from the detached \CIV\ absorption components). 
This would produce the two distinct observed \lya\ emission regions. We will test this picture in Sect.~\ref{model}.

Considering a DLA galaxy half-light radius of a few kpc (see Sect.~\ref{impact_sfr}), the velocity dispersion 
(measured in Sect.~\ref{dplya}) provides us with a rough estimate of the virial mass of 
the galaxy to be of order of 10$^{10}$~M$_{\odot}$, similar to that of the SMC \citep[e.g.][]{Bekki09}. 
The DLA studied here seems therefore to arise from an intermediate mass galaxy. The mass of neutral gas within the galaxy radius 
can also be estimated --at best at the order of magnitude level-- to be a few $10^{9}$~M$_{\odot}$ (considering a 
uniform column density over a projected region of a few kpc$^2$).
Simulations have revealed the existence of an anti-correlation between the mass of the galaxy and 
the escape fraction \citep[see e.g. Fig. 9 of][]{Laursen09b}. According to these simulations, the low virial 
mass of the DLA galaxy here is well consistent with the large \lya\ escape fraction. Finally, the high 
star formation rate and the low virial mass are favourable to starburst-driven outflowing winds 
\citep[e.g.][]{Samui08}. 

There are evidences that strong \MgII\ systems are associated to star-forming galaxies at intermediate redshifts. 
For example, the correlation between the [\OII] luminosity and the \MgII\ equivalent width \citep{Noterdaeme10oiii} 
indicates that 
stronger \MgII\ systems are found closer to their host galaxy. While this correlation does not necessarily imply 
the presence of winds \citep[see][]{Lopez12}, \citet{Nestor11} observed an association between the most extreme 
absorbers and starburst galaxies. 
These authors further showed that galactic outflows are necessary to account for their large velocity spread. 
Therefore, the extremely 
large \MgII\ equivalent equivalent width measured here could further support an association with a starburst 
galaxy. 
This is also in agreement with the suggestion by \citet[][]{Bouche11} that strong \MgII\ absorbers could trace 
star formation-driven 
winds in low-mass haloes ($M_h \le 5\times 10^{10} M_{\odot}$). Indeed, the \MgII\ profile is spread beyond the 
bulk of low-ionisation metals and, as seen for \CIV, presents detached components that likely arise from the low column 
density outflowing gas (see Fig.~\ref{vplot}).

\subsection{Stellar continuum}

We can obtain further information on the nature of the DLA galaxy by constraining its stellar UV continuum
at the bottom of saturated \lya\ lines at $z\ge z_{\rm gal}$. 
These lines arise from IGM clouds that are located behind the galaxy and therefore completely absorb the 
quasar flux but not that of the galaxy.
Conversely, saturated \lya\ lines at $z<z_{\rm gal}$ absorb in turn photons from both the QSO and the galaxy at the 
corresponding wavelength. 
In order to check the accuracy of the sky subtraction, we measured the mean residual zero-flux 
level at the bottom of a saturated line at $\lambda\approx3792$~{\AA} (i.e. at  $z<z_{\rm gal}$) to be 
$0.8\pm7.0\times10^{-19}$~erg\,s$^{-1}$\,cm$^{-2}$\,{\AA}$^{-1}$.

\begin{figure*}
\centering
\includegraphics[bb=200 90 385 740,angle=90,width=0.98\hsize]{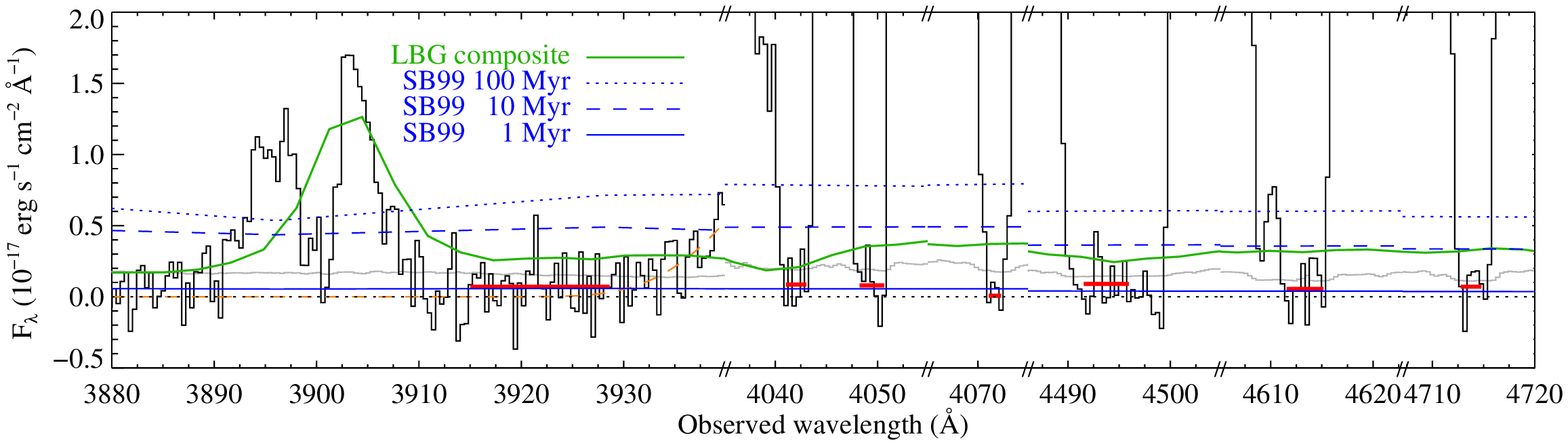}
\caption{X-shooter spectrum in the bottom of saturated lines at $z>z_{\rm dla}$ (black, with error spectrum in grey). 
The orange dashed curve raising 
at $\sim 3930$-3940~{\AA} shows the bottom of synthetic Voigt profile with $\log N(\HI)=22.10$. The green spectrum 
is the LBG composite spectrum from 
\citet{Kornei10}, scaled to match the integrated \lya\ emission line. Note that the \lya\ emission line is commonly 
diminished in the blue 
for LBGs, resulting in the apparent shift of the composite \lya\ line towards redder wavelengths.
Blue lines are synthetic stellar continua calculated with {\sc starburst99} 
\citep{Leitherer99}, using a Salpeter IMF, metallicity $Z=0.004$, and continuous SFR~=~25~M$_\odot$\,yr$^{-1}$ since 
respectively, 100~Myr (dotted), 10~Myr (dashed), 1~Myr (solid).  Horizontal segments illustrate the mean fluxes 
in the corresponding regions. Note that in the case of the line at $\sim$4495~{\AA}, the \lya\ absorption is partially 
blended with \SiIV\,$\lambda$1402 from the DLA. The blended region (4497-4500\,{\AA}) is therefore 
not considered to compute the mean.\label{bottomuv}}
\end{figure*}

In Fig.~\ref{bottomuv}, we plot portions of the X-shooter spectrum around saturated \lya\ lines at  $z \ge z_{\rm gal}$ 
(including the red side of the DLA line as in \citealt{Moller04}). While the UVES spectrum is far too 
noisy\footnote{The error level is 
around 4.5$\times$10$^{-17}$~erg\,s$^{-1}$\,cm$^{-2}$\,{\AA}$^{-1}$.} to detect the \lya\ emission 
--and {\sl a fortiori} the continuum--, we used its higher 
spectral resolution to confirm that the lines are truly saturated and not blends of lines. 
In the figure, we compare the X-shooter line-flux residuals with the LBG composite spectrum from \citet{Kornei10} scaled to 
match the observed Lyman-$\alpha$ flux. The observed continuum flux is lower than what is expected from a typical 
Lyman-break galaxy {\sl with the same \lya\ flux}, or equivalently, the \lya\ equivalent width is larger in the 
DLA galaxy than in the LBG composite.
This could indicate that the DLA galaxy has a more recent stellar population than typical LBGs, as expected in the 
case of a young starburst galaxy.  
If the star formation is recent in the DLA galaxy, we can expect that it has not yet reached the plateau in the luminosity 
at $\sim1500$~{\AA} (with contribution from both massive and intermediate-mass stars) but that it has a higher relative 
flux further in the UV (mostly from massive-stars), which triggers the \Ha\ line.

Using {\sc starburst99} \citep{Leitherer99}, we estimate that the stellar UV luminosity of the galaxy would reach 
about $7\times10^{41}$~erg\,s$^{-1}$\,{\AA}$^{-1}$  at 1500~{\AA} for a Salpeter IMF, a metallicity $Z=0.004$ (close to what 
is observed both in the emitting and absorbing gas) and the \Ha-derived SFR of 25~M$_\odot$\,yr$^{-1}$ in about 100\,Myr. 
This conclusion is little dependent on the adopted IMF and exact value of the metallicity.
The UV luminosity plateau is much higher than what is observed, indicating that the star formation in the galaxy is 
very recent. The predicted spectrum at $10^7$ yrs is still {significantly} higher than the observed spectrum while 
models at $10^6$~yrs predict a flux below the observed one (see Fig.~\ref{bottomuv}). We recall however that we cannot 
correct for dust extinction because 
the \Hb\ line is not available, meaning that models can overpredict the observed flux for a given age.
Since the \lya\ escape fraction is large, the metallicities are low and the amount of dust along the QSO line of sight 
(i.e., very close 
to the star-forming region) is small, we can assume that the effect of dust on the UV continuum is probably not 
very large. 
The mean flux in the saturated lines is 11$\pm$1.1$\times10^{-19}$~erg\,s$^{-1}$\,cm$^{-2}$\,{\AA}$^{-1}$, while the 
systematic error due to sky subtraction is probably much lower than that estimated above (from a single line in a bluer 
region of the spectrum).
This means that we likely detect the galaxy continuum, as can 
also be seen directly from Fig.~\ref{bottomuv} where the red segments are systematically above zero (dotted line).
This exercise therefore gives us a rough estimate of the age of the star-forming activity. 
Even allowing for a factor of 5 underestimate of the UV luminosity (due to dust extinction and uncertainties 
in the sky subtraction), we find that the 
star formation has been triggered $\la 10^7$~yrs before the rest-frame time of the observation, with a enhanced 
population of massive stars. This age is typical of what is seen in \lya-selected LAEs at high redshift 
\citep[e.g.][]{Pirzkal07,Gawiser07,Lai08,Finkelstein09}.


\section{Modelling the galaxy counterpart  \label{model}}

Unlike non-resonant lines like \Ha, the \lya\ line is not
readily fitted by, e.g., a Gaussian. Many processes dictate the shape of the
line, and as discussed in Sect.~\ref{dplya}, analytical fits are only available
for the very simplest models, such as a spherical, homogeneous, isothermal,
dustless, static blobs of gas \citep{Dijkstra06a}.
We must turn to numerical simulations to build more realistic models. 
Resonant scattering being stochastic in nature, so-called Monte Carlo simulations are an obvious
choice, where the journey of individual photons are traced as they scatter
their way out of a simulated or modelled galaxy
\citep[e.g.][]{Zheng02a,Tasitsiomi06,Verhamme06,Laursen07}.  

To this end, we apply the \lya\ radiative transfer code {\sc MoCaLaTA}
\citep{Laursen09a}, including the effects of dust \citep{Laursen09b}, to a
semi-realistic model galaxy. 
As discussed in Sect.~\ref{dplya}, \lya\ photons escaping from an expanding agglomeration of gas tend 
to diffuse in frequency to the red side of the line centre. 
Conversely, for collapsing gas the photons escape more easily after having
diffused to the blue side \citep[see e.g.][]{Dijkstra06a}.
Both scenarios are met in astrophysical contexts, although by far most observed
\lya\ lines are characterised by an asymmetric red wing
\citep[e.g.][]{Venemans05,Nilsson07,Tapken07,Grove09}.

\begin{figure}
\centering
\includegraphics[width=0.8\hsize]{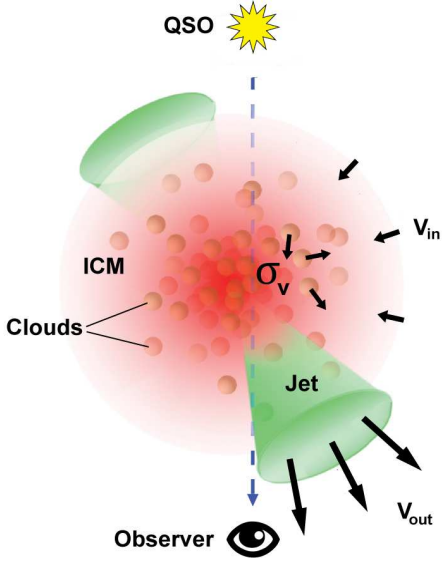}
\caption{Graphical representation of the model galaxy: A number of cool,
high-density, and dusty clouds (\emph{brown}) are randomly dispersed in a
hotter and
more tenuous inter-cloud medium of exponentially decreasing density (\emph{red}).
Two jets of high ionisation fraction emerge in opposite directions
(\emph{green}) and the QSO is observed along the line of sight passing within
$\sim 1$ kpc of the galactic centre (\emph{dashed blue line}).
\label{fig_galmod}}
\end{figure}

Here, we interpret the spatial separation of the blue 
and the red part of the spectrum found in Sect.~\ref{sec_spa_lya} as being caused by a 
smooth, overall cold gas accretion, with a high, chiefly central star formation giving rise to two
jets of outflowing gas. 
The galaxy is hence modelled as a sphere of gas existing in two phases; warm, spherical 
neutral clouds of high gas and dust density, dispersed in a hotter and more tenuous inter-cloud 
medium (ICM). Both the gas density of the ICM and the number density of clouds decrease 
exponentially from the centre.
Two jets of higher ionisation fraction are assumed to expel
gas at high velocities (as seen from the detached \CIV\ components, see Fig.~\ref{vplot}) 
from near the centre and outward in opposite directions,
while the rest of the galaxy is characterised by a net small overall infall of gas. 
We assume that there are no neutral clouds inside the jets. 
The model is illustrated in Fig.~\ref{fig_galmod}.

A model galaxy is constructed on a grid of $512^3$ cells, each cell being
characterised by the following parameters: neutral hydrogen density $\nhi$,
dust density $n_{\mathrm{d}}$, gas temperature $T$, and three-dimensional bulk
velocity $\vec{v}$. The parameters are uniform within a given cell.
In essence, a large number of \lya\ radiative transfer (RT) calculations are conducted, and the
favoured model is then the one that at the same time reproduces reasonably well
not only the observed spectrum, but also the other observables we have at hand:
\emph{i)} a SFR of $\sim 25 M_\odot$ yr$^{-1}$, 
\emph{ii)} a \lya\ escape fraction of $\sim 0.20$,
\emph{iii)} the location of blue and red photons on the sky seen in
         Figs.~\ref{spa_lya} and \ref{im_lya} with respect to the galactic
         centre given by the [\OIII] emission,
\emph{iv)}  a large \HI\ column density $\sim 10^{22}$ \cmsq\ and colour excess
         $E(B-V) \sim 0.04$ through the whole system at the sightline toward
         the quasar, and
\emph{v)} a few ($\sim$ 8) clouds intercepted by the QSO line of sight, spanning a velocity 
         range $\sim [-100,100]$ \kms, as inferred from the metal absorption lines
         (Fig.~\ref{metalsUVES}).

Many observational constraints we have on the physical parameters of the
galaxy allow us to make a qualified estimate of some of the input
parameters of the model:

The inferred metallicity of [Zn/H] $\sim -1.1$, together with the depletion
ratios from Sect.~\ref{sec_met}, give us an idea about the amount of dust.
Assuming again an SMC extinction law, and scaling the extinction in each cell
to the appropriate metallicity, provides $n_{\mathrm{d}}$.

The dust is assumed to scale also with $N(\HI)$, i.e. with a reduced density 
in the intercloud medium (ICM). In this picture, dust is destroyed in regions 
where hydrogen is ionised. Furthermore, the jet is assumed to be devoid of dust. 
Note, however, that since the largest part of the
photons' journey takes place inside the high density clouds, the presence of
dust outside the clouds does not have a strong impact on the RT.

Distributing the total \HI\ column density to several clumps implies an average cloud column density 
of the order of a few $10^{21}$\cmsq. 
This provides a relation between cloud size and neutral hydrogen density inside the clouds 
(e.g. for $\nhi \sim$ a few cm$^{-3}$, the clouds should be a few hundred parsec across).
The extent of the [\OIII] emission (of the order of 3 kpc, see
Sect.~\ref{impact_sfr}) constrains the physical size of the system.

The rest of the variables are estimated from known typical values.
To find the best-fitting model, in principle an $n$-dimensional grid of the $n$
different parameters controlling the RT could be constructed, running the RT
calculations for every model, and finding the best one through $\chi^2$
minimisation. This approach has already been followed, albeit for an 
outflowing shell model of four parameters only. \citet{Verhamme08} and \citet{Schaerer11} fitted
observed \lya\ profiles by modelling the galaxies as a thin shell characterised
by a column density of homogeneously mixed gas and dust with some temperature
and expansion velocity. In our model, however, a much higher number of
parameters may be varied, and searching a parameter space this big is not
feasible.
Furthermore, as made clear above we aim not only the fit the spectrum, but to
reproduce several observables. Consequently,
instead, in practice we run a large number of RT calculations with different
initial conditions, inspect the results visually, change slightly one or a few
parameters which from experience are believed to be the appropriate ones before
running a new series, until a satisfactory set of results are obtained.

\subsection{Results from modelling}
The values used for the preferred model are given in Table~\ref{tab_params}.
With such a large number of parameters, of which many are to a large extent
degenerate, we have no aspirations of matching perfectly all of the mentioned
observed quantities. Nevertheless, it is striking that a relatively simple
model manages to give meaningful results.

\begin{table}
\centering
\caption{Model input parameters \label{tab_params}}
\begin{tabular}{ll}
\hline
\hline
{\large \strut}Parameter                           & Value                         \\
\hline
{\large \strut}Box size                            & 10 kpc                        \\
Galaxy radius                       & 3 kpc                         \\
Number of clouds                    & 4000                          \\
Cloud radii                         & 150 pc                        \\
Cloud temperature                   & $10^4$ K                      \\
Cloud density\tablefootmark{a}      & 5 cm$^{-3}$                   \\
Cloud/ICM metallicity               & 0.08 Z$_\odot$                 \\
Cloud velocity dispersion           & 60 \kms                       \\
Overall infall velocity             & 60 \kms                       \\
Central ICM density\tablefootmark{a}           & $10^{-2}$ cm$^{-3}$           \\
ICM density\tablefootmark{a} scale length      & 2 kpc                         \\
ICM temperature                     & $5\times10^4$ K               \\
Jet inner radius\tablefootmark{c}              & 0.5 kpc                       \\
Jet outer radius\tablefootmark{c}              & 5   kpc                       \\
Jet terminal\tablefootmark{c} velocity         & 400 km s$^{-1}$               \\
Total jet column density\tablefootmark{a}      & $5\times10^{19}$ cm$^{-2}$    \\
Jet temperature                     & $5\times10^5$ K               \\
Jet opening angle                   & 50$^\circ$                    \\
Jet direction\tablefootmark{d}                 & $(0.25, -0.25, 1)$            \\
\lya\ emission scale length         & 1.5 kpc                       \\
\hline
\end{tabular}\\
\tablefoot{
\tablefoottext{a}{All densities refer to \emph{neutral} hydrogen only.}
\tablefoottext{c}{The gas accelerates from 0 km s$^{-1}$ to the terminal
velocity going from the inner to the outer radius of the jet.}
\tablefoottext{d}{Jet direction is in the coordinate system where the observer
is located at $z = \infty$ (or actually $z = d_L$).}
}
\end{table}

\subsubsection*{Spectrum}

\begin{figure}
\centering
\includegraphics[width=0.95\hsize]{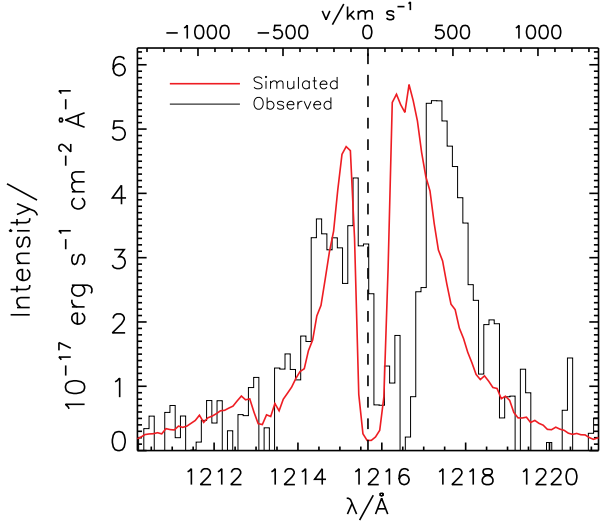}
\caption{Comparison of the simulated (\emph{red}) and observed (\emph{black}) Lyman-$\alpha$ emission profiles. 
The spectra are shown in the rest frame of the DLA at $z=2.2066$. \label{fig_Lyafit}}
\end{figure}

The synthetic spectrum of the best model is seen in Fig.~\ref{fig_Lyafit}.
Although qualitatively the spectrum matches the observed spectrum reasonably,
we have not been able to reproduce the apparent shift of the 
trough between the \lya\ peaks and the centroid of other emission lines of around
200~\kms. 
Since \lya\ and \Ha\ are produced in the same regions, this shift is
rather puzzling. However similar shifts have recently been observed in several 
systems by \citet{Kulas12}. As discussed by these authors, complex geometry and 
velocity fields are needed to explain the observed profile. This means that our 
model is probably still too simplistic but increasing its complexity (and hence 
the number of parameters) would be beyond the scope of this paper.

Taking into account 
the simulated escape fraction in the direction of the observer (0.23), a SFR of 
24~$M_\odot$\,yr$^{-1}$ was needed to fit the observed spectrum. These values 
are surprisingly well consistent with the observed ones (see Table~\ref{line_fluxes}).

\subsubsection*{Spatial separation of the red and blue photons}

\begin{figure*}
\centering
\includegraphics[width=\hsize]{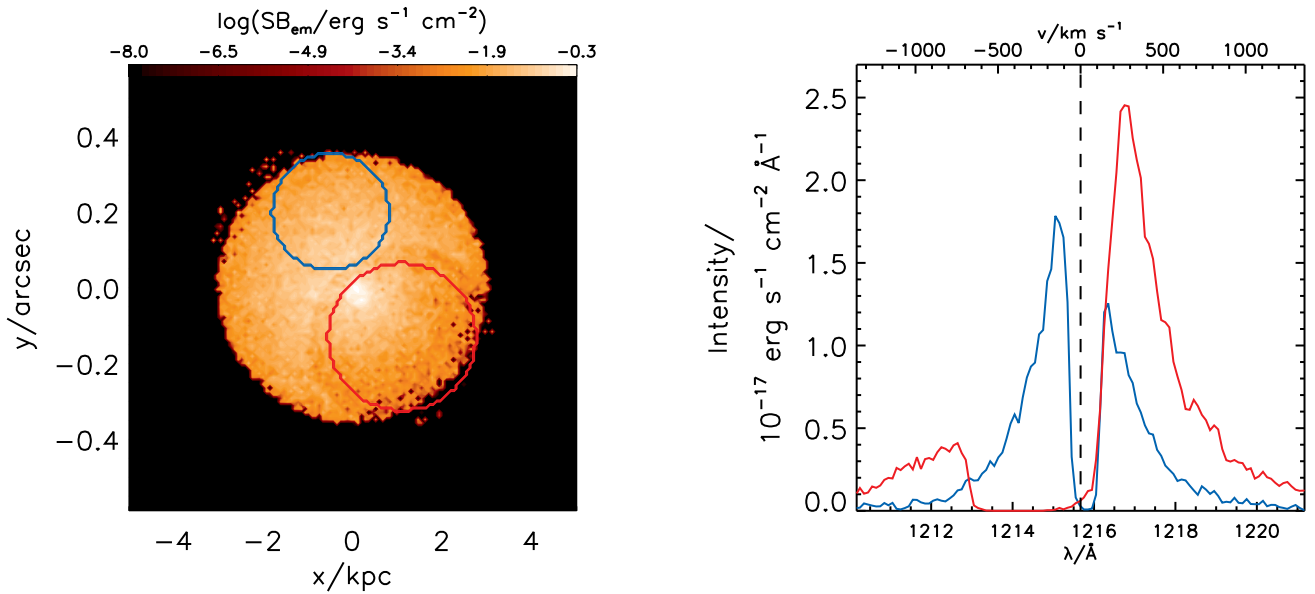}
\caption{Surface brightness map of the model galaxy (\emph{left}) and spectra
(\emph{right}) of the \lya\ light within the two apertures indicated on the left in
\emph{blue} and \emph{red}.
More red \lya\ photons emerges from the jet, whereas the
galaxy itself emits bluer \lya\ light.}
\label{fig_Aperture}
\end{figure*}

Figure \ref{fig_Aperture} shows the surface brightness (SB) map of the simulated galaxy, along with
the spectra extracted from two different apertures.

The jet covers the lower right portion of the galaxy, and it is seen that this
effectively blocks blue \lya\ photons from escaping. On the other hand, the slight
infall onto the rest of the galaxy skews the spectrum from that part toward
being bluer. Consequently, the different parts of the spectrum indeed seem
to escape at different locations on the sky. This SB map matches at least
qualitatively the spatial configuration inferred in Sect.~\ref{sec_spa_lya}. 

\begin{figure}
\centering
\renewcommand{\tabcolsep}{1pt}
\begin{tabular}{cc}
\includegraphics[width=0.45\hsize]{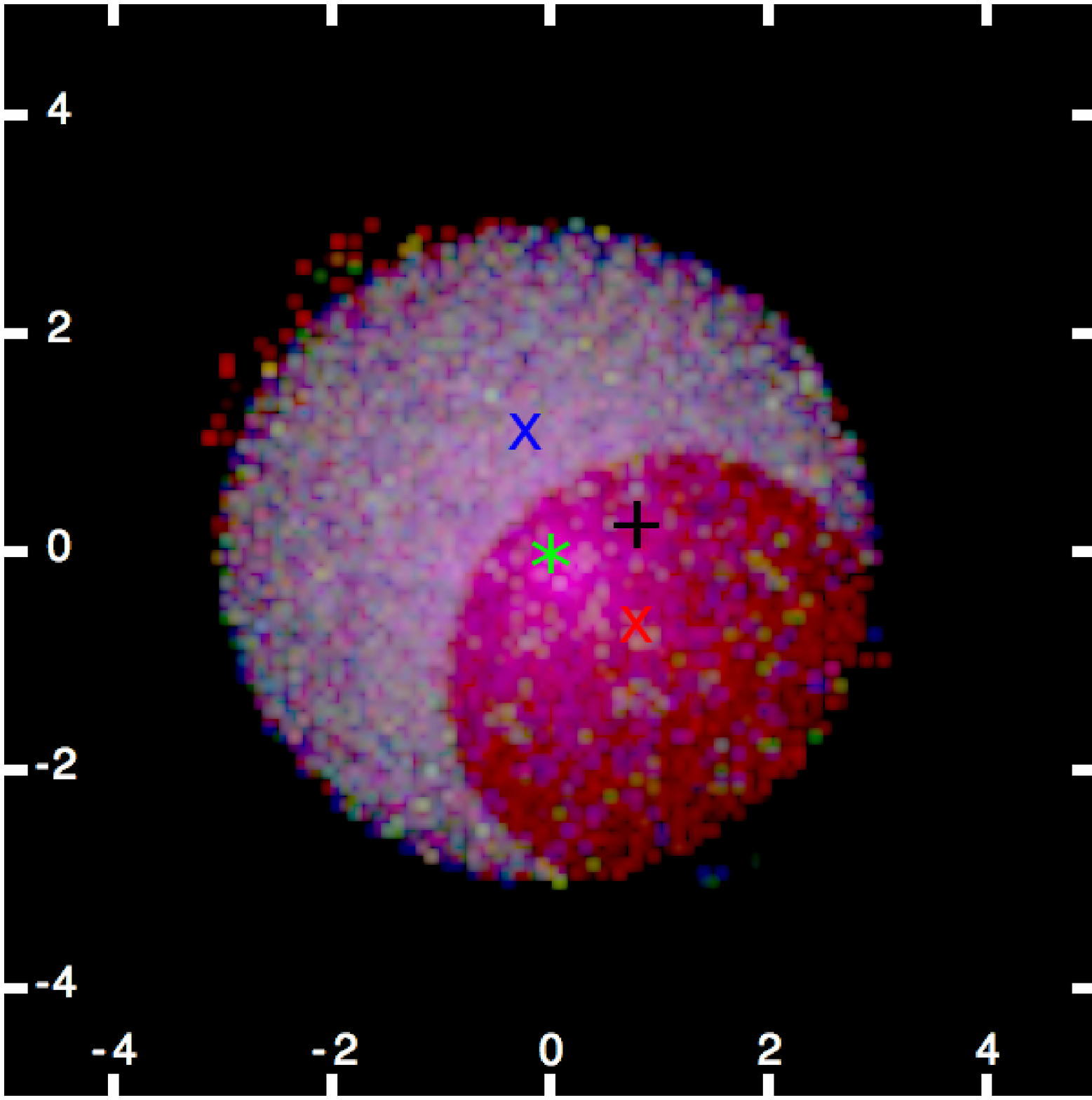}  \ \ \ &  
\includegraphics[width=0.45\hsize]{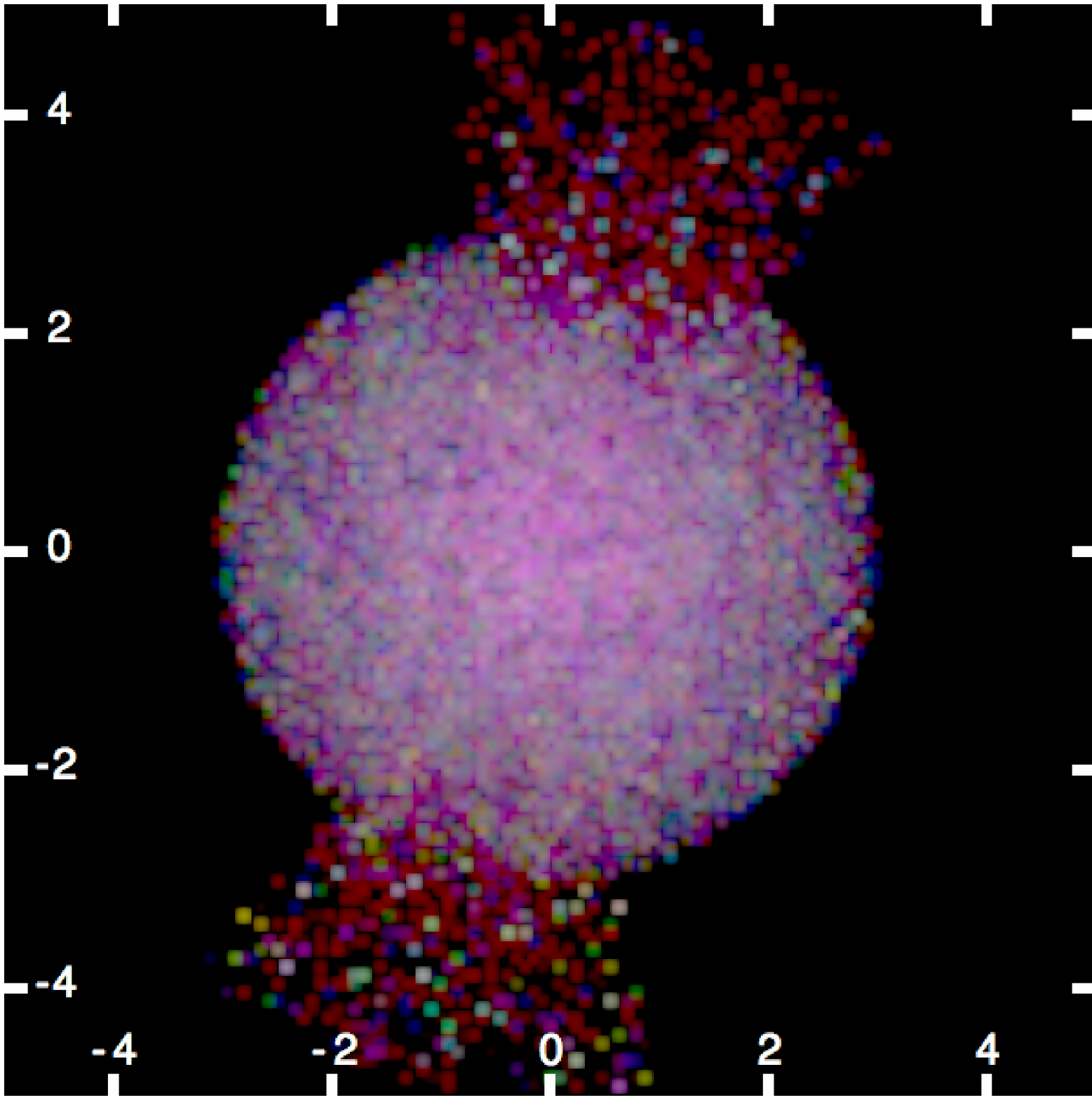} \\
\end{tabular}
\caption{
RGB SB map of the system, as viewed from our direction (\emph{left}), as well
as from the point of view of a fictional observer located at 90$^\circ$ with
respect to us. Photons escaping in the wavelength regions
$[1210,   1215.25]$ {\AA},
$[1215.25,1215.75]$ {\AA}, and
$[1215.75,1222]$    {\AA} are displayed in \emph{blue}, \emph{green}, and
\emph{red}, respectively. The image is centred on the centre of the galaxy
(\emph{green asterisk}, observationally given by the \OIII~ emission), and the
position of the QSO is marked by a \emph{black plus sign}. The positions on the
sky of the red and blue photons of \qso~ as inferred from triangulation are
indicated by the \emph{red} and \emph{blue crosses}, respectively.
The axis scale is in kpc.
}
\label{fig_RGB}
\renewcommand{\tabcolsep}{6pt}
\end{figure}

These effects are also seen in Fig.~\ref{fig_RGB}, where for comparison the
system is also shown as it would look if observed from the side.
Here, it becomes obvious that an outflowing jet partially covering the galaxy
may be responsible for the separation of the two parts of the spectrum. 
Interestingly, the dynamical age of the winds is found consistent with the age of 
the star formation activity.

\subsubsection*{Absorption properties along the QSO line of sight}
Due to the stochasticity of the exact positions of the clouds in the model
galaxy, the line of sight towards the quasar may or may not be representative
of the system as a whole. To this end, 
200 realisations of the favoured model have been carried out. This way, we
obtain confidence limits for the quantities ``observed'' along the line of
sight:

The number of clouds $N_{\mathrm{cl}}$ intercepted (either through the centre or the 
outskirts of a cloud) by the sightline is $5.2\pm2.2$. The corresponding 
total total column density through the galaxy is then $\log N(\HI) = 21.9\pm0.2$ 
and the colour excess is $E(B-V) = 0.037\pm0.015$. These values are well consistent with the 
observed ones.

By picking a few random realisation, having total hydrogen column density close to the 
measured value and uniform metallicity, we generated simulated absorption profiles of
\SiII$\lambda1808$. 
These patterns 
resemble well the observed profile with well defined components of comparable strengths 
in a rather compact configuration (Fig.~\ref{metalsUVES}). 
This exercise further supports the view that the neutral gas in DLAs may well be distributed 
over the velocity components seen in the metal absorption profiles. Because H$_2$ is usually 
detected associated to only a few components that constitute the metal profile 
\citep[e.g.][]{Petitjean02,Noterdaeme07lf}, this means that 
measured molecular fractions \citep[e.g.][]{Petitjean06} are lower limits to the actual value in 
the H$_2$-bearing clumps \citep[see also][]{Noterdaeme10co, Srianand11}.

\begin{figure}
\centering
\renewcommand{\tabcolsep}{1pt}
\begin{tabular}{cc}
\includegraphics[bb=219 230 393 628, clip=, angle=90, width=0.48\hsize]{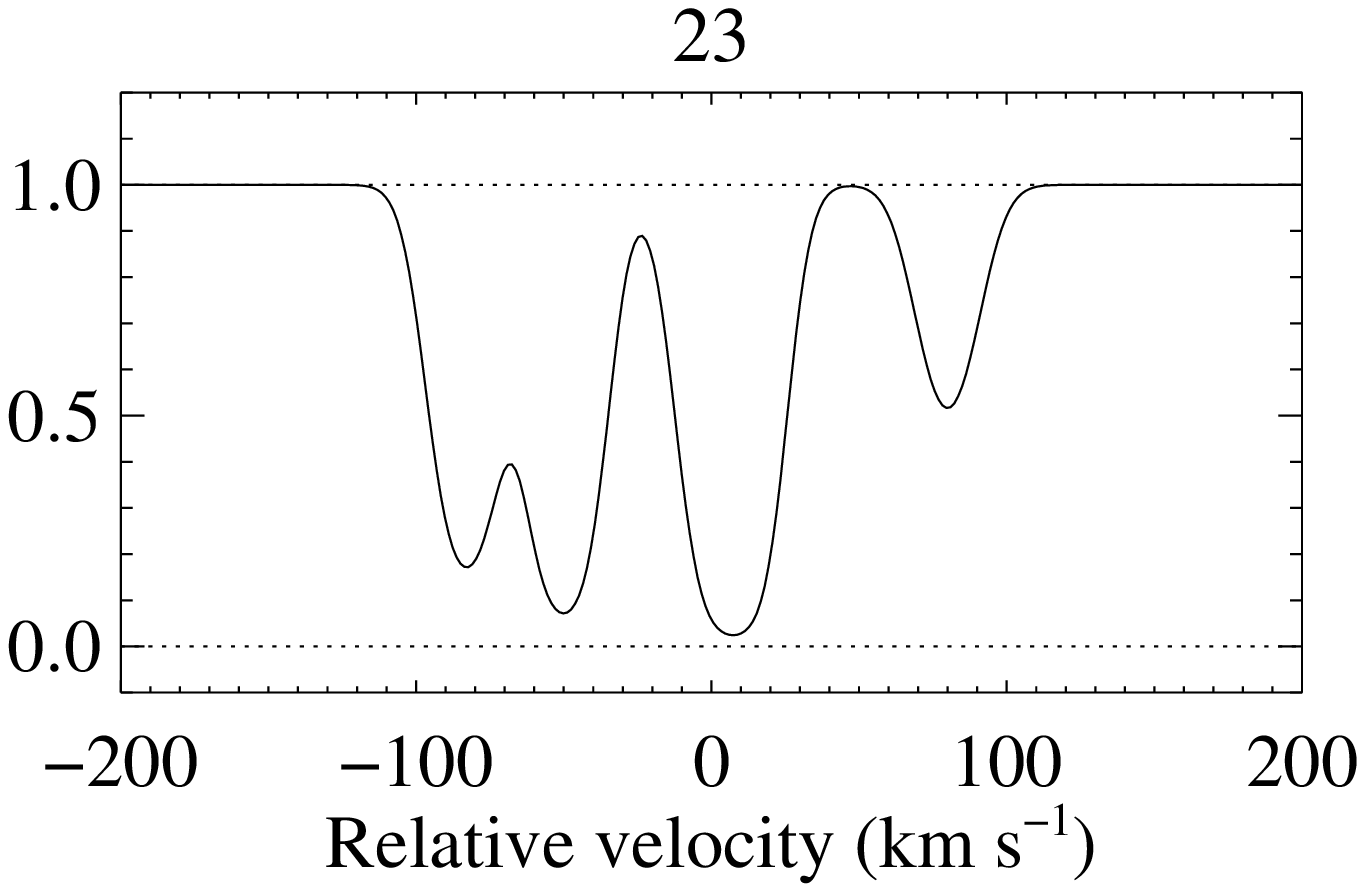}&
\includegraphics[bb=219 230 393 628, clip=, angle=90, width=0.48\hsize]{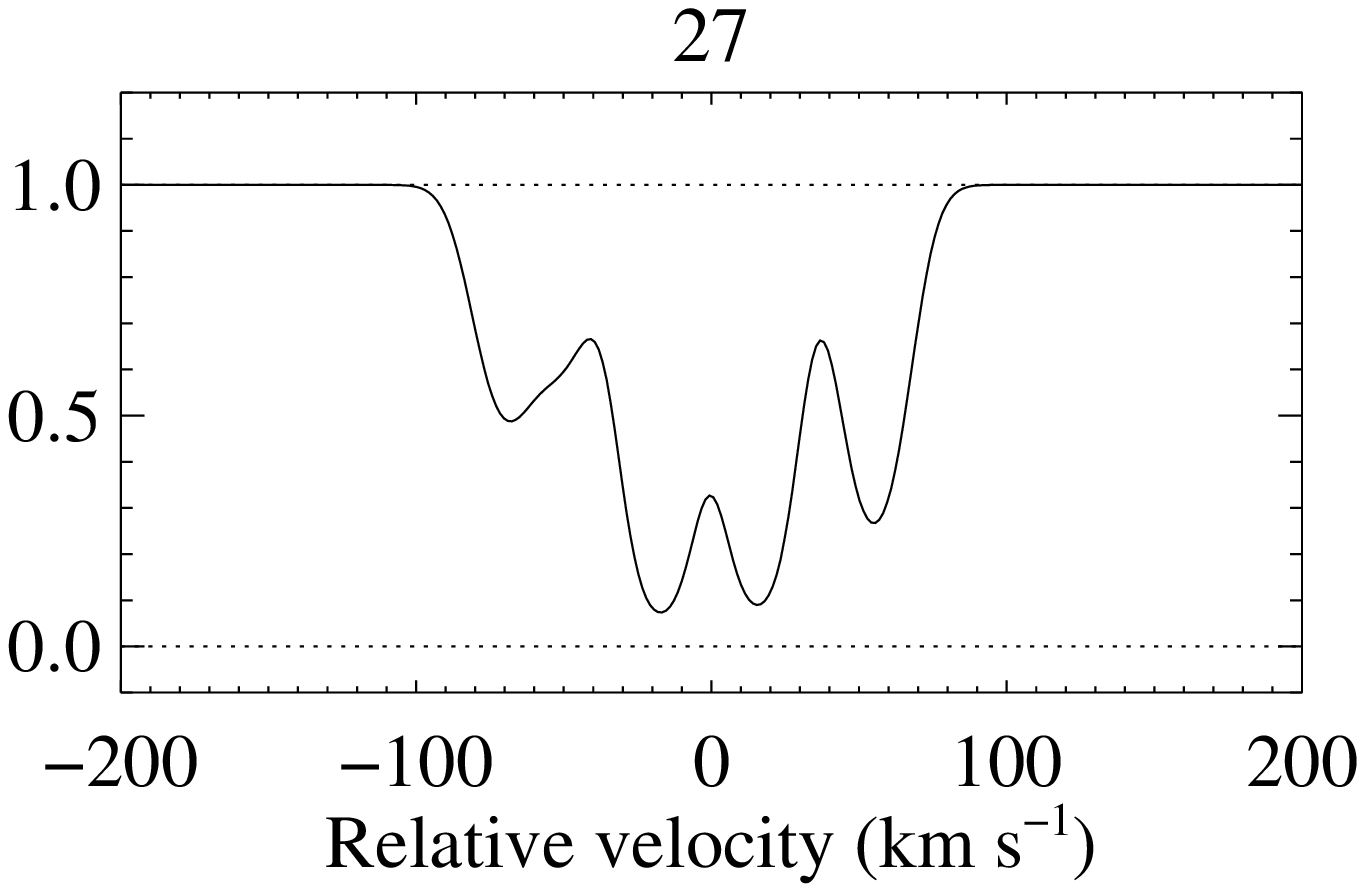}\\
\includegraphics[bb=219 230 393 628, clip=, angle=90, width=0.48\hsize]{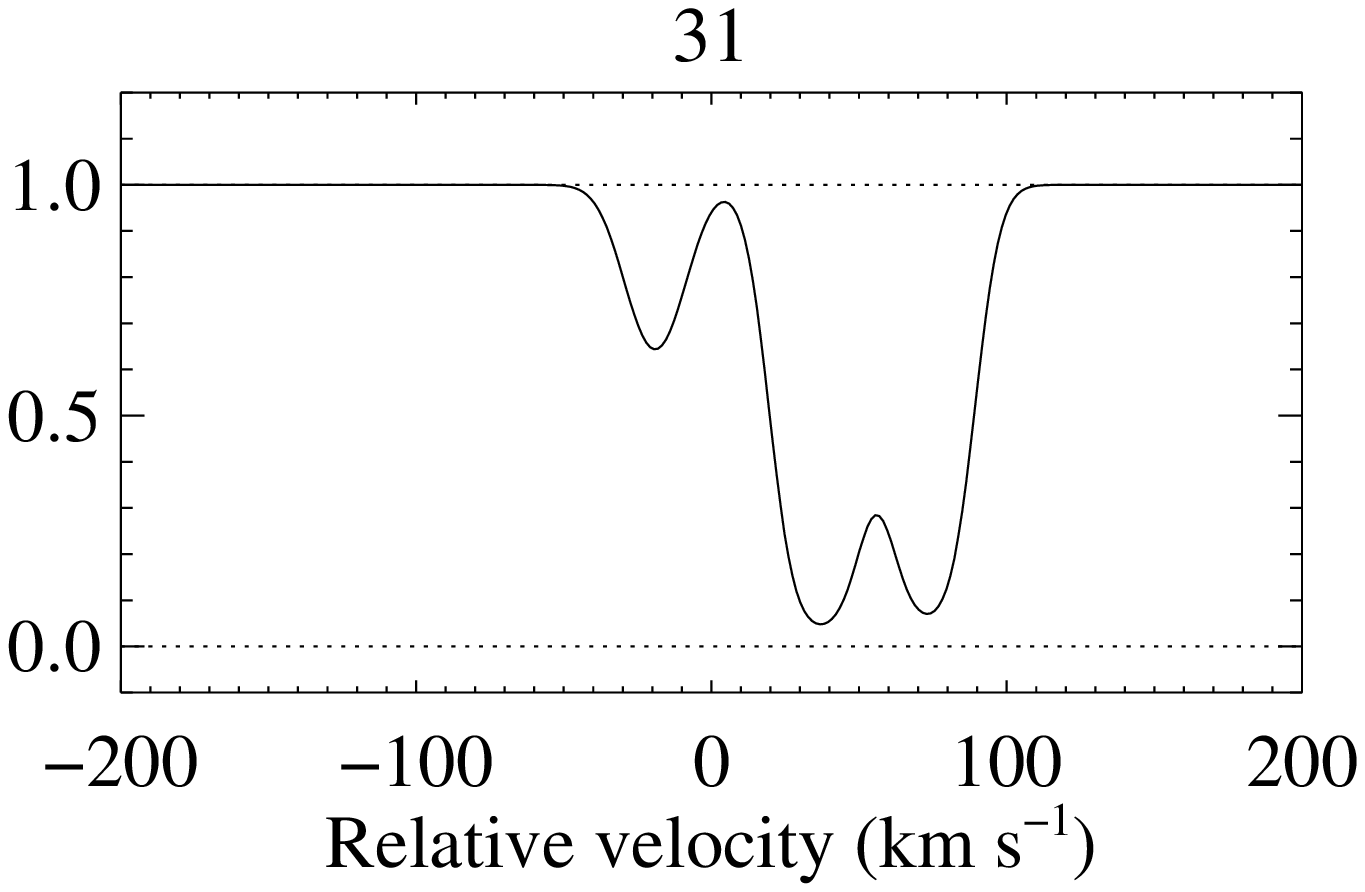}&
\includegraphics[bb=219 230 393 628, clip=, angle=90, width=0.48\hsize]{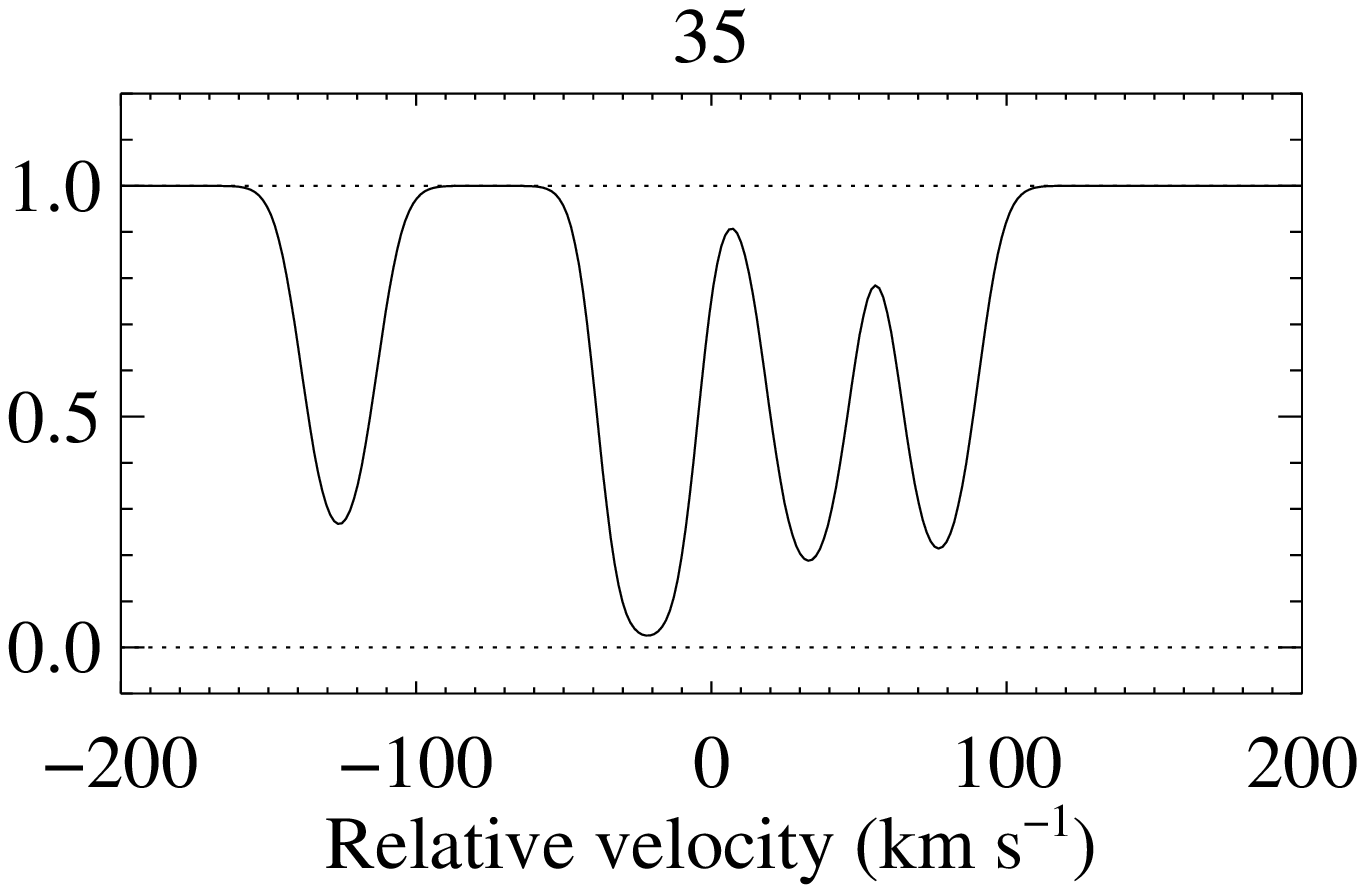}\\
\includegraphics[bb=165 230 393 628, clip=, angle=90, width=0.48\hsize]{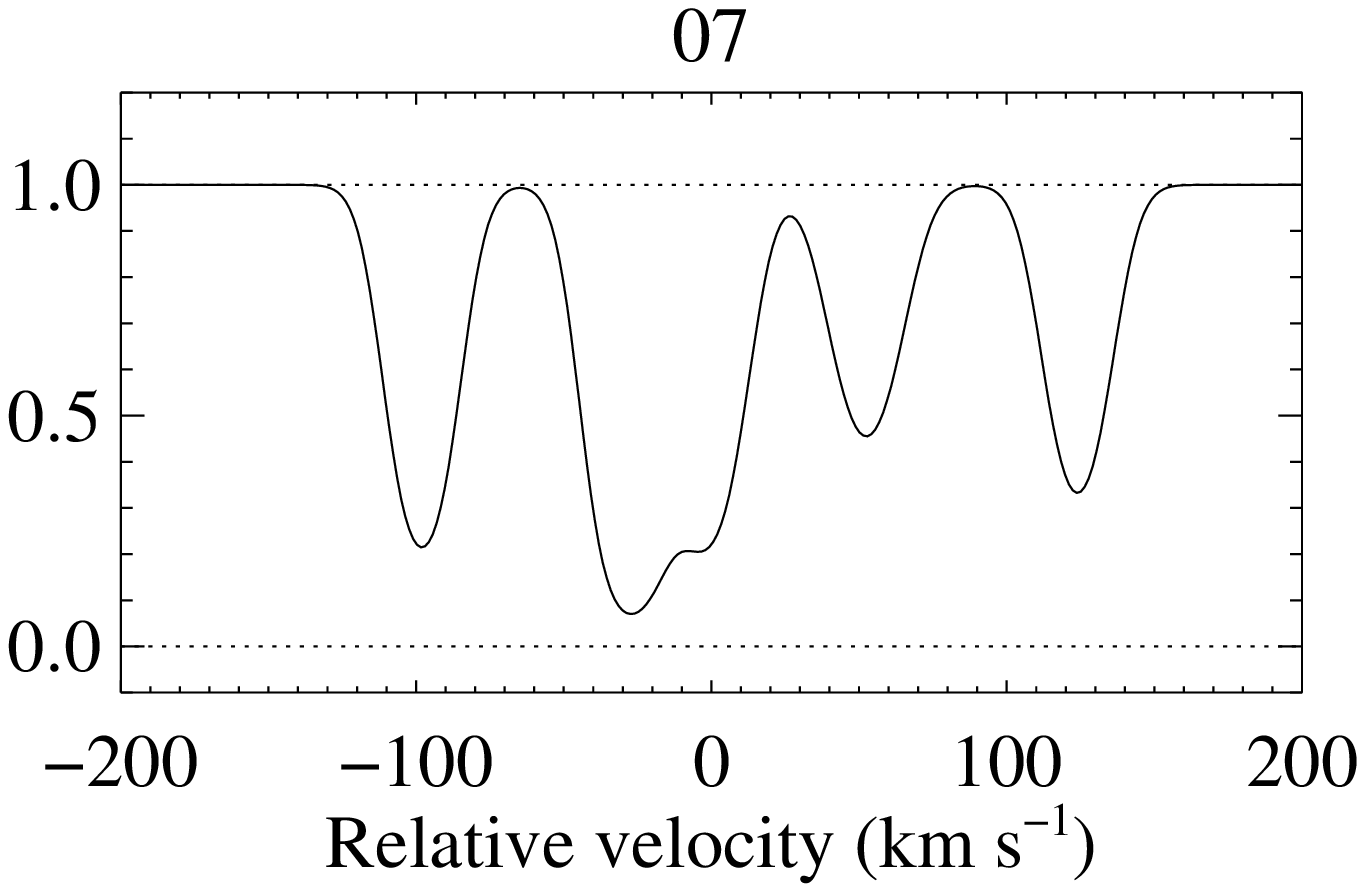}&
\includegraphics[bb=165 230 393 628, clip=, angle=90, width=0.48\hsize]{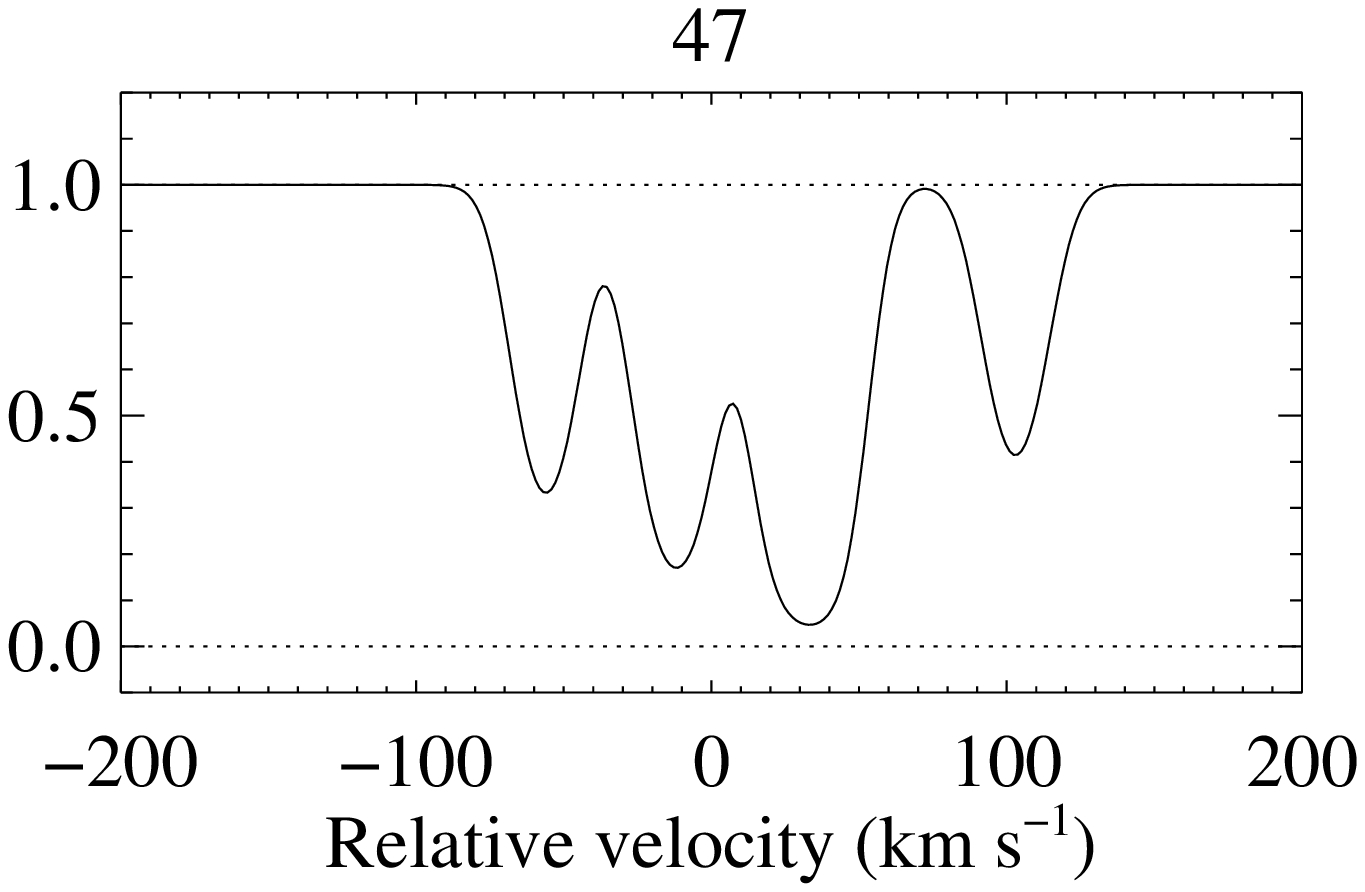}\\
\end{tabular}
\renewcommand{\tabcolsep}{6pt}
\caption{Random realisations of the \SiII$\lambda1808$ {\AA} line observed through
the simulated galaxy along the line of sight towards the quasar. A common
$b$-parameter of 10 \kms\ has been assumed for all clouds.}
\label{fig_SiII}
\end{figure}

\subsection{Degeneracies in the model}

Obviously, since we have not scanned the full parameter space, we cannot be
entirely sure that we have found a unique solution for the model, and indeed
some parameters are to a certain extent degenerate. However, in the process of
finding the model that simultaneously matches all our observables we have
tested several hundred different models, and in the end it became apparent that
it was not possible to vary a given parameter much before one or more of the
fits was clearly poor. For instance, the same value of $N(\HI)$ could be
obtained using a smaller (larger) number of higher (lower) density clouds.
However, not only does this affect $n_{\mathrm{cl}}$, but since most of the
scatterings take place inside the clouds, this will also affect the shape of
the emerging spectrum. Similarly, the amplitude difference of the two peaks
could also be obtained with larger (smaller) velocity \emph{difference} between
the jet outflow and the overall infall of gas. However, this will affect both
the velocity pattern of the clouds and the \lya\ escape fraction.

Finally, we note that a separation of the blue and red photons on the sky may
in principle also be caused by rotation of the whole system \citep[see Fig.~8 of][]{Zheng02a}. 
However, we aimed here at testing the picture first derived from the observed absorption 
and emission properties of the DLA galaxy (all consistent with a young outflowing 
starburst galaxy, relatively compact and of low mass), and not to test possible geometries that only 
explain the \lya\ emission alone.

\section{Conclusion: implications for the DLA-galaxy connection \label{concl}}
At low and intermediate redshifts, galaxies of all spectral types have been found associated to QSO 
absorption systems \citep[e.g.][]{Lebrun97,Bergeron91,Steidel95}, which were generally selected upon 
their \MgII\ absorption lines.
While strong \MgII\ are also likely DLAs \citep{Rao06}, little is known 
about the galaxies associated to bona-fide DLAs at high redshift. Because of their selection upon 
the cross-section of neutral gas only, we can expect that DLA arise from galaxies with a large diversity 
of morphological types \citep[see][for recent results at low redshift]{Battisti12}. 
It is thus very important to search for the associated galaxies and look for possible trends between the 
properties of the absorbers and their associated galaxies to start drawing a coherent picture of DLA galaxies.
 
Considering the velocity width of low-ionisation metal profiles, \citet{Ledoux06a} proposed that DLAs with 
higher metallicities are associated to more massive galaxies. This interpretation led \citeauthor{Fynbo10} to 
select high metallicity DLAs to search for associated galaxies. In three cases, DLA galaxies have been found 
at respectively 6, 16 and 4~kpc (\citealt{Fynbo10,Fynbo11}, Krogager et al., in prep), demonstrating the 
efficiency of the selection. Since there is some delay between the peak of star formation and the injection of 
metals in the ISM, we can expect that DLA galaxies selected this way have already passed through an epoch of 
star formation (interestingly, the abundances of nitrogen and oxygen reveal a picture in which DLAs undergo 
successive star bursts \citep{Petitjean08}). In addition, the large \HI\ extent of massive galaxies can 
explain the relatively large impact parameters found for high-metallicity selected DLA galaxies 
(Krogager et al., in prep).

The velocity spread of low-ionisation metal lines ($\Delta v \approx 180$~\kms) and the metallicity 
([Zn/H]$\approx -1.1$) place the DLA towards \qso\ in a intermediate regime on the metallicity-velocity 
correlation observed by \citet{Ledoux06a}. This is consistent with the intermediate mass derived for the 
DLA galaxy.
We measure a high star formation rate of $\sim25$~M$_\odot$\,yr$^{-1}$ at very small impact parameter 
from the QSO line of sight ($b\approx$~0.9~kpc). The SFR is much above the values measured in a population 
of extended, low surface brightness \lya\ emitters, recently proposed to be the optical counterparts of DLAs 
\citep{Rauch08}. The \lya\ luminosity is also much higher than the average \lya\ flux from DLAs obtained by 
stacking several hundreds of DLA fibre spectra from the SDSS \citep{Rahmani10} and the measured SFR is also 
well above the limit obtained from searches of extended continuum emission of DLAs by \citet{Wolfe06}. 
Clearly, the DLA galaxy studied here lays at the high-luminosity end of the general DLA-galaxy population. 
It is therefore important to understand how the absorption properties of the DLA can {reveal} the presence of 
an actively star-forming galaxy at small impact parameter.

From the absorption point of view, the DLA studied here has a relative metal and dust content typical of 
the general DLA population. Clearly, what makes the DLA towards \qso\ different 
from other DLAs is the large \HI\ column density  
which is, to our knowledge, the largest ever measured along a QSO line of sight. 
If intervening absorption systems follow some kind of Schmidt-Kennicutt law \citep[see e.g.][]{Chelouche10}, then we 
can expect that high column density systems are, among different classes of absorbers, those most closely 
associated with star-forming regions. Indeed, high surface density is expected to trigger star formation 
\citep{Schaye04} since the high column density \HI\ gas can collapse into cold and molecular gas 
subsequently feeding the star formation. According to \citet{Schaye01}, this could provide a natural 
explanation\footnote{We note however that while a transition is seen in the molecular fraction at $\log N(\HI)=21$ in 
our galaxy \citep{savage77}, this transition is not readily seen in DLAs \citep{Noterdaeme08} nor in SMC 
\citep{Tumlinson02}, where it probably occurs at higher column densities. See also \citet{Erkal12}.} 
to the steep slope seen in the \HI-frequency distribution at high column density \citep{Noterdaeme09dla}, 
though a dust bias against high column density systems is not excluded \citep[][]{Pei95}. 

In addition, high \HI\ column density systems are frequently seen 
at the redshift of long-duration gamma ray burst (GRB) afterglows 
\citep[e.g.][]{Vreeswijk04,Chen05,Berger06,Jakobsson06,Ledoux09,Fynbo09}. 
Because there is accumulating evidence that these phenomena are connected to the death of massive stars 
\citep[see e.g.][]{Galama98,Bloom99,Fruchter06}, it is expected that GRB explosions occur in star-forming regions.  
Interestingly, the DLA galaxy studied here also appears to share some properties with at least 
a fraction of GRB host galaxies: namely being a young compact star-forming galaxy 
\citep[see e.g.][]{Savaglio09,Levesque10}.

Selecting the most gas-rich DLAs could therefore be an efficient way to detect the associated galaxies 
at small impact parameter and during a period of active star formation. With the steadily increasing number 
of known DLAs, especially thanks to the SDSS \citep[see][]{Prochaska05,Prochaska09a,Noterdaeme09dla} and soon 
to the ongoing Baryon Oscillation Spectroscopy Survey \citep{Eisenstein11}, we can hope to test this hypothesis 
further in the near future.

\acknowledgement{
We warmly thank the Chilean National Time Allocation Committee for allowing 
us to perform the observations with MagE and Las Campanas Observatory staff for support. 
We are grateful to the ESO Director Discretionary Time allocation committee and 
the ESO Director General, Tim de Zeeuw, for allowing us to carry out the X-shooter observations. 
We thank Paranal Observatory staff for carrying out the X-shooter observations in service mode, 
Guido Cupani for help with treating the X-shooter data and Lorena Fuentealba for help with 
graphical issues.
We made use of the ESO Science Archive Facility and of the Sloan Digital Sky Survey database.
PN acknowledges support from CONICYT-CNRS and kind hospitality by the {\sl Departamento de Astronom\'ia de 
la Universidad de Chile} during the time part of this work was done. PL acknowledges funding from the 
Villum Foundation.
PPJ and RS gratefully acknowledge support from the Indo-French Centre for the Promotion of Advanced 
Research ({\sl Centre Franco-Indien pour la promotion de la recherche avanc\'ee}) under Project N.4304-2. 
MJM was supported by a Master scholarship through FONDECYT grant No. 1100214 and is now supported by 
a Fulbright-CONICYT PhD fellowship. 
JPUF acknowledges support from the ERC-StG grant EGGS-278202. 
The Dark cosmology centre is funded by the DNRF. SL was supported by FONDECYT grant No. 1100214.}



\end{document}